\documentstyle[12pt]{article}

\catcode`\@=11
\long\def\@makefntext#1{ 
\protect\noindent \hbox to 3.2pt {\hskip-.9pt
$^{{\ninerm\@thefnmark}}$\hfil}#1\hfill} 

\def\thefootnote{\fnsymbol{footnote}}

 \def\@makefnmark{\hbox to 0pt{$^{\@thefnmark}$\hss}}  

\def\ps@myheadings{\let\@mkboth\@gobbletwo
\def\@oddhead{\hbox{} 
\rightmark\hfil\ninerm\thepage}
\def\@oddfoot{}\def\@evenhead{\ninerm\thepage\hfil 
\leftmark\hbox{}}\def\@evenfoot{}
\def\sectionmark##1{}\def\subsectionmark##1{}}

\def\CVector#1#2#3{
\left\vert\begin{array}{c}
                 #1 \\
                 #2
                \end{array}, #3\right>}

\def\AVector#1#2#3{\left|  #1, #2; #3 \right>}

\begin{document}

\newcommand{\symbolfootnote}{\renewcommand{\thefootnote}
        {\fnsymbol{footnote}}}
\renewcommand{\thefootnote}{\fnsymbol{footnote}}
\newcommand{\alphfootnote}
        {\setcounter{footnote}{0}
         \renewcommand{\thefootnote}{\sevenrm\alph{footnote}}}

\newcounter{sectionc}\newcounter{subsectionc}\newcounter{subsubsectionc}
\renewcommand{\section}[1] {\vspace{0.6cm}\addtocounter{sectionc}{1}
\setcounter{subsectionc}{0}\setcounter{subsubsectionc}{0}\noindent
        {\bf\thesectionc. #1}\par\vspace{0.4cm}}
\renewcommand{\subsection}[1] {\vspace{0.6cm}\addtocounter{subsectionc}{1}
        \setcounter{subsubsectionc}{0}\noindent
        {\it\thesectionc.\thesubsectionc. #1}\par\vspace{0.4cm}}
\renewcommand{\subsubsection}[1]
{\vspace{0.6cm}\addtocounter{subsubsectionc}{1}
        \noindent {\rm\thesectionc.\thesubsectionc.\thesubsubsectionc.
        #1}\par\vspace{0.4cm}}
\newcommand{\nonumsection}[1] {\vspace{0.6cm}\noindent{\bf #1}
        \par\vspace{0.4cm}}

\newcounter{appendixc}
\newcounter{subappendixc}[appendixc]
\newcounter{subsubappendixc}[subappendixc]
\renewcommand{\thesubappendixc}{\Alph{appendixc}.\arabic{subappendixc}}
\renewcommand{\thesubsubappendixc}
        {\Alph{appendixc}.\arabic{subappendixc}.\arabic{subsubappendixc}}

\renewcommand{\appendix}[1] {\vspace{0.6cm}
        \refstepcounter{appendixc}
        \setcounter{figure}{0}
        \setcounter{table}{0}
        \setcounter{equation}{0}
        \renewcommand{\thefigure}{\Alph{appendixc}.\arabic{figure}}
        \renewcommand{\thetable}{\Alph{appendixc}.\arabic{table}}
        \renewcommand{\theappendixc}{\Alph{appendixc}}
        \renewcommand{\theequation}{\Alph{appendixc}.\arabic{equation}}
        \noindent{\bf Appendix \theappendixc #1}\par\vspace{0.4cm}}
\newcommand{\subappendix}[1] {\vspace{0.6cm}
        \refstepcounter{subappendixc}
        \noindent{\bf Appendix \thesubappendixc. #1}\par\vspace{0.4cm}}
\newcommand{\subsubappendix}[1] {\vspace{0.6cm}
        \refstepcounter{subsubappendixc}
        \noindent{\it Appendix \thesubsubappendixc. #1}
        \par\vspace{0.4cm}}

\def\abstracts#1{{
        \centering{\begin{minipage}{30pc}\tenrm\baselineskip=12pt\noindent
        \centerline{\tenrm ABSTRACT}\vspace{0.3cm}
        \parindent=0pt #1
        \end{minipage} }\par}}

\newcommand{\bibit}{\it}
\newcommand{\bibbf}{\bf}
\renewenvironment{thebibliography}[1]
        {\begin{list}{\arabic{enumi}.}
        {\usecounter{enumi}\setlength{\parsep}{0pt}
\setlength{\leftmargin 1.25cm}{\rightmargin 0pt}
         \setlength{\itemsep}{0pt} \settowidth
        {\labelwidth}{#1.}\sloppy}}{\end{list}}

\topsep=0in\parsep=0in\itemsep=0in
\parindent=1.5pc

\newcounter{itemlistc}
\newcounter{romanlistc}
\newcounter{alphlistc}
\newcounter{arabiclistc}
\newenvironment{itemlist}
        {\setcounter{itemlistc}{0}
         \begin{list}{$\bullet$}
        {\usecounter{itemlistc}
         \setlength{\parsep}{0pt}
         \setlength{\itemsep}{0pt}}}{\end{list}}

\newenvironment{romanlist}
        {\setcounter{romanlistc}{0}
         \begin{list}{$($\roman{romanlistc}$)$}
        {\usecounter{romanlistc}
         \setlength{\parsep}{0pt}
         \setlength{\itemsep}{0pt}}}{\end{list}}

\newenvironment{alphlist}
        {\setcounter{alphlistc}{0}
         \begin{list}{$($\alph{alphlistc}$)$}
        {\usecounter{alphlistc}
         \setlength{\parsep}{0pt}
         \setlength{\itemsep}{0pt}}}{\end{list}}

\newenvironment{arabiclist}
        {\setcounter{arabiclistc}{0}
         \begin{list}{\arabic{arabiclistc}}
        {\usecounter{arabiclistc}
         \setlength{\parsep}{0pt}
         \setlength{\itemsep}{0pt}}}{\end{list}}

\newcommand{\fcaption}[1]{
        \refstepcounter{figure}
        \setbox\@tempboxa = \hbox{\tenrm Fig.~\thefigure. #1}
        \ifdim \wd\@tempboxa > 6in
           {\begin{center}
        \parbox{6in}{\tenrm\baselineskip=12pt Fig.~\thefigure. #1 }
            \end{center}}
        \else
             {\begin{center}
             {\tenrm Fig.~\thefigure. #1}
              \end{center}}
        \fi}

\newcommand{\tcaption}[1]{
        \refstepcounter{table}
        \setbox\@tempboxa = \hbox{\tenrm Table~\thetable. #1}
        \ifdim \wd\@tempboxa > 6in
           {\begin{center}
        \parbox{6in}{\tenrm\baselineskip=12pt Table~\thetable. #1 }
            \end{center}}
        \else
             {\begin{center}
             {\tenrm Table~\thetable. #1}
              \end{center}}
        \fi}

\def\@citex[#1]#2{\if@filesw\immediate\write\@auxout
        {\string\citation{#2}}\fi
\def\@citea{}\@cite{\@for\@citeb:=#2\do
        {\@citea\def\@citea{,}\@ifundefined
        {b@\@citeb}{{\bf ?}\@warning
        {Citation `\@citeb' on page \thepage \space undefined}}
        {\csname b@\@citeb\endcsname}}}{#1}}

\newif\if@cghi
\def\cite{\@cghitrue\@ifnextchar [{\@tempswatrue
        \@citex}{\@tempswafalse\@citex[]}}
\def\citelow{\@cghifalse\@ifnextchar [{\@tempswatrue
        \@citex}{\@tempswafalse\@citex[]}}
\def\@cite#1#2{{$\null^{#1}$\if@tempswa\typeout
        {IJCGA warning: optional citation argument
        ignored: `#2'} \fi}}
\newcommand{\citeup}{\cite}

\def\fnm#1{$^{\mbox{\scriptsize #1}}$}
\def\fnt#1#2{\footnotetext{\kern-.3em
        {$^{\mbox{\sevenrm #1}}$}{#2}}}

\font\twelvebf=cmbx10 scaled\magstep 1
\font\twelverm=cmr10 scaled\magstep 1
\font\twelveit=cmti10 scaled\magstep 1
\font\elevenbfit=cmbxti10 scaled\magstephalf
\font\elevenbf=cmbx10 scaled\magstephalf
\font\elevenrm=cmr10 scaled\magstephalf
\font\elevenit=cmti10 scaled\magstephalf
\font\bfit=cmbxti10
\font\tenbf=cmbx10
\font\tenrm=cmr10
\font\tenit=cmti10
\font\ninebf=cmbx9
\font\ninerm=cmr9
\font\nineit=cmti9
\font\eightbf=cmbx8
\font\eightrm=cmr8
\font\eightit=cmti8


\centerline{\tenbf QUANTUM ALGEBRAS IN NUCLEAR STRUCTURE}
\vspace{0.8cm}
\centerline{\tenrm Dennis BONATSOS}
\baselineskip=13pt
\centerline{\tenit ECT$^*$, Villa Tambosi, Strada delle Tabarelle 286}
\baselineskip=12pt
\centerline{\tenit I-38050 Villazzano (Trento), Italy}
\vspace{0.3cm}
\centerline{\tenrm C. DASKALOYANNIS}
\baselineskip=13pt
\centerline{\tenit Department of Physics, Aristotle University of Thessaloniki}
\baselineskip=12pt
\centerline{\tenit GR-54006 Thessaloniki, Greece}
\vspace{0.3cm}
\centerline{\tenrm P. KOLOKOTRONIS, D. LENIS}
\baselineskip=13pt
\centerline{\tenit Institute of Nuclear Physics, NCSR ``Demokritos''}
\baselineskip=12pt
\centerline{\tenit GR-15310 Aghia Paraskevi, Attiki, Greece}
\vspace{0.9cm}
\abstracts{
Quantum algebras are a mathematical tool which provides us with a class of
symmetries wider than that of Lie algebras, which are contained in the former
as a special case. After a self-contained introduction to the necessary
mathematical tools ($q$-numbers, $q$-analysis, $q$-oscillators, $q$-algebras),
the su$_q$(2) rotator model and its extensions, the construction of deformed
exactly soluble models (Interacting Boson Model, Moszkowski model), the use of
deformed bosons in the description of pairing correlations, and the symmetries
of the anisotropic quantum harmonic oscillator with rational ratios of
frequencies, which underly the structure of superdeformed and hyperdeformed
nuclei, are discussed in some detail. A brief description of similar
applications to molecular structure and an outlook are also given.
}

\vfil
\rm\baselineskip=14pt
\section{Introduction}

Quantum algebras \cite{Dri798,Jim247,Jim10,KR101,Skl262} (also called quantum
groups) are deformed versions of the
usual Lie algebras, to which they reduce when the deformation parameter
$q$ is set equal to unity. From the mathematical point of view they are
Hopf algebras \cite{Abe1977}.
Their use in physics became popular with the
introduction of the $q$-deformed harmonic oscillator (sec. 10) as a tool for
providing a boson realization of the quantum algebra su$_q$(2) (sec. 14),
although
similar mathematical structures had already been known (sec. 11).
Initially used for solving the quantum Yang--Baxter equation \cite{Jim10},
quantum algebras
have subsequently found applications in several branches of physics, as, for
example, in the description of spin chains, squeezed states, rotational
and vibrational nuclear and molecular spectra, and in conformal
field theories. By now several kinds of generalized deformed oscillators
(sec. 12)
and generalized deformed su(2) algebras (sec. 17)  have been introduced.

It is clear that quantum algebras provide us with a class of symmetries
which is richer than the class of Lie symmetries, which are contained
in the former as a special case. It is therefore conceivable that quantum
algebras can turn out to be appropriate for describing symmetries of
physical systems which are outside the realm of Lie algebras.

Here we shall confine ourselves to applications of quantum algebras in nuclear
structure physics. The structure of this review is as follows:
 In order to make this review self-contained, we are going
first to give a brief account of the necessary tools: $q$-numbers and
$q$-analysis (secs 2--9), $q$-deformed oscillators (secs 10--13), $q$-deformed
su(2) algebras (secs 14--18).
The remainder will be devoted to specific applications in nuclear
structure problems, starting with phenomenology and advancing towards more
microscopic subjects.  The su$_q$(2) rotator
model (secs 19--23) and its extensions (secs 24--26),
as well as  the formulation of deformed exactly soluble
models (Interacting Boson Model (secs 27--29), Moszkowski model (sec. 30))
will be covered in some detail. Subsequently, the use of quantum algebraic
techniques for the description of pairing correlations in nuclei (secs
31--33),
as well as the symmetries of the anisotropic quantum harmonic oscillators
with rational ratios of frequencies (sec. 34) will also be considered in some
detail. The latter are of current interest in connection with the symmetries
underlying superdeformed and hyperdeformed nuclear bands (sec. 34).
Finally, a brief account of applications of the same techniques to molecular
structure (sec. 35) and an outlook (sec. 36) will be given.

\section{$q$-numbers}

The
{\sl $q$-number} corresponding to the ordinary number $x$  is defined as
$$ [x]= {q^x-q^{-x}\over q-q^{-1}},\eqno(2.1)$$
where $q$ is a parameter. The same definition holds if $x$ is an operator.
We remark that $q$-numbers remain invariant under the substitution
$q\rightarrow q^{-1}$.

If $q$ is real, $q$-numbers can easily be put in the form
$$ [x] = {\sinh(\tau x)\over \sinh(\tau)}, \eqno(2.2)$$
where $q=e^{\tau}$ and $\tau$ is real.

If $q$ is a phase factor, $q$-numbers can be written as
$$ [x]= {\sin(\tau x)\over \sin(\tau)},\eqno(2.3)$$
where $q=e^{i\tau}$ and $\tau$ is real.

In both cases it is clear that in the limit $q\to 1$ (or, equivalently,
$\tau\to 0$) $q$-numbers (or operators) tend to the ordinary numbers
(or operators):
$$ \lim_{q\to 1} [x] = x .\eqno(2.4)$$

A few examples of $q$-numbers are given here:
$$ [0]=0,\qquad [1] =1, \qquad [2]= q+q^{-1}, \qquad [3] = q^2 +1+q^{-2}.
\eqno(2.5)$$

Identities between $q$-numbers exist. They are, however, different from the
familiar identities between usual numbers. As an exercise one can show
(using the definition of $q$-numbers) that
$$ [a] [b+1]-[b] [a+1] = [a-b].\eqno(2.6)$$

The {\sl $q$-factorial} of an integer $n$ is defined as
$$ [n]!=[n] [n-1] \ldots [2] [1].\eqno(2.7)$$

The {\sl $q$-binomial coefficients} are defined as
$${m \brack n} ={[m]!\over [m-n]! [n]!},\eqno(2.8)$$
while the {\sl $q$-binomial expansion} is given by
$$ [a \pm b]^m = \sum_{k=0}^m {m\brack k} a^{m-k} (\pm b)^k.\eqno(2.9)$$

In the limit $q\to 1$ we obviously have
$$ [n]!\to n! \qquad {\rm and} \qquad {m\brack n} \to {m\choose n},
\eqno(2.10)$$
where $n!$ and ${m\choose n}$ are the standard factorial and binomial
coefficients respectively.

It should be noticed that two-paremeter deformed numbers have also been
introduced
$$ [x]_{p,q}= {q^x-p^{-x}\over q-p^{-1} }.\eqno(2.11)$$
In the special case $p=q$ they reduce to the usual $q$-numbers.

\section{$q$-deformed elementary functions}

In addition to $q$-deformed numbers and operators, $q$-deformed elementary
functions can be introduced. The {\sl $q$-exponential function} is defined as
$$ e_q (a x) = \sum_{n=0}^{\infty}  {a^n\over [n]!} x^n ,\eqno(3.1)$$
while the {\sl $q$-trigonometric functions} are defined as
$$ \sin_q (x)= \sum_{n=0}^{\infty} (-1)^n {x^{2n+1}\over [2n+1]!},\eqno(3.2)$$
$$ \cos_q (x)= \sum_{n=0}^{\infty} (-1)^n {x^{2n}\over [2n]!}.\eqno(3.3)$$
It should also be noticed that $q$-deformed polynomials, such as
$q$-deformed Hermite polynomials \cite{FV45,VdJ267,CGY1517}
and $q$-deformed Laguerre polynomials \cite{FV45}
also exist (see also subsec. 34.3).

\section{$q$-derivatives}

Proceeding along this path one can build a new differential calculus, based on
$q$-deformed quantities (see \cite{GN945,BAZG1379} for concise expositions).
For this purpose the {\sl $q$-derivative}
is defined as
$$ D^q_x f(x) = {f(qx)-f(q^{-1} x)\over (q-q^{-1})x}.\eqno(4.1)$$
The similarity between the present definition and the one of $q$-numbers
(eq.(2.1)) is clear.

Using the definition of the $q$-derivative one can easily see that
$$ D^q_x (a x^n) = a [n] x^{n-1},\eqno(4.2)$$
$$ D^q_x e_q(ax) = a e_q(ax).\eqno(4.3)$$
{}From the definition of the $q$-derivative one can further derive
the sum rule
$$ D_x^q (f(x)+g(x)) = D_x^q f(x)+ D_x^q g(x),\eqno(4.4)$$
as well as the rule
$$D^q_{x_2} [ a x_1 \pm b x_2]^m =\pm [m] b [a x_1 \pm b x_2]^{m-1},
\eqno(4.5)$$
where $a$ and $b$ are costants and $[a x_1 \pm b x_2]^m$ is given
by the $q$-binomial expansion (eq. (2.9)). One can also prove the
{\sl $q$-integration
by parts} formula
$$ D_x^q (f(x)g(x)) = { f(q x) g(q x) - f(q^{-1}x) g(q^{-1}x)\over
(q-q^{-1})x }.\eqno(4.6) $$
{}From this, the following two forms of the Leibnitz rule can be derived
$$ D_x^q (f(x)g(x))= (D_x^q f(x)) g(q^{-1}x)+ f(qx) (D_x^q g(x)),\eqno(4.7)$$
$$ D_x^q (f(x)g(x))= (D_x^q g(x)) f(q^{-1}x)+ g(qx) (D_x^q f(x)).\eqno(4.8)$$
In addition one can show the property
$$D^q_x f(qx) = q D^q_x f(x)\vert_{x=qx},\eqno(4.9)$$
and the chain rules
$$D^q_{ax} f(x) = {1\over a} D_x^q f(x)\qquad,\eqno(4.10)$$
$$ D^q_x f(x^n) = [n] x^{n-1} D^{q^n}_{x^n} f(x^n),\eqno(4.11)$$
where $a$ is a constant.
Another useful result is
$$ D^{q^n}_x f(x)= {1\over [n]} \sum_{k=0}^{n-1} D^q_x f(q^{2k-(n-1)} x).
\eqno(4.12)$$

\section{ $q$-integration}

The  $q$-integration (see \cite{GN945,BAZG1379} for concise expositions)
in the interval $[0,a]$ is defined by
$$\int_0^a f(x) d_qx = a(q^{-1}-q)\sum_{n=0}^{\infty} q^{2n+1} f(q^{2n+1}a),
\eqno(5.1)$$
while for the interval $[0,\infty)$ one has
$$\int_0^{\infty} f(x) d_qx = (q^{-1}-q)\sum_{n=-\infty}^{\infty}
q^{2n+1} f(q^{2n+1}).\eqno(5.2)$$
The {\sl indefinite $q$-integral} is defined as
$$ \int f(x) d_qx= (q^{-1}-q)\sum_{n=0}^{\infty}  q^{2n+1}  x f(q^{2n+1} x)
+ \rm{constant},\eqno(5.3)$$
where $0 < q <1$. For entire functions  $f(x)$ one can easily see that
this $q$-integral approaches the Riemann integral as $q\to 1$, and also
that the operators of $q$-dif\-fer\-ent\-iat\-ion and
$q$-integration are inverse
to each other
$$ D_x^q \int f(x) d_qx = f(x) = \int D_x^q f(x) d_qx .\eqno(5.4)$$
One can also easily see that
$$\int a x^{n-1}  d_qx = {1\over [n]} a x^n + \rm{constant}, \eqno(5.5)$$
$$ \int e_q (ax) d_qx= {1\over a} e_q(ax) +\rm{constant}.\eqno(5.6)$$
{}From (4.6) one can also prove the following formulae of integration
by parts
$$\int_0^a f(qx) (D_x^q g(x)) d_qx= f(x)g(x) \vert_{x=0}^{x=a} -
\int_0^a (D_x^q f(x)) g(q^{-1}x) d_q x,\eqno(5.7)$$
$$\int_0^a f(q^{-1}x) (D_x^q g(x)) d_q x = f(x) g(x) \vert_{x=0}^{x=a}
-\int_0^a (D_x^q f(x)) g(qx) d_qx.\eqno(5.8)$$

The following formulae can also be proved
$$\int f(x) d_{aq}x = a\int f(x) d_qx , \eqno(5.9)$$
$$ \int f(x^n)  d_{q^n} x^n = [n] \int x^{n-1} f(x^n) d_qx, \eqno(5.10)$$
$$\int f(x) d_qx= {1\over [n]} \sum_{k=0}^{n-1} q^{2k-(n-1)} \int
f(q^{2k-(n-1)}x) d_{q^n}x.\eqno(5.11)$$

The {\sl $q$-analogue for Euler's formula} for the function $\Gamma (x)$ is
$$\int_0^{\zeta} e_q(-x) x^n d_qx= [n] [n-1] [n-2] \ldots [1] =[n]!.
\eqno(5.12)$$
A proof of this formula can be found in \cite{GN945}.

\section{$Q$-numbers}

The definition of $q$-numbers given in sec. 2 is not the only possible one.
Much literature exists \cite{Exton,Andrews}
using the definition of {\sl $Q$-numbers}
$$ [x]_Q = {Q^x - 1 \over Q - 1}, \eqno(6.1)$$
where $x$ can be a number or an operator and $Q$ is a deformation parameter.
$Q$ is a real number ($Q\ne 0,1$). The notation $Q=e^T$, where $T$ a real
number, will be often used.
The subscript $Q$ will be used in this review in order to distinguish
deformed numbers defined as in eq. (6.1) from these defined by eq. (2.1).
It is clear that in the limit $Q\to 1$ (or, equivalently, $T\to 0$)
$Q$-numbers become ordinary numbers, i.e. $[x]_Q \to x$.

A few examples of $Q$-numbers are given here:
$$ [0]_Q=0, \qquad [1]_Q=1,\qquad [2]_Q= Q+1, \qquad [3]_Q= Q^2+Q+1.
\eqno(6.2)$$

$Q$-numbers clearly do not remain invariant under the substitution
$Q\rightarrow Q^{-1}$. One can easily prove that
$$ \left[x\right]_Q = Q^{x-1} \left[x\right]_{1/Q}.\eqno(6.3)$$

$Q$-numbers are connected to $q$-numbers through the relation \cite{GN945}
$$ [x]= q^{1-x}[x]_Q, \qquad {\rm with} \qquad Q=q^2.\eqno(6.4)$$

The definitions of $Q$-factorials and $Q$-binomial coefficients still look
like the ones given in eqs. (2.7)--(2.8):
$$ [n]_Q! = [n]_Q [n-1]_Q \ldots [1]_Q,\eqno(6.5)$$
$$ {m \brack n}_Q = {[m]_Q! \over [m-n]_Q! [n]_Q!}.\eqno(6.6)$$
As it can be easily seen from eq. (6.1) under the substitution $Q\rightarrow
Q^{-1}$ one obtains
$$ \left[n\right]_Q!= Q^{n(n-1)/2}\left[n\right]_{1/Q},\eqno(6.7)$$
and
$$\left[ \begin{array}{c} n\\ k\end{array}\right]_Q =Q^{k(n-k)}
\left[\begin{array}{c} n \\ k \end{array}\right]_{1/Q}.\eqno(6.8) $$
$Q$-factorials are connected to $q$-factorials by
$$[n]!= q^{-n(n-1)/2}[n]_Q!,  \qquad {\rm with} \qquad Q=q^2.\eqno(6.9)$$

\section{$Q$-deformed elementary functions}

The definitions of $Q$-deformed elementary functions \cite{Exton}
 look similar
to these given in sec. 3. The {\sl $Q$-deformed exponential function}
is defined as
$$ e_Q(ax)=\sum _{n=0}^{\infty} {a^n\over [n]_Q!} x^n,\eqno(7.1)$$
and satisfies the property
$$ e_Q(x) e_{1/Q}(-x) =1.\eqno(7.2)$$
(Notice that $e_Q(x) e_Q(-x) \neq 1.$)

The {\sl $Q$-deformed trigonometric functions} are defined as
$$ \sin_Q (x)= \sum_{n=0}^{\infty} (-1)^n {x^{2n+1}\over [2n+1]_Q!},
\eqno(7.3)$$
$$ \cos_Q (x)= \sum_{n=0}^{\infty} (-1)^n {x^{2n}\over [2n]_Q!}.\eqno(7.4)$$
One can easily show that
$$ \sin_Q(x)= {1\over 2i} \left( e_Q(ix)-e_Q(-ix) \right),\eqno(7.5)$$
$$ \cos_Q(x)= {1\over 2}  \left( e_Q(ix)+e_Q(-ix) \right).\eqno(7.6)$$
Instead of the familiar identity $\sin^2(x)+\cos^2(x)=1$ one has
$$ \sin_Q(x) \sin_{1/Q}(x)+\cos_Q(x)\cos_{1/Q}(x)=1.\eqno(7.7)$$

$Q$-deformed polynomials, such as $Q$-deformed Hermite polynomials and
$Q$-de\-fo\-rmed Laguerre polynomials also exist \cite{Exton}.

\vfill\eject
\section{ $Q$-derivative}

Given the function $f(x)$ one defines its $Q$-derivative $D^Q_x$
\cite{Exton} by
the expression
$$ D^Q_x f(x) = { f(Qx) -f(x) \over (Q-1)x}. \eqno(8.1)$$
The similarity between this definition and the one of $Q$-numbers (eq. (6.1))
is clear.

One can easily prove that
$$ D^Q_x x^n \equiv {Q^n x^n -x^n \over (Q -1)x} = [n]_Q x^{n-1}, \eqno(8.2) $$
which looks exactly like eq. (4.2). In addition one has
$$ D^Q_x e_Q(ax)= a e_Q(ax),\eqno(8.3)$$
$$ D^Q_x e_{1/Q} (ax)= a e_{1/Q} (aQx),\eqno(8.4)$$
$$ D^Q_x \sin_Q(ax)= a\cos_Q(ax), \eqno(8.5)$$
$$ D^Q_x \cos_Q(ax)=-a\sin_Q(ax).\eqno(8.6)$$
One can also easily see that $\sin_Q(ax)$ and $\cos_Q(ax)$ are the linearly
independent solutions of the $Q$-differential equation
$$ (D^Q_x)^2 u(x) +a^2 u(x) =0,\eqno(8.7)$$
while the functions $\sin_{1/Q} (ax)$ and $\cos_{1/Q}(ax)$ satisfy the
equation
$$ (D^Q_x)^2 u(x) + a^2 u(Q^2 x) =0.\eqno(8.8)$$

The following {\sl Leibnitz rules} can also be shown:
$$	D^Q_x\left(f_1(x) f_2(x)\right)= \left(D^Q_x f_1(x)\right)f_2(Qx) +
	f_1(x)\left(D^Q_x f_2(x)\right), \eqno(8.9)$$
$$	D^Q_x\left(f_1(x) f_2(x)\right)= \left(D^Q_x f_1(x)\right)f_2(x) +
	f_1(Qx)\left(D^Q_x f_2(x)\right).\eqno(8.10)$$
One can further obtain
$$ D^Q_x {f_1(x) \over f_2(x)} = {\left(D^Q_x f_1(x)\right) f_2(x) -f_1(x)
	\left(D^Q_x f_2(x) \right) \over f_2(Qx) f_2(x) }.\eqno(8.11) $$
	For the second derivative of $f(x)$ one has
$$ (D^Q_x)^2 f (x)= (Q-1)^{-2} Q^{-1} x^{-2}\left\{f(Q^2 x) -(Q+1) f(Q x)
+Q f(x) \right\}, \eqno(8.12)$$
	and by mathematical induction we obtain the general formula
$$ (D^Q_x)^n f(x) =(Q-1)^{-n} Q^{-n(n-1)/2} x^{-n} \sum_{k=0}^n\left[
\begin{array}{c}
n \\ k \end{array}\right]_Q (-1)^k Q^{k(k-1)/2} f(Q^{n-k}x). \eqno(8.13)$$

\section{  $Q$-integration}

In a way analogous to that of sec. 5
 the definite $Q$-integral of the function $f(x)$
in the interval $[0, 1]$ is defined \cite{Exton} as follows
$$ \int_0^1 f(x) d_Q x = (1 -Q) \sum_{s=0}^{\infty} f(Q^s) Q^s, \eqno(9.1)$$
	assuming that $Q$ is real and $|Q| < 1$, while
for the definite integral of $f(x)$ in the interval
	$[0, \infty]$, we have
$$ \int_0^{\infty} f(x) d_Q x = (1-Q) \sum_{s=-\infty}^{\infty} Q^s f(Q^s).
\eqno(9.2)$$
For the indefinite $Q$-integral of $f(x)$ one has
$$ \int f(x) d_Q x = (1-Q) x \sum_{s=-\infty}^{\infty} Q^s f(Q^sx) + const.
\eqno(9.3)$$
One can easily check that $Q$-differentiation and $Q$-integration are
operations inverse to each other
$$ D^Q \int f(x)d_Q x = f(x).\eqno(9.4)$$
The formula for $Q$-integration by parts reads
$$ \int\left(D^Q f_1(x)\right)f_2(x) d_Q x= f_1(x) f_2(x) -
	\int f_1(Qx) \left(D^Q f_2(x) \right) d_Q x. \eqno(9.5)$$

\section{The $q$-deformed harmonic oscillator}

The interest for possible applications of quantum algebras in physics has been
triggered in 1989 by the introduction of the $q$-deformed harmonic oscillator
\cite{Bie873,Mac4581,SF983},
of which earlier equivalent versions existed \cite{AC524,Kur111}.

 The $q$-deformed harmonic oscillator
\cite{Bie873,Mac4581,SF983,Song821,Ng1023,Yan459}  is defined
in terms
of the creation and annihilation operators $a^\dagger$ and $a$ and the number
operator $N$, which satisfy the commutation relations
$$ [N, a^\dagger] = a^\dagger, \quad [N, a] =-a, \eqno(10.1)$$
$$ a a^\dagger - q^{\mp 1} a^\dagger a = q^{\pm N}  .\eqno(10.2)$$
In addition the following conditions of hermitian conjugation hold
$$ (a^\dagger)^\dagger =a, \qquad N^\dagger =N.\eqno(10.3)$$
Eq. (10.1) is the same as in ordinary quantum mechanics, while eq. (10.2) is
modified by the presence of the deformation parameter $q$.
For $q\to 1$ it is clear that eq. (10.2) goes to the usual
boson commutation relation $[a, a^\dagger]=1$.
 An immediate consequence of (10.2) is that
$$ a^\dagger a = [N], \quad a a^\dagger = [N+1]. \eqno(10.4)$$
Thus the number operator $N$ is {\it not} equal to $a^\dagger a$, as in the
ordinary case. The operators $a^\dagger$ and $a$ are referred to as
{\sl  $q$-deformed boson creation and annihilation operators} respectively.

The basis of the Fock space is defined by repeated action of the creation
operator $a^\dagger$ on the vacuum state, which is annihilated by $a$:
$$ a\vert 0\rangle=0,\qquad    |n> = {(a^\dagger)^n\over\sqrt{ [n]!}} |0>.
\eqno(10.5)$$
The action of the operators on the
basis is given by
$$N |n> = n |n>, \eqno(10.6)$$
$$a^\dagger |n> = \sqrt{[n+1]} |n>, \eqno(10.7)$$
$$a |n> = \sqrt{[n]} |n-1>. \eqno(10.8)$$
We remark that these equations look very similar to the ones of the
ordinary case, the only difference being that $q$-numbers appear under
the square roots instead of usual numbers.

The Hamiltonian of the $q$-deformed harmonic oscillator is
$$H={\hbar \omega \over 2} (a a^\dagger + a^\dagger a), \eqno(10.9)$$
and its eigenvalues in the basis given above are
$$ E(n)= {\hbar \omega \over 2} ([n]+[n+1]). \eqno(10.10)$$
One can easily see that for $q$ real the energy eigenvalues increase more
rapidly than the ordinary case, in which the spectrum is equidistant, i.e.
the spectrum gets ``expanded''.
In contrast, for $q$ being a phase factor ($q=e^{i\tau}$ with $\tau$ real)
the eigenvalues of the energy increase less rapidly than the ordinary
(equidistant) case, i.e. the spectrum is ``compressed''.
In particular, for $q$ real ($q=e^\tau$) the eigenvalues can be written as
$$ E(n) ={\hbar \omega \over 2} {\sinh \left(\tau\left(n+{1\over 2}\right)
\right) \over \sinh{\tau \over 2}} ,\eqno(10.11)$$
while for $q$ being a phase factor ($q=e^{i\tau}$) one has
$$ E(n) = {\hbar\omega\over 2} {\sin\left(\tau\left( n+{1\over 2}\right)
\right) \over \sin{\tau \over 2}}.\eqno(10.12)$$
In both cases in the limit $q\to 1$ ($\tau \to 0$) the ordinary expression
$$ E(n)= \hbar \omega \left( n+{1\over 2}\right) \eqno(10.13)$$
is recovered.

In addition, the following commutation relation holds
$$ [a, a^\dagger] = [N+1] -[N]. \eqno(10.14)$$
For $q$ being a phase factor, this commutation relation takes the form
$$[a, a^\dagger] = {\cos{(2N+1)\tau\over 2}\over\cos{\tau\over 2}}.
\eqno(10.15)$$

It is useful to notice that the $q$-deformed boson operators $a^\dagger$ and
$a$ can be expressed in terms of usual boson operators $\alpha^\dagger$ and
$\alpha$ (satisfying $[\alpha, \alpha^\dagger]=1$ and $N=\alpha^\dagger
\alpha$) through the relations \cite{Song821,KD415}
$$a=\sqrt{[N+1]\over N+1} \alpha =\alpha\sqrt{[N]\over N}, \qquad
a^\dagger=\alpha^\dagger \sqrt{[N+1]\over N+1}=\sqrt{[N]\over N}
\alpha^\dagger.\eqno(10.16)$$
The square root factors in the last equation have been called {\sl
$q$-deforming functionals}.

For $q$ being a primitive root of unity, i.e. $q=e^{2\pi i/k}$
($k=2$, 3,~\dots), it is clear the the representation of eqs (10.5)--(10.8)
becomes
finite-dimensional and has dimension $k$, since only the vectors $|0>$,
$|1>$, \dots, $|k-1>$ can be present. This case has been related to the
system of two anyons \cite{FT163}. In what follows we are going to assume
that $q$ is {\sl not} a primitive root of unity.

\section{The $Q$-deformed harmonic oscillator}

A different version of the deformed harmonic oscillator can be
obtained by defining \cite{KD415,BF171,Jan91} the operators $b$, $b^+$ through
the equations
$$ a= q^{1/2} b q^{-N/2}, \quad a^\dagger=q^{1/2}  q^{-N/2} b^\dagger .
\eqno(11.1)$$
Eqs. (10.1) and  (10.2) then give
$$ [N, b^\dagger ] = b^\dagger, \quad [N, b]=-b , \eqno(11.2)$$
$$ b b^\dagger -  q^2 b^\dagger b =1 . \eqno(11.3)$$
This oscillator has been first introduced by Arik and Coon \cite{AC524}
and later considered also by Kuryshkin \cite{Kur111}.
One then easily finds that
$$b^\dagger b= [N]_Q, \quad b b^\dagger = [N+1]_Q, \eqno(11.4)$$
where $Q=q^2$ and $Q$-numbers are defined in (6.1).
The basis is defined by
$$ b\vert 0\rangle=0, \qquad   |n>= {(b^\dagger)^n\over \sqrt{[n]_Q!}}|0>,
\eqno(11.5)$$
while the action of the operators on the basis is given by
$$ N |n> = n |n>, \eqno(11.6)$$
$$ b^\dagger |n>= \sqrt {[n+1]_Q} |n+1>, \eqno(11.7)$$
$$ b |n> = \sqrt{[n]_Q} |n-1> .\eqno(11.8)$$
The Hamiltonian of the corresponding deformed harmonic oscillator
has the form
$$H = {\hbar \omega\over 2} (b b^\dagger + b^\dagger b), \eqno(11.9)$$
the eigenvalues of which are
$$E(n)={\hbar \omega \over 2} ([n]_Q+[n+1]_Q). \eqno(11.10)$$
One can easily see that for $Q=e^T$, where $T>0$ and real, the spectrum
increases more rapidly than the ordinary (equidistant) spectrum, while
for $Q=e^T$, with $T<0$ and real, the spectrum is increasing less rapidly
than the ordinary (equidistant) case.

{}From the above relations, it is clear that the following
commutation relation holds
$$[b, b^\dagger] = Q^N.\eqno(11.11)$$

\section{The generalized deformed oscillator}

In addition to the oscillators described in the last two sections,
many kinds of deformed oscillators have been introduced in the literature
(see \cite{BD100} for a list).
All of them can be accommodated within the common mathematical framework
of the {\it generalized deformed oscillator} \cite{Das789,DY4157},
which is defined
as the algebra generated by the operators $\{1, a, a^\dagger, N\}$ and the
{\sl structure function} $\Phi(x)$, satisfying the relations
$$ [a, N]=a, \qquad [a^\dagger, N]=-a^\dagger, \eqno(12.1)$$
$$ a^\dagger a =\Phi(N) =[N], \qquad aa^\dagger = \Phi(N+1) =[N+1],
\eqno(12.2)$$
where $\Phi(x)$ is a positive analytic function with $\Phi(0)=0$ and $N$ is
the number operator.
{}From eq. (12.2) we conclude that
$$N=\Phi^{-1} (a^\dagger a),\eqno(12.3)$$
and that the following commutation and anticommutation relations are
obviously satisfied:
$$ [a, a^\dagger]=[N+1]-[N], \qquad \{a,a^\dagger\}=[N+1]+[N].\eqno(12.4)$$
The {\sl structure function} $\Phi(x)$ is characteristic to the deformation
scheme. In Table 1 the structure functions corresponding to different
deformed oscillators are given. They will be further discussed at the end of
this section.

It  can be proved that
the generalized deformed algebras possess a Fock space of
 eigenvectors
$|0>,|1>,\ldots,|n>,\ldots$
of the number operator $N$
$$N|n>=n|n>,\quad <n|m>=\delta_{nm}, \eqno(12.5) $$
if the {\it vacuum state} $|0>$ satisfies the following relation:
$$ a|0>=0. \eqno(12.6)$$
 These eigenvectors are generated by the formula:
 $$ \vert n >= {1 \over \sqrt{ [n]!}} {\left( a^\dagger \right)}^n \vert 0 >,
\eqno(12.7) $$
where
 $$[n]!=\prod_{k=1}^n [k]= \prod_{k=1}^n \Phi(k). \eqno(12.8) $$
The generators  $a^\dagger$ and $a$ are the creation and
annihilation operators of this deformed oscillator algebra:
$$a\vert n> = \sqrt{[n]} a\vert n-1>,\qquad
 a^\dagger \vert n> = \sqrt{[n+1]} a\vert n+1>. \eqno(12.9) $$

These eigenvectors are also eigenvectors of the energy operator
$$H={\hbar \omega \over 2} (aa^\dagger +a^\dagger a), \eqno(12.10)$$
corresponding to the eigenvalues
$$ E(n)= {\hbar\omega \over 2} (\Phi(n)+\Phi(n+1)) =
{\hbar \omega \over 2} ([n]+[n+1]).\eqno(12.11)$$

For $$\Phi(n)=n  \eqno(12.12)$$
one obtains the results for the ordinary harmonic oscillator.
For $$\Phi(n)= {q^n -q^{-n}\over q-q^{-1}}=[n]  \eqno(12.13)$$
one has the results for the $q$-deformed harmonic oscillator, while the
choice
$$ \Phi(n) ={Q^n-1\over Q-1}= [n]_Q  \eqno(12.14)$$
leads to the results of the $Q$-deformed harmonic oscillator. Many more
cases are shown in Table 1, on which the following comments apply:


\begin{table}[bt]
\begin{center}
\caption{ Structure functions of special deformation schemes}
\bigskip
\begin{tabular}{|c c p{2.0 in}|}
\hline
\ & $\Phi(x)$ & Reference \\
\hline\hline
\romannumeral 1 & $x$ &
harmonic  oscillator, bosonic algebra\\[0.05in]
\romannumeral 2&  ${ {q^x- q^{-x} }  \over {q- q^{-1} } } $ &
$q$-deformed harmonic oscillator \cite{Bie873,Mac4581}
\\[0.05in]
\romannumeral 3& ${ {q^x- 1 } \over {q- 1 } } $ & Arik--Coon,
Kuryshkin, or $Q$-deformed oscillator \cite{AC524,Kur111} \\[0.05in]
\romannumeral 4&  ${ {q^x- p^{-x} }  \over {q- p^{-1} } } $ &
2-parameter deformed oscillator \cite{BJM820,JBM775,CJ711}
\\[0.05in]
\romannumeral 5& $x(p+1-x)$ & parafermionic oscillator
 \cite{OK82} \\[0.05in]
\romannumeral 6& $ { \sinh (\tau x) \sinh (\tau (p+1-x) )}\over
{ \sinh^2(\tau) } $ & $q$-deformed parafermionic oscillator
\cite{FV1019,OKK591} \\[0.05in]
\romannumeral 7& $x\cos^2(\pi x/2) + (x+p-1)\sin^2(\pi x /2)  $&
parabosonic oscillator \cite{OK82} \\[0.05in]
\romannumeral 8&
$\begin{array}{c}
\frac{\sinh(\tau x)}{\sinh(\tau)}
\frac{\cosh(\tau (x+2N_0-1))}{\cosh(\tau)} \cos^2 (\pi x/2) +\\
+ \frac{\sinh(\tau (x+2N_0-1))}{\sinh(\tau)}
\frac{\cosh(\tau x)}{\cosh(\tau)} \sin^2 (\pi x/2)
\end{array}$
 & $q$-deformed parabosonic oscillator
 \cite{FV1019,OKK591} \\[0.05in]
\romannumeral 9 &
$\sin^2 {\pi x/2}$ & fermionic algebra \cite{JBSZ123} \\[0.05in]
\romannumeral 10 & $ q^{x-1} \sin^2 {\pi x/2}$ &
$q$-deformed fermionic algebra \cite{Hay129,CK72,FSS179,Gang819,Sci219,PV613}
 \\[0.05in]
\romannumeral 11&
$\frac{1-(-q)^x}{1+q}$ & generalized $q$-deformed fermionic algebra
 \cite{VPJ335} \\[0.05in]
\romannumeral 12& $x^n$ & \cite{Das789} \\[0.05in]
\romannumeral 13& ${ {sn(\tau x)} \over {sn(\tau )} }$ & \cite{Das789}
 \\[0.05in]
\hline
\end{tabular}
\end{center}
\end{table}

i) Two-parameter deformed oscillators have been introduced
\cite{BJM820,JBM775,CJ711}, in analogy to the one-parameter deformed
oscillators.

ii) Parafermionic oscillators \cite{OK82} of order $p$ represent particles
of which the maximum number which can occupy the same state is $p$.
Parabosonic oscillators \cite{OK82} can also be introduced.

iii) $q$-deformed versions of the parafermionic and parabosonic oscillators
have also been introduced \cite{FV1019,OKK591}.

iv) $q$-deformed versions of the fermionic algebra \cite{JBSZ123} have also
been introduced \cite{Hay129,CK72,FSS179,Gang819,Sci219,PV613}, as well as
$q$-deformed versions
of generalized $q$-deformed fermionic algebras \cite{VPJ335}. It has been
proved, however, that $q$-deformed fermions are fully equivalent to the
ordinary fermions \cite{BD1589,JX891,Kong92}.

\section{The physical content of deformed harmonic oscillators}

In order to get a feeling about the physical content of the various
deformed harmonic oscillators
it is instructive to construct potentials giving spectra similar to these of
the oscillators.

\subsection{Classical potentials equivalent to the $q$-oscillator}

Let us consider the $q$-deformed harmonic oscillator first. For small values
of $\tau$ one can
 take Taylor expansions
of the functions appearing there and thus find an expansion of the $q$-number
$[n]$ of eq. (2.1) in terms of powers of $\tau^2$. The final result is
$$[n]= n\pm {\tau^2  \over 6} (n-n^3) +{\tau^4 \over 360} (7n -
10 n^3 + 3 n^5) \pm {\tau^6\over 15120} (31 n -49 n^3 +21 n^5 -3 n^7)
+\ldots, \eqno(13.1)$$
where the upper (lower) sign corresponds to $q$ being a phase (real).
Using this expansion the energy of the $q$-deformed harmonic oscillator
of eq. (10.10) can be rewritten as
$$E(n)/(\hbar \omega)=(n+{1\over 2}) (1\pm {\tau^2 \over 24}) \mp {\tau^2 \over
6}
(n+{1\over 2})^3 +\ldots \eqno(13.2)$$
On the other hand, one can consider the potential
$$V(x)=V_0+k x^2 +\lambda x^4 + \mu x^6 +\xi x^8 +\ldots.\eqno(13.3)$$
If $\lambda$, $\mu$, $\xi$ are much smaller than $k$, one can
consider this potential as a harmonic oscillator potential plus some
perturbations and calculate the corresponding spectrum through
the use of perturbation theory \cite{BDK6153} (see also subsec. 32.2).
 In order to keep the subsequent formulae simple, we measure
$x$ in units of $(\hbar/(2m\omega))^{1/2}$.
 Using standard first order perturbation theory one finds that the
corresponding spectrum up to the order considered is
$$E(n)= E_0+(2\kappa +25 \mu) (n+{1\over 2}) + (6\lambda+245 \xi)
(n+{1\over 2})^2 +20 \mu (n+{1\over 2})^3 + 70 \xi (n+{1\over 2})^4.
\eqno(13.4)$$
By equating coefficients of the same powers of $(n+{1\over 2})$
in eqs. (13.2) and (13.4), we can determine the coefficients
appearing in the expansion of the potential given in eq.
(13.3). The final result for the potential, up to the order
considered, is
$$V(x)= ({1\over 2} \pm {\tau^2\over 8}) x^2 \mp
{\tau^2 \over 120} x^6 .\eqno(13.5)$$
We see therefore that to lowest order one can think of the $q$-oscillator
with a small value of the parameter $\tau$ as a harmonic oscillator
perturbed by a $x^6$ term.
It is clear that if we go to higher order, the next term to appear
will be proportional to $ \tau^4 x^{10}$.

The results of this subsection are corroborated by an independent study
of the relation between the $q$-deformed harmonic oscillator and the
ordinary anharmonic oscillator with $x^6$ anharmonicities \cite{AZB2555}.

\subsection{Classical potentials equivalent to the $Q$-deformed oscillator}

Similar considerations can be made for the $Q$-oscillator.
Defining $Q=e^T$ it is instructive to construct the expansion of
the $Q$-number of eq. (6.1) in powers of $T$. Assuming that $T$ is
small and taking Taylor expansions in eq. (6.1) one finally has
$$[n]_Q= n+{T\over 2} (n^2 -n) +{T^2\over 12} (2n^3-3n^2+1) +
{T^3\over 24} (n^4-2n^3+n^2) +\ldots\eqno(13.6)$$
Then the corresponding expansion of the energy levels of the
oscillator of eq. (11.10) is
$$E(n)/(\hbar \omega)= E'_0 + (1-{T\over 2} +{T^2\over 8}
-{T^3\over 16}+\ldots) (n+{1\over 2}) + ({T\over 2} -{T^2\over 4}
+{5 T^3\over 48} -\ldots) (n+{1\over 2})^2   $$
$$+({T^2\over 6}-{T^3\over 12}+\ldots)  (n+{1\over 2})^3
+({T^3\over 24} -\ldots) (n+{1\over 2})^4  .\eqno(13.7)$$
Comparing this expansion to eq. (13.4) and equating
equal powers of $(n+{1\over 2})$ we arrive at the following
expression for the potential
$$V(x)={T^2\over 12} + ({1\over 2} -{T\over 4}-{T^2\over 24}+{T^3
\over 48}) x^2 + ({T\over 12}-{T^2\over 24}-{T^3\over 144})x^4 $$
$$+ ({T^2\over 120} -{T^3\over 240}) x^6 + {T^3\over 1680} x^8 .
\eqno(13.8)$$
Thus to lowest order one can think of the $Q$-oscillator with small
values of the parameter $T$ as a harmonic oscillator perturbed by a
$x^4$ term. A similar expression is found, for example, by Taylor expanding
the modified P\"oschl--Teller potential \cite{PT1933},
 which, among
other applications, has been recently used in the description of
hypernuclei \cite{GLM283,LGM2098,LGM303,GLMP391}.
The modified P\"oschl--Teller potential (see also subsec. 32.2) has the
form
$$ V(x)_{PT}= -{A\over \cosh(ax)^2}. \eqno(13.9)$$
Its Taylor expansion is
$$ V(x)_{PT}= A(-1+ a^2 x^2 -{2\over 3}a^4 x^4 +{17\over 45} a^6 x^6 -\ldots).
\eqno(13.10)$$
We remark that this expansion contains the same powers of $x$
as the expansion (13.8).  Furthermore, the signs of the coefficients
of the same powers of $x$ in the two expansions are the same for
$T<0$.

\subsection{WKB-EPs for the $q$-deformed oscillator}

The potentials obtained above are only rough lowest order estimates. More
accurate methods exist for constructing WKB equivalent potentials (WKB-EPs)
giving (within the limits of WKB approximation) the same spectrum as the
above mentioned oscillators. A method by which this can be achieved has been
given by Wheeler \cite{Whe1976} and is described by Chadan and Chabatier
\cite{CS1977}. Applying this method to the $q$-deformed oscillator (sec. 10)
with $q$ being a phase factor one finds the potential \cite{BDK795,BDKJMP}
$$V(x)=\left( {\tau\over 2\sin(\tau/2)}\right)^2 {m \omega^2 \over 2} x^2$$
$$\left[ 1-{8\over 15} \left({x\over 2 R_e}\right)^4 +{4448\over 1575}
\left({x\over 2R_e}\right)^8 -{345344\over 675675} \left({x\over 2R_e}
\right)^{12} +\dots\right],\eqno(13.11)$$
where
$$R_e= {1\over \tau} \left( {\hbar^2\over 2 m}\right)^{1/2}
\left( {2\sin(\tau/2)\over \hbar \omega}\right)^{1/2},\eqno(13.12)$$
while for the $q$-deformed oscillator with $q$ real one has
$$V(x)=  \left({\tau\over 2\sinh(\tau/2)}\right)^2 {m \omega^2\over 2} x^2$$
$$\left[ 1+{8\over 15}\left({x\over 2 R_h}\right)^4+{4448\over 1575} \left(
{x\over 2 R_h}\right)^8 +{345344\over 675675}\left( {x\over 2 R_h}\right)^{12}
+\dots\right],\eqno(13.13)$$
where
$$ R_h ={1\over \tau} \left({\hbar^2\over 2 m}\right)^{1/2}
\left( 2\sinh (\tau/2) \over \hbar \omega\right)^{1/2}.\eqno(13.14)$$
The results of this subsection are corroborated by an independent study of
the relation between the $q$-deformed harmonic oscillator and the ordinary
anharmonic oscillator with $x^6$ anharmonicities\cite{AZB2555}.

\subsection{WKB-EPs for the $Q$-deformed oscillator}

Using the same technique one finds that the WKB
equivalent potential
for the $Q$-deformed harmonic oscillator (sec. 11) takes the form
\cite{Ioann93}
$$ V(x) = V_{min} + {(\ln Q)^2 \over Q} \left( {Q+1\over Q-1}\right) ^2
{1\over 2} m \omega^2 x^2 $$
$$\left[ 1 -{2\over 3} \left( {x\over R'}\right)^2
+ {23\over 45} \left( {x\over R'}\right)^4 -{134\over 315} \left( {x\over R'}
\right)^6 + {5297\over 14172} \left( {x\over R'}\right)^8 -\ldots \right],
\eqno(13.15)$$
where
$$V_{min} = {\hbar \omega (\sqrt{Q}-1) \over 2 \sqrt{Q} (\sqrt{Q}+1)},
\eqno(13.16)$$
and
$$ R'= \left( {\hbar \sqrt{Q} (Q-1) \over \omega m (Q+1) }\right)^{1/2}
{(\ln Q)^{-1} \over \sqrt{2}}.\eqno(13.17) $$

We remark that this WKB-EP contains all even powers of $x$, in contrast to
the WKB-EPs for the $q$-oscillator (eqs (13.11), (13.13)), which contains
only the powers
$x^2$, $x^6$, $x^{10}$, \dots. This is in agreement with the lowest order
results obtained in subsecs 13.1 and 13.2.

\section{The quantum algebra su$_q$(2)}

Quantum algebras are generalized versions of the usual Lie algebras, to
which they reduce when the deformation parameter $q$ is set equal to unity.
A simple example of a quantum algebra is provided by su$_q$(2)
\cite{Bie873,Mac4581}, which
is generated by the operators
$J_+$, $J_0$, $J_-$, satisfying  the commutation relations
$$ [ J_0, J_{\pm} ] = \pm J_{\pm} , \eqno (14.1)$$
$$ [ J_+ , J_- ] = [ 2 J_0 ] , \eqno (14.2)$$
with $J_0^{\dagger} = J_0$, $(J_+)^{\dagger} = J_-$.
We remark that eq. (14.1) is the same as in the case of the ordinary su(2)
algebra, while eq. (14.2) is different, since in the usual su(2) case it
reads $$[J_+, J_-]=2J_0.\eqno(14.3)$$
In the rhs of eq. (14.3) one has the first power of the $J_0$ operator, while
in the rhs of eq. (14.2) one has the $q$-operator $[2J_0]$, defined in sec. 2.
Because of eqs. (2.2) and (2.3) it is clear that if one writes the rhs of eq.
(14.2) in expanded form all odd powers of $J_0$ will appear:
$$ [J_+,J_-]= {1\over \sinh(\tau)} \left( {2\tau J_0\over 1!}+
{(2\tau J_0)^3\over 3!} +{(2\tau J_0)^5\over 5!}+\dots\right) \qquad {\rm for}
\qquad q=e^\tau,\eqno(14.4)$$
$$ [J_+,J_-]= {1\over \sin(\tau)} \left( {2\tau J_0\over 1!} -
{(2\tau J_0)^3\over 3!} +{(2\tau J_0)^5\over 5!} -\dots\right) \qquad
{\rm for} \qquad q=e^{i\tau}.\eqno(14.5)$$
Thus su$_q$(2)
can be loosely described as a nonlinear generalization of the usual su(2):
While in usual Lie algebras the commutator of two generators is always
producing a linear combination of generators, in the case of quantum
algebras the commutator of two generators can contain higher powers
of the generators as well.

The irreducible representations (irreps) $D^J$ of su$_q$(2)
(which have dimensionality $2J+1$) are
determined by highest weight states with $J=0, {1\over 2}, 1,
\ldots$. The basic states $|J, M>$ (with $-J\leq M \leq J$) are
connected with highest weight states $|J, J>$ as follows
$$|J, M> = \sqrt{[J+M]!\over [2J]! [J-M]!} (J_-)^{J-M} |J,J>,
\eqno (14.6)$$
with $J_+|J,J> =0$ and $<J,J|J,J>=1$.
The action of the generators of the algebra on these basic vectors is
given by
$$ J_0 |J,M>= M |J,M>,\eqno(14.7)$$
$$ J_{\pm} |J,M>=\sqrt{ [J\mp M] [J\pm M+1]} |J,M\pm 1>.\eqno(14.8)$$
These expressions look similar to the ones of the usual su(2) algebra, the
only difference being that $q$-numbers appear under the square root instead
of ordinary numbers.

The second order Casimir operator of su$_q$(2) is determined from the condition
that it should commute with all of the generators of the algebra. The resulting
operator is
$$C^q_2 = J_- J_+ +[J_0] [J_0+1] = J_+J_- +[J_0][J_0-1]. \eqno (14.9)$$
Its eigenvalues in the above mentioned basis are given by
$$C^q_2 |J, M> = [J] [J+1] |J, M> , \eqno(14.10)$$
while for the usual su(2) the eigenvalues of the Casimir operator are $J(J+1)$.
One can easily check that for real $q$ ($q=e^\tau$ with $\tau$ real)
the eigenvalues $[J][J+1]$ produce
a spectrum increasing more rapidly than $J(J+1)$ (an expanded spectrum),
while for $q$ being a phase factor ($q=e^{i\tau}$ with $\tau$ real) the
eigenvalues $[J][J+1]$ correspond to a spectrum increasing less rapidly
than $J(J+1)$ (a compressed spectrum).

It should be noticed that the generators $J_+$, $J_0$, $J_-$ of su$_q$(2)
are connected to the generators $j_+$, $j_0$, $j_-$ of the usual su(2),
which satisfy the commutation relations
$$[j_0, j_{\pm}]=\pm j_{\pm}, \qquad [j_+, j_-]=2j_0, \eqno(14.11)$$
through the relations \cite{CZ237,CGZ676}
$$ J_0=j_0, \qquad J_+=\sqrt{[j_0+j] [j_0-1-j]\over (j_0+j) (j_0-1-j)}j_+,
         \qquad J_-=j_-\sqrt{[j_0+j][j_0-1-j]\over [j_0+j][j_0-1-j]},
\eqno(14.12)$$
where $j$ is determined by the relation for the second order Casimir operator
of su(2)
$$ C= j_-j_++j_0(j_0+1) = j_+ j_-+j_0(j_0-1) = j(j+1).\eqno(14.13)$$

\vfill\eject
\section{Realization of su$_q$(2) in terms of $q$-deformed bosons}

Realizations of Lie algebras in terms of (ordinary) bosons are useful not only
as a convenient mathematical tool, but also because of their applications in
physics \cite{KM375}. In the case of quantum algebras it turns out
that boson realizations are possible in terms of the $q$-deformed boson
operators already introduced in sec. 10.

In the case of su$_q$(2) the generators can be mapped onto $q$-deformed
bosons as follows \cite{Bie873,Mac4581}
$$ J_+= a_1^\dagger a_2, \qquad J_-=a_2^\dagger a_1, \qquad J_0={1\over 2}
(N_1-N_2), \eqno(15.1)$$
where $a_i^\dagger$, $a_i$ and $N_i$ are $q$-deformed boson creation,
annihilation and number operators as these introduced in sec. 10.
One can easily prove that the boson images satisfy the commutation
relations (14.1) and (14.2). For example, one has
$$[J_+,J_-]= a_1^\dagger a_2 a_2^\dagger a_1-a_2^\dagger a_1 a_1^\dagger a_2=
[N_1] [N_2+1]-[N_1+1][N_2] = [N_1-N_2]=[2J_0],$$
where use of the identity (2.6) has been made.

In the $q$-boson picture the normalized highest weight vector is
$$|JJ> ={(a_1^\dagger)^{2J}\over \sqrt{[2J]!}} |0>,\eqno(15.3)$$
while the general vector $|JM>$ is given by
$$ |JM> = {(a_1^\dagger)^{J+M}\over \sqrt{[J+M]!}} {(a_2^\dagger)^{J-M}
\over \sqrt{[J-M]!}} |0>.\eqno(15.4)$$

It should be  noticed that it was the search for a boson realization of
the su$_q$(2) algebra that led to the introduction of the $q$-deformed
harmonic oscillator in 1989 \cite{Bie873,Mac4581}.

Starting from su$_q$(2) one can formulate a $q$-deformed version of angular
momentum theory. Some references are listed here:

i) Clebsch-Gordan coeeficients for su$_q$(2) can be  found in
\cite{KK443,GKK2769,Rue1085,Nom1954,KK4009,STK959,AM1139,AS981}.

ii) 6-j symbols for su$_q$(2) can be found in
\cite{KK2717,STK1746,RR3761,Mae2598}.

iii) 9-j symbols for su$_q$(2) can be found in \cite{STK2863}.

iv) The $q$-deformed version of the Wigner--Eckart theorem can be found
in \cite{Kli2919,Nom2345}.

In addition, it should be noticed that a two-parameter deformation of su(2),
labelled as su$_{p,q}$(2) has been introduced
\cite{BJM820,CJ711,BH165,BH629,Kib9234,Jing543,JVdJ2831}. Clebsch-Gordan
coefficients for su$_{p,q}$(2) have been discussed in
\cite{SW5563,DMM43,MM5177,Que19,KAS9439}.

The way in which the algebra su$_q$(2) can be realized in terms of the
$Q$-deformed bosons of sec. 11 is given in subsec. 33.1 (see eqs
(33.13)--(33.16)).

\section{The quantum algebra su$_q$(1,1)}

In this section we shall give a brief account of the algebra su$_q$(1,1)
\cite{KD415,UA237,Aiz1115}.

In the classical case the so(2,1) generators satisfy the
commutation relations \cite{LSK87}
$$ [K_1, K_2] =-i K_3, \quad\quad [K_2, K_3]= i K_1, \quad\quad
[K_3, K_1] =i K_2, \eqno(16.1)$$
which differ from the classical so(3) commutation relations in the sign
of the r.h.s. of the first commutator.
Defining
$$ K_+=K_1 +i K_2, \quad\quad K_-=K_1-i K_2, \quad\quad
K_3=K_z, \eqno(16.2)$$
one obtains the su(1,1) commutation relations
$$[K_z, K_{\pm}] = \pm K_{\pm}, \quad\quad [K_+, K_-]=-2 K_z,
\eqno(16.3)$$
which differ from the familiar su(2) commutation relations in
the sign of the r.h.s. of the last commutator.
The generators of su(1,1) accept the following boson representation
\cite{AIG27}
$$K_+= a_1^\dagger a_2^\dagger, \quad\quad K_-=a_1 a_2, \quad\quad
K_z = {1\over 2} (a_1^\dagger a_1 + a_2^\dagger a_2 +1) , \eqno(16.4)$$
where $a_1^\dagger$, $a_1$, $a_2^\dagger$, $a_2$ satisfy usual boson
commutation relations.

The second order Casimir operator of so(2,1) is \cite{LSK87}
$$C_2[{\rm so}(2,1)]= -(K_1^2+K_2^2-K_3^2),\eqno(16.5)$$
while for su(1,1) one has
$$C_2 [{\rm su}(1,1)]= [K_0] [K_0-1] -K_+ K_- = [K_0] [K_0+1] -K_- K_+.
\eqno(16.6)$$

In the quantum case, the generators of su$_q$(1,1) satisfy the
commutation relations \cite{KD415,UA237,Aiz1115}
$$[K_0, K_{\pm}]=\pm K_{\pm}, \quad\quad [K_+, K_-]=-[2 K_0].
\eqno(16.7)$$

The generators of su$_q$(1,1) accept the following boson
representation
$$K_+= a_1^\dagger a_2^\dagger, \quad\quad K_-=a_1 a_2, \quad \quad
K_0= {1\over 2} (N_1 +N_2 +1), \eqno(16.8)$$
where the bosons $a_i^\dagger$, $a_i$ ($i=1,2$) satisfy the usual $q$-boson
commutation relations.

The second order Casimir operator of su$_q$(1,1) is
$$C_2[{\rm su}_q(1,1)]=[K_0][K_0-1]-K_+K_-=[K_0][K_0+1]-K_-K_+. \eqno(16.9)$$
Its eigenvalues are \cite{UA237,Aiz1115}
$$C_2[{\rm su}_q(1,1)] |\kappa \mu> = [\kappa] [\kappa-1] |\kappa \mu>,
\eqno(16.10)$$
where
$$\kappa ={1+|n_1 -n_2|\over 2}, \quad
\mu ={1+n_1+n_2\over 2}, \eqno(16.11)$$
since the basis has the form
$| \kappa \mu> = | n_1 > | n_2 >$, with
$$|n_i> = {1 \over \sqrt{[n_i]!}} (a^\dagger_i)^{n_i} |0>. \eqno(16.12)$$
In this basis the possible values of $\mu$ are given by
$$\mu=\kappa, \kappa+1, \kappa+2, \ldots,\eqno(16.13)$$
up to infinity, while $\kappa$ may be any positive real number.
The action of the generators on this basis is given by
$$ K_0 \vert \kappa \mu \rangle = \mu \vert \kappa \mu \rangle ,\eqno(16.14) $$
$$ K_{\pm} \vert \kappa \mu \rangle = \sqrt{ [\mu\pm \kappa]
[\mu\mp \kappa\pm 1]} \vert \kappa \mu\pm 1\rangle.\eqno(16.15)$$

Clebsch-Gordan and Racah coefficients for su$_q$(1,1) can be found in
\cite{STK959,STK1746,Mae2598,Aiz1937,Alv92}.
Furthemore, a two-parameter deformed version of
su$_q$(1,1), labelled as su$_{p,q}$(1,1), has been introduced
\cite{CJ711,Kib9234,BBCH341,Jing495,Kli9339}.

\section{Generalized deformed su(2) algebras}

In the same way that in addition to the $q$-deformed oscillators one can
have generalized deformed oscillators, it turns out that
generalized deformed su(2) algebras, containing su$_q$(2) as a special case
and  having representation theory similar
to that of the usual su(2),  can be constructed \cite{BDK871,Pan5065}. It has
been proved that it is possible to construct an algebra
$$ [J_0, J_{\pm}]=\pm J_{\pm}, \qquad [J_+,J_-]=\Phi(J_0(J_0+1))-
\Phi(J_0(J_0-1)),\eqno(17.1)$$
where $J_0$, $J_+$, $J_-$ are the generators of the algebra and
$\Phi(x)$ is any increasing entire function defined for $x\geq -1/4$.
Since this algebra is characterized by the function $\Phi$, we use for it
the symbol su$_{\Phi}$(2). The appropriate basis $|l,m>$ has the
properties
$$ J_0|L,M> = M |L,M>, \eqno(17.2)$$
$$ J_+|L,M> = \sqrt{\Phi(L(L+1))-\Phi(M(M+1))} |L,M+1>, \eqno(17.3)$$
$$ J_-|L,M> = \sqrt{\Phi(L(L+1))-\Phi(M(M-1))} |L, M-1>,\eqno(17.4) $$
where
$$ L=0, {1\over 2}, 1, {3\over 2}, 2, {5\over 2}, 3, \ldots,\eqno(17.5)$$
  and
$$ M= -L, -L+1, -L+2, \ldots, L-2, L-1, L.\eqno(17.6)$$
The Casimir operator is
$$ C= J_-J_+ +\Phi(J_0(J_0+1))=J_+J_-+\Phi(J_0(J_0-1)),
\eqno(17.7)$$
its eigenvalues indicated by
$$C |L,M> = \Phi(L(L+1)) |L,M>.\eqno(17.8)$$
The usual su(2) algebra is recovered for
$$ \Phi(x(x+1))= x(x+1),\eqno(17.9)$$
while the quantum algebra su$_q$(2)
$$ [J_0, J_{\pm}]=\pm J_{\pm}, \qquad [J_+, J_-]=[2 J_0]_q,\eqno(17.10) $$
occurs for
$$ \Phi(x(x+1))= [x]   [x+1]  .\eqno(17.11)$$

The su$_{\Phi}$(2) algebra occurs in several cases, in which the rhs of the
last equation in (17.1) is an odd function of $J_0$ \cite{BDK2197}. It can be
seen that other
algebraic structures, like the quadratic Hahn algebra QH(3) \cite{GLZ217}
and the finite W algebra  $\bar{\rm W}_0$ \cite{Bow945}
 can be brought into the su$_\Phi$(2) form, the
advantage being that the representation theory of su$_\Phi$(2) is already
known. It can also be proved that several physical systems, like
the isotropic oscillator in a 2-dim curved space with constant curvature
\cite{Hig309,Lee489}, the Kepler system in a 2-dim curved space with constant
curvature \cite{Hig309,Lee489}, and the system of two identical particles in
two dimensions \cite{LM3649} can also be put into
an su$_\Phi$(2) form. More details can be found in \cite{BDK2197}.

\section{Generalized deformed parafermionic oscillators}

It turns out that the generalized deformed su$_\Phi$(2) algebras mentioned
in the last section are related to
 generalized deformed parafermionic oscillators, which we will therefore
describe here.

 It has been proved \cite{Que245} that any generalized
deformed parafermionic algebra of order $p$ can be written as a generalized
oscillator (sec. 12) with structure function
$$ F(x)= x (p+1-x) (\lambda +\mu x+\nu x^2 +\rho x^3 +\sigma x^4 +\ldots),
\eqno(18.1)$$
where $\lambda$, $\mu$, $\nu$, $\rho$, $\sigma$, \dots are real constants
satisfying the conditions
$$ \lambda + \mu x + \nu x^2 + \rho x^3 + \sigma x^4 +\ldots > 0, \qquad
x \in \{ 1,2,\ldots, p\}.\eqno(18.2)$$

Considering an su$_{\Phi}$(2) algebra \cite{BDK871} with structure function
$$ \Phi(J_0(J_0+1))= A J_0(J_0+1) + B (J_0(J_0+1))^2 + C (J_0(J_0+1))^3,
\eqno(18.3) $$
and making the correspondence
$$ J_+ \to A^{\dag}, \qquad J_-\to A, \qquad J_0\to N,\eqno(18.4)$$
one finds that the su$_{\Phi}$(2) algebra is equivalent to a generalized
deformed parafermionic  oscillator of the form
$$F(N)= N (p+1-N)$$ $$ [ -(p^2(p+1)C +p B)+ (p^3 C +(p-1)B) N+
((p^2-p+1)C +B) N^2+ (p-2) C N^3 + C N^4], \eqno(18.5)$$
if the condition
$$ A+ p(p+1) B + p^2 (p+1)^2 C =0 \eqno(18.6)$$
holds. The condition of eq. (18.2) is always satisfied for $B>0$ and $C>0$.

In the special case of $C=0$ one finds that the su$_{\Phi}$(2) algebra
with structure function
$$ \Phi(J_0(J_0+1))= A J_0(J_0+1) + B (J_0(J_0+1))^2\eqno(18.7)$$
is equivalent to a generalized deformed parafermionic oscillator
characterized by
$$ F(N)= B N (p+1-N) (-p+(p-1)N+ N^2),\eqno(18.8)$$
if the condition
$$ A+ p(p+1) B=0\eqno(18.9)$$
is satisfied. The condition of eq. (18.2) is satisfied for $B>0$.

Including higher powers of $J_0(J_0+1)$ in eq. (18.3) results in higher powers
of $N$ in eq. (18.5) and higher powers of $p(p+1)$ in eq. (18.6). If, however,
one sets $B=0$ in eq. (18.7), then eq. (18.8) vanishes, indicating that no
parafermionic oscillator equivalent to the usual su(2) rotator can be
constructed.

It turns out that several other mathematical structures, like the finite W
algebras $\bar {\rm W}_0$ \cite{Bow945}
and W$^{(2)}_3$ (see subsec. 34.5) can be put into the generalized
deformed parafermionic oscillator form. The same is true for several
physical systems, such as the isotropic oscillator and the Kepler problem
in a 2-dim curved space with constant curvature \cite{Hig309,Lee489}, and the
Fokas--Lagerstrom \cite{FL325}, Smorodinsky--Winternitz \cite{SW444},
and Holt \cite{Holt1037} potentials. Further details can be found in
\cite{BDK2197}.

\section{The su$_q$(2) rotator model}

It has been suggested by Raychev, Roussev and Smirnov \cite{RRS137}
 and independently by Iwao \cite{Iwao363,Iwao368} that
rotational spectra of deformed nuclei can be described by the $q$-deformed
rotator, which corresponds to the 2nd order
Casimir operator of the quantum algebra su$_q$(2), already studied in
sec. 14. We shall show here that this assumption works and discuss the
reasons behind this success, as well as the relation \cite{PLB251}
 between the
su$_q$(2) model and the Variable Moment of Inertia (VMI) model (see sec. 20).

The $q$-deformed rotator corresponds to the  Hamiltonian
$$H = {1\over 2 I} C_2({\rm su}_q(2)) + E_0, \eqno (19.1)$$
where $I$ is the moment of inertia and $E_0$ is the bandhead energy
(for ground state bands $E_0=0$).
For $q$ real, i.e. with $q=e^{\tau}$ with $\tau$ real, the
energy levels of the $q$-rotator are
$$E(J) = {1\over 2I} [J] [J+1] +E_0 = {1\over 2I}
{\sinh(\tau J) \sinh(\tau (J+1))\over  \sinh^2(\tau) } +E_0.\eqno(19.2)$$
For $q$ being a phase, i.e. $q=e^{i\tau}$ with $\tau$ real,
 one obtains
$$E(J) ={1\over 2I} [J] [J+1] +E_0= {1\over 2I}
{\sin (\tau J) \sin(\tau(J+1))\over \sin^2 (\tau)} +E_0. \eqno (19.3)$$

 Raychev {\it et al.} \cite{RRS137} have  found that good fits of rotational
spectra of
even--even rare earths and actinides are obtained with eq. (19.3).
It is easy to check that eq. (19.2) fails in describing such spectra.
In order to understand  this difference, it is useful to make
Taylor expansions of the  quantities in the numerator of eq. (19.2)
(eq. (19.3)) and collect together the terms containing the same
powers of $J(J+1)$ (all other terms cancel out), finally summing up
the coefficients of each power. In the first case the final
result is
$$E(J) = E_0 + {1\over 2I}{1\over (\sqrt{\pi \over 2\tau} I_{1/2}(\tau)
)^2} (\sqrt{\pi\over 2\tau} I_{1/2}(\tau) J(J+1) +\tau
\sqrt{\pi\over 2\tau} I_{3/2}(\tau) (J(J+1))^2 $$
$$+{2\tau^2\over 3} \sqrt{\pi \over 2\tau} I_{5/2}(\tau) (J(J+1))^3
+{\tau^3\over 3} \sqrt{\pi\over 2\tau} I_{7/2} (\tau) (J(J+1))^4
+\ldots \eqno(19.4)$$
where $\sqrt{\pi\over 2\tau} I_{n+{1\over 2}} (\tau)$ are the
modified spherical Bessel functions of the first kind \cite{AS72}.

In the second case (eq. (19.3)) following the same procedure one
obtains
$$E(J) =E_0+ {1\over 2I} {1\over (j_0(\tau))^2} (j_0(\tau) J(J+1)
 -\tau j_1(\tau) (J(J+1))^2 $$ $$+{2\over 3} \tau^2 j_2(\tau) (J(J+1))^3
-{1\over 3} \tau^3  j_3(\tau) (J(J+1))^4+{2\over 15} \tau^4
j_4(\tau) (j(j+1))^5-\ldots), \eqno(19.5)$$
where $j_n(\tau)$ are the spherical Bessel functions of the first
kind \cite{AS72}.

Both  results are of the form
$$ E(J)= AJ(J+1)+B(J(J+1))^2+C (J(J+1))^3+D(J(J+1))^4+\dots.\eqno(19.6)$$
Empirically it is known that nuclear rotational spectra do show such
a behaviour, the coefficients $A$, $B$, $C$, $D$, \dots having alternating
signs (starting with $A$ positive) and magnitudes dropping by about
3 orders of magnitude each time one moves to the next higher power of $J(J+1)$
\cite{deVoigt949,XWZ2337}.

It is interesting to check if the empirical characteristics of
the coefficients $A$, $B$, $C$, $D$ are present in the case of the
expansions of eqs. (19.2), (19.3), especially for small values of
$\tau$. (Since we deal with rotational spectra, which
are in first order approximation described by the usual algebra
su(2), we expect $\tau$  to be relatively small, i.e.
the deviation of su$_q$(2) from su(2) to be small. This is in
agreement to the findings of \cite{RRS137}, where $\tau$ is
 found to be around 0.03.)

One can easily see that in eq. (19.2) it is impossible to get alternating
signs, while in eq. (19.3) the condition of alternating signs is readily
fulfilled. This fact has as a result that the energy levels given by eq. (19.2)
increase more rapidly than the levels given by the $J(J+1)$ rule,
while the levels given by eq. (19.3) increase less rapidly than $J(J+1)$.
In order to check the order of magnitude of the coefficients
for small values of $\tau$, it is useful to expand the spherical
Bessel functions appearing in eq. (19.3) and keep only the lowest order
term in each expansion. The result is
$$E(J) = E_0+{1\over 2I} (J(J+1)-{\tau^2\over 3} (J(J+1))^2
+{2\tau^4\over 45} (J(J+1))^3 $$
$$-{\tau^6\over 315} (J(J+1))^4 +{2\tau^8\over 14175}
(J(J+1))^5 -\ldots ).
\eqno(19.7)$$
We remark that each term contains a factor $\tau^2$ more
than the previous one. For $\tau$ in the area of 0.03, $\tau^2$
is of the order of $10^{-3}$, as it should. We conclude therefore that
eq. (19.6) is suitable for fitting rotational spectra, since its
coefficients have the same characteristics as the empirical
coefficients of eq. (19.6).

Extended comparisons of the su$_q$(2) predictions to experimental data for
ground state bands of rare earth and actinide nuclei can be found in
\cite{RRS137,PLB251,CGST180,HVW1337}.
More recently, the su$_q$(2) formalism has been used for the
description of $\beta$- and $\gamma$-bands of deformed rare earths and
actinides, with satisfactory results \cite{MDRRB95}.

It is necessary for $E(J)$ to be an increasing function of $J$. In order
to guarantee this in eq. (19.3) one must have
$$ \tau (J+1) \leq {\pi \over 2}.\eqno(19.8)$$
In the case of $\tau=0.036$ (as in $^{232}$U in \cite{PLB251}),
one finds $J\leq 42$, this limiting value being larger than the highest
observed $J$ in ground state bands in the actinide region \cite{Sakai84}.
Similarly, for $\tau=0.046$ (as in $^{178}$Hf in \cite{PLB251}),
one finds $J\leq 32$, this limiting value being again higher than the highest
observed $J$ in ground state bands in the rare earth region \cite{Sakai84}.

\section{ Comparison of the su$_q$(2) model to other models}

\subsection{The Variable Moment of Inertia (VMI) model}

In lowest order approximation rotational nuclear spectra can be described by
the formula
$$E(J)={J(J+1)\over 2\Theta} ,\eqno(20.1)$$
where $\Theta$ is the moment of inertia of the nucleus, which is assumed to
be constant. However, in order to get closer
agreement to experimental data, one finds that he has to include higher
order terms in this formula, as shown in eq. (19.6).

Another way to improve agreement with experiment is to let the moment of
inertia $\Theta$ to vary as a function of the angular momentum $J$.
One thus obtains the {\sl Variable Moment of Inertia (VMI) model}
\cite{MGB1864}. In this model the
levels of the ground state band are given by
$$ E(J) = {J(J+1)\over 2\Theta(J)} + {1\over 2} C (\Theta(J)-\Theta_0)^2,
\eqno(20.2)$$
where $\Theta(J)$ is the moment of inertia of the nucleus at the state with
angular momentum $J$, while $C$ and $\Theta_0$ are the two free parameters
of the model, fitted to the data. The parameter $\Theta_0$ corresponds to
the ground state moment of inertia, while instead of the parameter $C$
it has been found meaningful to use the parameter combination
$$ \sigma= {1\over 2 C \Theta_0^3},\eqno(20.3)$$
which is related to the softness of the nucleus. The moment of inertia
at given $J$ is determined through the variational condition
$$ {\partial E(J)\over \partial \Theta(J)} \vert _J = 0,\eqno(20.4)$$
which is equivalent to the cubic equation
$$ \Theta(J)^3 -\Theta(J)^2 \Theta_0 -{J(J+1) \over 2C} =0. \eqno(20.5) $$
This equation has only one real root, which can be written as
$$ \Theta(J) = \root 3 \of { {J(J+1)\over 4C} + {\Theta_0^3\over 27}
+ \sqrt{ {(J(J+1))^2\over 16 C^2} + { \Theta_0^3 J(J+1)\over 54C} } } $$
$$ + \root 3 \of { {J(J+1)\over 4C} + {\Theta_0^3\over 27} -
\sqrt{ {(J(J+1))^2 \over 16C^2} + {\Theta_0^3 J(J+1)\over 54C} } } +
{\Theta_0 \over 3}.\eqno(20.6)$$
Expanding the roots in this expression one obtains
$$ \Theta(J) = \Theta_0 ( 1+ \sigma J(J+1) -2 \sigma^2 (J(J+1))^2 $$
$$+7 \sigma^3 (J(J+1))^3 -30 \sigma^4 (J(J+1))^4 +\ldots ). \eqno(20.7)$$
Using eq. (20.7) in eq. (20.2) one obtains the following expansion for the
energy
$$E(J) = {1\over 2\Theta_0} ( J(J+1) -{1\over 2} \sigma (J(J+1))^2$$
$$+\sigma^2 (J(J+1))^3 -3 \sigma^3 (J(J+1))^4 +\dots ). \eqno(20.8)$$
Empirically it is known \cite{MGB1864} that for rotational
nuclei the softness parameter $\sigma$ is of the order of $10^{-3}$.
Therefore the expansion of eq. (20.8) has the same characteristics as the
expansion of eq. (19.6) (alternating signs, successive coefficients
falling by about 3 orders of magnitude).

\subsection{Comparison of the su$_q$(2) model to the VMI and related models}

We now turn to the comparison of the expansion of eq. (19.5) to
the Variable Moment of Inertia (VMI) model, discussed in the previous
subsection.
 Comparing eqs (19.5) and (20.8) we see that both expansions have
the same form. The moment of inertia parameter $I$ of (19.5)
corresponds to the ground state moment of inertia $\Theta_0$
of (20.8). The small parameter of the expansion is $\tau^2$
in the first case, while it is the softness parameter
$1/(2C\Theta_0^3)$ in the second. However, the numerical
coefficients in front of each power of $J(J+1)$ are not the same.

In \cite{PLB251}  a comparison is made between the
 parameters obtained by fitting the same spectra by the su$_q$(2)
and VMI formulae.
The agreement between $1/(2I)$ and $1/(2\Theta_0)$ is very good,
as it is the agreement between $\tau^2$ and $\sigma$ as well.
Therefore the known \cite{MGB1864}
 smooth variation of $\Theta_0$ and
$\sigma$ with the ratio $R_4=E(4)/E(2)$ is expected to hold
for the parameters $I$ and $\tau^2$ as well. This is indeed seen
in \cite{PLB251}.

 The difference between the expansions of eqs (19.5) (or (19.7)) and (20.8) is
also demonstrated by forming the dimensionless ratios
$AC/(4 B^2)$ and $A^2D/(24 B^3)$ in eq. (19.6).  In the case of eq.
(20.8) both quantities are equal to 1, as expected, since it is known
that  the VMI is equivalent \cite{DDK235,KDD333}
to the Harris expansion \cite{Harris65}, in which
both quantities are known to be equal to 1. In the case of
eq. (19.7) the corresponding values are $AC/(4B^2)=1/10$ and
$A^2 D/ (24 B^3)=1/280$. According to the empirical values
of these ratios given in \cite{XWZ2337,MWZ2545}, the ratios given from eq.
(19.7) are better than the ratios given by eq. (20.8), especially
the second one.

\subsection{ The hybrid model}

The hybrid model of nuclear collective motion
\cite{Mosh156,Mosh257,CMV605,CFHM1367}
has been introduced in order to provide a link
between the two successful ways of describing low-lying nuclear
excitations: the extended form of the Bohr-Mottelson model (BMM)
\cite{GG449,HSMG147}
and the Interacting Boson Model (IBM) \cite{IA1987} (see secs 27, 29
for more details).
The hybrid model combines the
advantages of both models, i.e. the geometrical significance of the
collective coordinates inherent in the extended BMM, and the use of
group theoretical concepts characterizing IBM. In the framework of
the rotational limit of the hybrid model, associated with the
chain u(6)$\supset$su(3),  Partensky and Quesne \cite{PQ340,PQ2837},
starting from the fact that in the geometrical description the square
of the deformation is proportional to the moment of inertia of the
ground state band, proved that  the energy levels of the
ground state band are given by
$$ E(J)= {A\over J(J+1)+B} J(J+1),\eqno(20.9)$$
with $A$ being a free parameter and $B$ given by
$$ B= 8 N^2 +22 N-15, \eqno(20.10)$$
where $N$ is the sum of the number of valence proton pairs (or proton-hole
pairs, when more than half of the proton valence shell is filled) $N_{\pi}$
and the number of the valence neutron pairs (or neutron-hole pairs, when more
than half of the neutron valence shell is filled) $N_{\nu}$. We remark
that in the framework of this model the moment of inertia
$$ \Theta(J,N) = {J(J+1)+B \over 2A}\eqno(20.11)$$
depends on both the angular momentum $J$ and the valence pair number $N$.

It is instructive to expand $E(J)$ in powers of J(J+1) \cite{MRR67}
$$ E(J) = A \sum_{k=0}^{\infty} (-1)^k \left( {J(J+1)\over B} \right)^{k+1} =
{A\over B} ( J(J+1) $$ $$- {1\over B} (J(J+1))^2 + {1\over B^2}
(J(J+1))^3 -{1\over B^3} (J(J+1))^4 +\ldots). \eqno(20.12)$$
Comparing the present expansion to the one of eq. (19.7) for the su$_q$(2)
model one has
$$ \tau = \sqrt{3\over B} = \sqrt{3\over 8 N^2 + 22 N-15},\eqno(20.13)$$
obtaining thus a connection between the $\tau$ parameter of the su$_q$(2)
model and a microscopic quantity, the valence nucleon pair number.
{}From this equation it is clear that $\tau$ decreases for increasing $N$.
Thus the minimum values of $\tau$ are expected near the midshell regions,
where the best rotators are known to be located. This is a reasonable
result, since $\tau$ describes the deviations from the su(2) (rigid
rotator) limit. It is clear that the minimum deviations should occur in the
case of the best rigid rotators.

{}From eq. (20.13) one can obtain for each nucleus the value of $\tau$ from
the number of valence nucleon pairs present in it. These predictions for
$\tau$ have been compared in \cite{MRR67} to the values of
$\tau$ found empirically by fitting the corresponding spectra, with good
results in both the rare earth and the actinide regions.  In addition eq.
 (20.13) indicates that in a given shell nuclei characterized by the same
valence nucleon pair number $N$ will correspond to the same value of
$\tau$. Such multiplets have been studied in the framework of the hybrid
model by \cite{Bon351}.

It is worth noticing that fits of $\gamma_1$-bands in the rare earth region
(Er, Yb isotopes) \cite{MDRRB95}
give $\tau$ parameters very similar to the ones coming
from fitting the corresponding ground state bands, in addition exhibiting
the same as the one mentioned above behaviour of $\tau$ as a function of $N$.

Taking further advantage of the above connection between the su$_q$(2)
model and the hybrid model, the parameter $\tau$ has been connected
\cite{MRR557} to
the nuclear deformation parameter $\beta$, as well as to the
electromagnetic transition probabilities B(E2:$2_1^+\rightarrow 0_1^+$).
Since both $\beta$ and B(E2:$2_1^+\rightarrow 0_1^+$) are known to increase
with increasing collectivity, $\tau$ is expected to decrease with increasing
$\beta$ or increasing B(E2:$2_1^+\rightarrow 0_1^+$). This expectation is
corroborated by the results reported in \cite{MRR557}.

\subsection{ Other models}

It should be noticed that an empirical formula very similar to that of
eq. (19.3) has been recently proposed by Amal'ski\u{\i} \cite{Amal70}
 on completely different physical grounds. The formula reads
$$ E(J) = A \sin^2\left( {\pi J \over B}\right) ,\eqno(20.14)$$
where $A$ and $B$ are free parameters. This formula should be compared
to eq. (19.3), with which it is almost identical.

A different formula, also giving very good fits of
rotational spectra, has been introduced by Celeghini, Giachetti, Sorace
and Tarlini \cite{CGST180}, based on the $q$-Poincar\'e rotator.

\section{Electromagnetic transitions in the su$_q$(2) model}

We have already seen that the su$_q$(2) formalism provides an alternative
to the VMI model, the deformation parameter $q$ being connected to the
softness parameter of the VMI model.

The stretching effect present in rotational energy levels,
which can equally well be described in terms of the VMI model
and the su$_q$(2) symmetry, should also manifest itself in the
B(E2) transition probabilities among these levels.
If deviations from the su(2) symmetry are observed in the
energy levels of a band, relevant deviations should also
appear in the B(E2) transitions connecting them.
In the case of the VMI model no way has been found for
making predictions for the B(E2) transition probabilities
connecting the levels of a collective band. The su$_q$(2)
symmetry naturally provides such a link. Before studying the su$_q$(2) case,
though, it is useful to recall the predictions of other models on this matter.

\subsection{The collective model of Bohr and Mottelson}

In rotational
bands one has \cite{BM75}
$$B(E2:J+2\rightarrow J) = {5\over 16\pi} Q_0^2
|C^{J+2,2,J}_{K,0,-K}|^2, \eqno(21.1)$$
i.e. the B(E2) transition probability depends on the relevant
Clebsch-Gordan coefficient of su(2), while  $Q_0^2$ is
the intrinsic electric quadrupole moment
and $K$ is the projection of the angular momentum $J$
on the symmetry axis of the nucleus in the body-fixed frame.
For $K=0$ bands, as the ground state bands,  one then has
\cite{BM75}
$$  B(E2:J+2\rightarrow J)= {5\over 16\pi} Q_0^2
{3\over 2} {(J+1)(J+2)\over (2J+3)(2J+5)}.\eqno(21.2)$$
It is clear that the B(E2) values should saturate
with increasing $J$.

\subsection{The Interacting Boson Model (IBM)}

It is also instructive to mention what happens in the case of the
Interacting Boson Model (IBM) \cite{IA1987}, the successful
algebraic model of nuclear structure with which we are going to deal
in more detail later (see sections 27, 29).
In the case of the su(3) limit of the IBM,
which is the limit applicable to deformed nuclei, the
corresponding expression is \cite{AI201}
$$B(E2:J+2\rightarrow J)= {5\over 16\pi} Q_0^2 {3\over 2}
{(J+1)(J+2)\over (2J+3)(2J+5)} {(2N-J)(2N+J+3)\over (2N+3/2)^2},
\eqno(21.3)$$
where $N$ is the total number of bosons. Instead of saturation
one then gets a decrease of the B(E2) values at high $J$,
which finally reach zero at $J=2N$. This is a well known
disadvantage of the simplest version of the model (IBM-1)
 due to the small number of collective bosons ($s$ ($J=0$) and
$d$ ($J=2$)) taken into account. It can be corrected by the
inclusion of higher bosons ($g$ ($J=4$), $i$ ($J=6$), etc),
which approximately restore saturation (see \cite{IA1987,Bon88} for full list
of references).

Another way to avoid the problem of decreasing B(E2)s in the
su(3) limit of IBM at high $J$ is the recently proposed \cite{Muk229}
transition from the compact su(3) algebra to the noncompact
sl(3,R) algebra. The angular momentum at which this transition
takes place is fitted to experiment. In this way an increase of the
B(E2) values at high $J$ is predicted, which agrees well with
the experimental data for $^{236}$U.

\subsection{The su$_q$(2) model}

In order to derive a formula similar to (21.2) in the su$_q$(2)
case, one needs to develop an su$_q$(2) angular
momentum theory. As mentioned in sec. 15, this has already been achieved.
It turns out that an equation similar to (21.1)
 holds in the q-deformed case,
 the only difference being that the Clebsch-Gordan
coefficient of the su$_q$(2) algebra must be used instead.
These coefficients have the form \cite{STK959}
$$ _q C^{J+2,2,J}_{K,0,-K} = q^{2K} \sqrt{
[3] [4] [J-K+2] [J-K+1] [J+K+1] [J+K+2]\over [2] [2J+2] [2J+3]
[2J+4] [2J+5]}. \eqno(21.4)$$
 For $K=0$ bands one then has
$$B_q(E2;J+2\rightarrow J)={5\over 16\pi} Q_0^2
{[3] [4] [J+1]^2 [J+2]^2 \over [2] [2J+2] [2J+3] [2J+4] [2J+5]}.
\eqno(21.5)$$
For $q=e^{i\tau}$ this equation takes the form
$$B_q(E2; J+2\rightarrow J) = {5\over 16\pi} Q_0^2
{\sin(3 \tau) \sin(4\tau) \over \sin(2\tau) \sin(\tau)}$$
$$ {(\sin(\tau(J+1)))^2 (\sin(\tau(J+2)))^2\over
\sin(\tau(2J+2)) \sin(\tau(2J+3))
\sin(\tau(2J+4))\sin(\tau(2J+5))}.\eqno(21.6)$$
  It is
useful to get an idea of the behaviour of this expression as a
function of $J$, especially for the small values of $\tau$
found appropriate for the description of ground state spectra.
Expanding all functions and keeping corrections of the leading
order in $\tau$ only, one has
$$B_q(E2, J+2\rightarrow J) = {5\over 16\pi} Q_0^2 {3\over 2}
{(J+1)(J+2)\over (2J+3)(2J+5)}
(1+{\tau^2\over 3}(6 J^2 +22 J+12)).\eqno(21.7)$$
We see that the extra factor, which depends on $\tau^2$,
contributes an extra increase with $J$, while the usual  su(2)
expression reaches saturation at high $J$ and IBM even predicts
a decrease.

\subsection{Comparison to experiment}

Is there any experimental evidence for such an increase? In order
to answer this question one should discover cases in which
the data will be consistent with the su$_q$(2) expression but
inconsistent with the classical su(2) expression. (The opposite
cannot happen, since the classical expression is obtained from
the quantum expression for the special parameter value $\tau=0$.)
Since error bars of B(E2) values are usually large, in most cases
both symmetries are consistent with the data. One should expect
the differences to show up more clearly in two cases:

i) In rare earth nuclei not very much deformed (i.e. with an
$R_4=E(4)/E(2)$ ratio around 3.0). These should be deformed enough
so that the su$_q$(2) symmetry will be able to describe them,
having, however, at the same time values of $\tau$ not very small.
Since in several of these nuclei backbending (or upbending) occurs
at $J=14$ or 16, one can expect only 5 or 6 experimental points
to compare the theoretical predictions with.

ii) In the actinide region no backbending occurs up to around
$J=30$, so that this is a better test ground for the two
symmetries.
 However, most nuclei in this region are well deformed, so
that small values of $\tau$ should be expected, making the
distinction between the two theoretical predictions difficult.

A few characteristic examples (4 rare earths and an actinide)
are given in \cite{BFRRS3275}.
In all cases it is clear that the su$_q$(2) curve follows the
experimental points, while the su(2) curve has a different shape
which cannot be forced to go through all the error bars.
Several comments are now in place:

i) For a given nucleus the value of the parameter $\tau$ obtained from
fitting the B(E2) values among the levels of the ground state band
should be equal to the value obtained from fitting the energy levels
of the ground state band. In \cite{BFRRS3275}  it is clear that both
 values
are similar, although in most cases the value obtained from the
B(E2)s is smaller than the value obtained from the spectra. It
should be taken into account, however, that in most cases the
number $n'$ of levels fitted is different (larger) than the number
$n$ of the B(E2) values fitted. In the single case ($^{184}$W)
in which $n=n'$, the two $\tau$ values are almost identical, as
they should.

ii) One can certainly try different fitting procedures. Using
the value of $\tau$ obtained from the B(E2) values for fitting
the spectrum one gets a reasonably good description of it,
although the squeezing of the spectrum is not as much as it should
have been (with the exception of $^{184}$W). Using the value of
$\tau$ obtained from the spectrum for fitting the B(E2) values
one obtains an increase more rapid than the one shown by the data
(again with the exception of $^{184}$W). One can also try to make an
overall fit of spectra and B(E2)s using a common value of $\tau$.
Then both the squeezing of the spectrum and the rise of the B(E2)s
can be accounted for reasonably well although not exactly.
 One should notice, however,
that the experimental uncertainties of the B(E2)s are much higher
than the uncertainties of the energy levels.

iii) Concerning energy levels, the rigid rotator model and the
su(3) limit of the IBM
predict a $J(J+1)$ increase, while the su$_q$(2) model and the VMI
model predict squeezing, which is seen experimentally.

iv) Concerning the B(E2) values, the VMI makes no prediction,
 the rigid rotator predicts saturation
at high $J$, the su(3) limit of the IBM predicts decrease,
while the su$_q$(2) model predicts an increase. The evidence
presented in \cite{BFRRS3275}
 supports the su$_q$(2) prediction, but
clearly much more work, both experimental and analytical,
is needed before final conclusions can be drawn. The modified
su(3) limit of IBM described in \cite{Muk229}  also supports the
increase of the B(E2) values at high $J$.
Increasing BE(2) values are also predicted in the framework of the
Fermion Dynamical Symmetry Model \cite{PCWF2224}.
There is also empirical evidence for increasing B(E2) values in the recent
systematics by Zamfir and Casten \cite{ZC1280}.

v) It is clear that much further work is needed as far as comparisons of the
su$_q$(2) predictions to experimental BE(2) values are concerned
for safe conclusions to be reached.

vi) Since the quadrupole operator is not a member of the symmetry algebra
su$_q$(2) under consideration, it is clear that the B(E2) values studied
here do not contain any dynamical deformation effects, but only the kinematical
ones (through the use of the $q$-deformed Clebsch-Gordan coefficients).
A more complete approach to the problem will be the construction of a
larger algebra, of which the quadrupole operator will be a member and it will
also be an irreducible tensor under su$_q$(2) or so$_q$(3). Work in this
direction is in progress.

\section{ Superdeformed bands}

One of the most impressive experimental discoveries in nuclear phy\-sics
during the last decade was that of superdeformation \cite{Twin811}
(see \cite{Hod365,NT533,JK321}
 for relevant reviews). The energy levels of superdeformed bands follow
the $J(J+1)$ rule much more closely than the usual rotational bands.
Levels with $J$ larger than 60 have been observed. A compilation of
superdeformed bands has been given in \cite{HW43}.
 The best examples have been found in the
A$\approx$150 mass region, while additional examples have been found in the
A$\approx$80,
 A$\approx$130 and A$\approx$190 regions. It is understood that the
superdeformed bands in the A$\approx$150 region correspond to elongated
ellipsoidal shapes with an axis ratio close to 2:1, while in the A$\approx$130
and A$\approx$190 regions the  ratios 3:2 and 1.65:1 respectively
appear closer to reality.

Since the su$_q$(2) model has been found suitable for describing normal
deformed bands, it is plausible that it will also be successful in describing
superdeformed bands as well. A test has been performed in \cite{BDRRS67}.
The su$_q$(2) model has
been found to give good results in all mass regions, the deformation
parameter $\tau$ being smaller than in the case of normal deformed bands,
thus indicating smaller deviations from the su$_q$(2) symmetry.
In particular, $\tau$ has been found to obtain values about 0.01 in the
A$\approx$130 and A$\approx$190 regions, while it obtains even smaller
values, around 0.004, in the A$\approx$150 region, which contains the best
examples of superdeformed bands observed so far. These results should be
compared to the values of $\tau$ around 0.03 obtained in the case of
normal deformations.

Concerning the corresponding B(E2) values, the experimental information
is still quite poor for allowing a meaningful comparison of the su$_q$(2)
predictions to experiment.

\section{ The physical content of the su$_q$(2) model}

{}From the above it is clear that the su$_q$(2) model offers a way of
describing
nuclear stretching, i.e. the departure of deformed nuclei from the su(2)
symmetry of the rigid rotator, similar to the one of the VMI model.
 The parameter $\tau$ describes this departure quantitatively, vanishing
in the rigid rotator limit. Therefore the deformation parameter $\tau$
should not be confused with nuclear deformation; it is in fact related
to nuclear softness, as already discussed in sec. 20.

On the other hand, the increase of the moment of inertia with increasing $J$
in the framework of the VMI model means that collectivity gets increased
\cite{BM75}.
The su$_q$(2) model is an alternative way for describing this increase
in collectivity. But increased collectivity implies increased B(E2)
transitions. Therefore it is not surprising that the su$_q$(2) model
predicts B(E2) values increasing with $J$.

Given the su$_q$(2) generators $J_+$, $J_-$, $J_0$, it is instructive to
define as usual the operators $J_x$, $J_y$, $J_z$ by
$$    J_+= J_x +i J_y, \qquad J_-=J_x-iJ_y, \qquad J_0=J_z.\eqno(23.1)$$
The su$_q$(2) commutation relations can then be rewritten in the form
$$ [J_x, J_y ] = {i\over 2} [2 J_z], \qquad [J_y, J_z]=i J_x, \qquad
[J_z, J_x] = i J_y,\eqno(23.2)$$
which is a generalization of the so(3) commutation relations, obtained
in the limit $q \rightarrow 1$. We remark that while in the classical
so(3) case the three commutation relations have exactly the same form,
in the quantum case the first commutation relation differs (in the right
hand side) from the other two, thus indicating that in the framework of
the problem under study the z-direction is not any more equivalent
to the x- and y- directions. This is of course a phenomenological way
to describe the softness of deformed nuclei by adding the appropriate
perturbations to the pure su(2) Hamiltonian and has nothing to do with
the isotropy of space, as implied in \cite{BNW375}.

It is worth remarking at this point that the Casimir operator of su$_q$(2)
is also invariant under the usual su(2) \cite{CCZB163}.
Therefore the quantum number $J$ characterizing the irreps of su$_q$(2),
and as a result the nuclear levels, is exactly the same as the quantum
number $J$ used in the case of the usual su(2). Therefore there is no
reason for a ``total reformulation of quantum mechanics'', as implied
in \cite{BNW375}, the su$_q$(2) generators being connected
to their su(2) counterparts by the $q$-deforming functionals
of sec. 14.
In other words, one continues to believe in usual angular momentum theory
and usual quantum mechanics. All what is done in the framework of the
su$_q$(2) model is to add to the usual su(2) Hamiltonian
several perturbations which have a special form making them
suitable to be summed up, including the original su(2) term,
into the form of the su$_q$(2) Hamiltonian.

\section{The u$_{p,q}$(2) rotator model}

An extension of the su$_q$(2) model is the u$_{p,q}$(2) model
\cite{Kib9358,BMK13,BMK385}, which is based
on a two-parameter deformed algebra (see sec. 15 for a list of references).
For $p=q$ (using the definition of $(p,q)$-numbers of eq. (2.11))
this model reduces to the su$_q$(2) one.
This model has been successfully applied to superdeformed nuclear bands
\cite{BMK13,BMK385}. When Taylor expanded, it becomes clear that the
eigenvalues of the  Hamiltonian of this model (which is the second order
Casimir operator of u$_{p,q}$(2)) contain
terms of the form $J(J(J+1))^n$, in addition to the $(J(J+1))^n$ ones.
It is therefore closer to the modification of su$_q$(2) which will be
discussed in subsec. 26.3.

\section{ Generalized deformed su(2) models}

Another formula giving very good results for rotational spectra has been
introduced by Holmberg and Lipas \cite{HL552}, and rediscovered in
\cite{HWZ1617}. In this case the energy levels are given by
$$ E(J)= a \left[\sqrt{1+b J(J+1)}-1\right] .\eqno(25.1)$$
Taylor expansion of the square root immediately shows that the present
formula is a special case of eq. (19.6). This formula can be derived from the
collective model of Bohr and Mottelson \cite{BM75}.

It has been argued in \cite{MWZ2545}  that the Hamiltonian of eq. (25.1)
gives better agreement to rotational nuclear spectra than the one
coming from the su$_q$(2) symmetry. Using the techniques described in
detail in sec. 17 one can construct a generalized deformed algebra
 su$_{\Phi}$(2), characterized by a function $\Phi(J(J+1)$,  giving the
spectrum
of eq. (25.1) exactly. In this particular case the algebra
 is characterized by the structure function
$$\Phi(J(J+1)) = a \left[\sqrt{1+b J(J+1)}-1\right].\eqno(25.2)$$
It is of interest to check if this choice of structure function also
improves the agreement between theory and experiment in the case of
the electromagnetic transition probabilities connecting these energy
levels. In order to study this problem, one has to construct the relevant
generalized Clebsch-Gordan coefficients. This problem is still open.

\section{Quantum algebraic description of vibrational and transitional
nuclear spectra}

We have already seen  that the su$_q$(2) model describes successfully
deformed and superdeformed bands.
It is not surprising that the applicability of the su$_q$(2) formalism
is limited to the rotational region (where the ratio $R_4= E(4)/E(2)$
obtains values between 3.0 and 3.33), since it is based on a deformation
of the rotation algebra. For describing nuclear spectra in the vibrational
($2.0 \leq R_4 \leq 2.4$) and transitional ($2.4 \leq R_4 \leq 3.0$)
regions it is clear that an extension of the model is needed. In order to
be guided towards such an extension, we briefly review the existing
experience of other successful models.

\subsection{The Interacting Boson Model}

 In the rotational (su(3)) limit \cite{AI201} of the
 Interacting Boson Model (IBM) (see secs 27, 29 for more details)
the spectrum is described by a $J(J+1)$ expression, while
in the vibrational (u(5)) \cite{AI253} and transitional (o(6))
 \cite{AI468} limits
expressions of the form $J(J+c)$  with $c>1$ appear. In the u(5) limit,
in particular, the energy levels are given by
$$E(N, n_d, v, n_{\Delta}, J, M_J )=  E_0 + \epsilon n_d +\alpha n_d (n_d+4)
+\beta 2v (v+3) +\gamma 2J(J+1), \eqno(26.1)$$
where $N$ is the total number of bosons, $n_d$ is the number of d-bosons,
$v$ is the seniority, $n_{\Delta}$ is the ``missing'' quantum number
in the reduction from o(5) to o(3), $M_J$ is the third component of the
angular momentum $J$, while $E_0$, $\epsilon$, $\alpha$, $\beta$,
$\gamma$ are free parameters. The ground state band, in particular,
is characterized by quantum numbers $n_d=0$, 1, 2, \dots, $v=n_d$,
$n_{\Delta}=0$, $J=2 n_d$, so that the energy expression for it reads
$$E(J)= E_0+ {\epsilon\over 2} J + {\alpha\over 4} J(J+8) +
{\beta \over 2} J(J+6) + 2 \gamma  J(J+1). \eqno(26.2)$$
In the o(6) limit the energy is given by
$$E(N,\sigma,\tau,\nu_{\Delta},J,M_J)= E_0+\beta 2\tau(\tau+3) +
\gamma 2J(J+1) + \eta 2\sigma(\sigma+4), \eqno(26.3)$$
where $\sigma$ is the quantum number characterizing the irreducible
representations (irreps) of o(6), $\tau$ is the quantum number
characterizing the irreps of o(5), $\nu_{\Delta}$ is the missing
quantum number in the reduction from o(5) to o(3), while $E_0$, $\beta$,
$\gamma$, $\eta$ are free parameters. The ground state band is
characterized by the quantum numbers $\sigma=N$, $\tau=0$, 1, 2, \dots,
$\nu_{\Delta}=0$, $J=2\tau$, so that the relevant energy expression
takes the form
$$E(J)= E_0 + {\beta\over 2} J(J+6) +\gamma 2J(J+1) +\eta 2N(N+4).
\eqno(26.4)$$

The message from eqs (26.2) and (26.4) is that nuclear anharmonicities
are described by expressions in which $J$ and $J^2$ appear with
different coefficients, and not with the same coefficient as in $J(J+1)$.
The earliest introduction of this idea is in fact the Ejiri formula
\cite{Eji1189}
$$ E(J) = k J(J+1) + a J , \eqno(26.5)$$
which has been subsequently justified microscopically  in \cite{DDK632}.

\subsection{Generalized VMI}

 The two-parameter VMI model is known to continue giving good fits in the
transitional and even in the vibrational region. In these regions,
however, the accuracy of the model is substantially improved by adding
a third parameter, which essentially allows for treating $J$ and $J^2$
with a different coefficient \cite{BK27,BK1879,BS1014}.

The usual VMI model has been briefly reviewed in subsec. 20.1.
One of the (essentially equivalent) three-parameter extensions of the
model, which give improved fits of vibrational and transitional
spectra, is the generalized VMI (GVMI) model \cite{BK27,BK1879}
in which the energy  levels are described by
$$ E(J) = {J+x J(J-2)\over \Phi(J)} +{1\over 2} k (\Phi(J)-\Phi_0)^2,
\eqno(26.6)$$
which can be easily rewritten in the form
$$E(J) = {J(J+x')\over 2\Phi'(J)} +{1\over 2} k' (\Phi'(J)-\Phi'_0)^2,
\eqno(26.7)$$
where $x'=x^{-1} -2$. It is clear that for $x=1/3$ the GVMI reduces
to the usual VMI, while for transitional and vibrational nuclei $x$
obtains lower values, so that $x'$ becomes greater than 1.
 The variational condition determining the moment of
inertia still has the form of eq. (20.4), while the expansion of the
energy turns out to be
$$ E(J)= {1\over 2\Phi'_0} (J(J+x') -{\sigma'\over 2} (J(J+x'))^2
$$ $$+ (\sigma')^2 (J(J+x'))^3 -3 (\sigma')^3 (J(J+x'))^4 +\dots),
\eqno(26.8)$$
where
$$\sigma' = {1\over 2 k' (\Phi'_0)^3}.\eqno(26.9)$$
We remark that an expansion in terms of $J(J+x')$ is obtained, as compared
to an expansion in terms of $J(J+1)$ in the case of the usual VMI.
The physical content of the parameters is clear: the centrifugal
stretching effect is accounted for by the softness parameter $\sigma'$,
as in the case of the usual VMI, while anharmonicities, important in the
vibrational region, are introduced by $x' >1$. Since centrifugal
stretching and anharmonicities are two effects of different origins, it is
reasonable to describe them by two different parameters.

\subsection{Modification of the su$_q$(2) model}

The evidence coming from the IBM and the generalized VMI model described
above, suggests a model in which the spectrum is given by
$$ E(J)= {1\over 2I} [J]_q [J+c]_q , \eqno(26.10)$$
which contains 3 parameters: the moment of inertia $I$, the deformation
parameter $q$ and the new parameter $c$, which is expected to be 1 in
the rotational limit and larger than 1 in the vibrational and
transitional  regions. This energy expression can be expanded as
$$E(J) = {1\over 2I} {1\over (j_0(\tau))^2} ( j_0(\tau) J(J+c)
-\tau j_1(\tau) (J(J+c))^2 $$ $$+ {2\over 3} \tau^2 j_2(\tau) (J(J+c))^3
 -{1\over 3} \tau^3 j_3(\tau) (J(J+c))^4 + {2\over 15} \tau^4
j_4(\tau) (J(J+c))^5 -\dots ), \eqno(26.11)$$
which is  similar to eq. (19.5) with $J(J+1)$ replaced by $J(J+c)$.

It is expected that the deformation parameter $\tau$, which plays the
role of the small parameter in the expansion, as the softness parameter
does in the case of the VMI, will describe the centrifugal stretching
effect, while the parameter $c$ will correspond to the anharmonicity
effects. These expectations are corroborated from  fits of
the experimental data reported in \cite{BDFRR497}.
The following comments can be made:

i) The anharmonicity parameter $c$
is clearly decreasing with increasing $R_4$, i.e. with increasing
collectivity. It obtains high values (8-18) in the vibrational region, while
in the rotational region it stays close to 1. (It should be noted that
by fixing $c=1$ in the rotational region the fits are only very slightly
changed, as expected.) In the transitional region its values are close to
3.

ii) The deformation parameter $\tau$, which corresponds to the centrifugal
stretching, is known from the su$_q$(2) model to obtain values close
to 0.3-0.4 in the rotational region, a fact also seen here. The same
range of values appears in the vibrational region as well, while in the
transitional region $\tau$ reaches values as high as 0.6. It is not
unreasonable for this parameter, which is connected to the softness of
the nucleus, to obtain its highest values in the region of $\gamma$-soft
nuclei.

iii) It is worth remarking that eq. (26.10) coincides for
$q=1$ and $c$=integer=N  with
the eigenvalues of the Casimir operator of the algebra
o(N+2) in completely symmetric states \cite{DGI873,DGI123}.
 In the rotational region
the fits gave N=1, which corresponds to o(3), as expected,
while in the transitional region the fits gave approximately N=3, which
corresponds to o(5), which is a
subalgebra contained in both the u(5) and o(6) limits of the IBM.

iv) It is also worth remarking that a special case of the expression of eq.
(26.10) occurs in the q-deformed version of the o(6) limit of the Interacting
Boson Model, which will be reported below (see eq. (29.7)).

v) The su$_q$(2) symmetry is known to make specific predictions for
the deviation of the behaviour of the B(E2) values from the rigid
rotator model (subsec. 21.3). It will be interesting to connect the spectrum
of eq. (26.10) to some deformed symmetry, at least for special values of $c$,
and examine the implications of such a symmetry for the B(E2) values.
Such a study in the framework of the q-deformed version of the o(6)
limit of IBM, mentioned in iv), is also of interest.

vi) It is worth noticing that an expansion in terms of $J(J+c)$ can also be
obtained from a generalized oscillator (sec. 12)
 with a structure function
$$ F(J) = [ J(J+c)]_Q, \eqno(26.12)$$
where $[x]_Q$ stands for the $Q$-numbers introduced in sec. 6
and $Q=e^T$, with $T$ real. This is similar to an oscillator
successfully used
for the description of vibrational spectra of diatomic molecules
(see sec. 35).
It can also be considered as a deformation of the oscillator corresponding
to the Morse potential (see sec. 35).

We have therefore introduced an extension of the su$_q$(2) model
of rotational nuclear spectra, which is applicable in the vibrational
and transitional regions as well. This extension is in agreement with
the Interacting Boson Model and the Generalized Variable Moment of
Inertia model.
In addition to the overall scale
parameter, the model contains two parameters, one related to the
centrifugal stretching and another related to nuclear anharmonicities.
In the rotational region the model coincides with the usual su$_q$(2)
model, while in the transitional region an approximate o(5)
symmetry is seen. These results give additional motivation in pursueing
the construction of a deformed version of the Interacting Boson Model.
This problem will be discussed in the next sections.

\section{ A toy Interacting Boson Model with su$_q$(3) symmetry}

The Interacting Boson Model (IBM) \cite{IA1987} is a very popular algebraic
model of nuclear structure.
In the simplest version of IBM  low lying
collective nuclear spectra are described in terms of $s$ ($J=0$)
and $d$ ($J=2$) bosons, which are supposed to be correlated
fermion pairs.
 The symmetry of the simplest version
of the model is u(6), which contains u(5) (vibrational), su(3)
(rotational) and o(6) ($\gamma$-unstable) chains of subalgebras (see also
sec. 29).
A simplified version of the model, having the su(3) symmetry
with su(2) and so(3) chains of subalgebras also exists \cite{BSR719}.
It can be considered as a toy model for two-dimensional nuclei, but
it is very useful in demonstrating the basic techniques used in the full IBM.

In the present section we will construct the $q$-deformed version of
this toy model. Since this project requires the construction of a realization
of su$_q$(3) in terms of $q$-deformed bosons, we will also use this
opportunity to study su$_q$(3) in some detail.

\subsection{The su$_q$(3) algebra}

In the classical version of the toy IBM \cite{BSR719}
 one introduces
bosons with angular momentum $m = 0, \pm 2$, represented by
the creation (annihilation) operators $a^\dagger_0$, $a^\dagger_+$,
$a^\dagger_-$
($a_0$, $a_+$, $a_-$). They satisfy usual boson commutation
relations
$$[a_i, a_j^\dagger]=\delta_{ij}, \quad [a_i, a_j]=[a_i^\dagger,
a_j^\dagger]=0.\eqno(27.1)$$
The 9 bilinear operators
$$\Lambda_{ij} = a^\dagger_i a_j \eqno(27.2)$$
satisfy then the commutation relations
$$[\Lambda_{ij}, \Lambda_{kl}] = \delta_{jk}\Lambda_{il}
-\delta_{il} \Lambda_{kj}, \eqno(27.3)$$
which are the standard u(3) commutation relations.
The total number of bosons
$$N=\Sigma_i \Lambda_{ii} = {a^\dagger_0} {a_0} + {a^\dagger_+} {a_+}
+ {a^\dagger_-} {a_-}
\eqno(27.4)$$
is kept constant. Since we are dealing with a system of bosons, only the
totally symmetric irreps $\{N,0,0\}$ of u(3) occur.

In the quantum case one has the u$_q$(3) commutation relations given in Table
2 \cite{STK437}, where $A_{ij}$ are the generators of u$_q$(3)
and the $q$-commutator is defined as
$$[A, B]_q = AB -qBA.\eqno(27.5)$$

\begin{table}[bth]
\begin{center}
\caption{ u$_q$(3) commutation relations $^{167}$,
given in the form
$[A,B]_a =C$. $A$ is given in the first column, $B$ in the
first row.
$C$ is given at the intersection of the row containing $A$ with the
column containing $B$. $a$, when different from 1, follows $C$, enclosed
in parentheses.}
\bigskip

\begin{tabular}{|c| c c c c c c |}
\hline
  & $A_{11}$ & $A_{22}$ & $A_{33}$ & $A_{12}$ & $A_{23}$ &
    $A_{13}$  \\
\hline
$A_{11}$ &0& 0 & 0 & $A_{12}$ & 0 & $A_{13}$ \\
$A_{22}$ &0&0&0& $-A_{12}$ & $A_{23}$ & 0\\
$A_{33}$ &0&0&0&0& $-A_{23}$ & $-A_{13}$  \\
$A_{12}$ & $-A_{12}$ & $A_{12}$ &0&0& $A_{13}$ ($q$)& 0 ($q^{-1}$)
 \\
$A_{23}$ & 0 & $-A_{23}$ & $A_{23}$ & $-q^{-1} A_{13} (q^{-1})$ &
0 & 0 ($q$)  \\
$A_{13}$ & $-A_{13}$ & 0& $A_{13}$ & 0($q$) & 0 ($q^{-1}$) & 0
 \\
$A_{21}$ & $A_{21}$ & $-A_{21}$ & 0& $-[A_{11}-A_{22}]$ & 0 &
$A_{23}q^{A_{11}-A_{22}}$  \\
$A_{32}$ & 0 & $A_{32}$ & $-A_{32}$ & 0 & $-[A_{22}-A_{33}]$ &
$-q^{-A_{22}+A_{33}} A_{12}$ \\
$A_{31}$ & $A_{31}$ & 0& $-A_{31}$ & $-q^{-A_{11}+A_{22}}A_{32}$ &
$-A_{21} q^{A_{22}-A_{33}}$ & $-[A_{11}-A_{33}]$
\\
\hline
\end{tabular}

\vskip 0.2truein

\begin{tabular}{| c | c c c | }
\hline
  &
    $A_{21}$ & $A_{32}$ & $A_{31}$  \\
\hline
$A_{11}$ & $-A_{21}$ &0& $-A_{31}$\\
$A_{22}$ & $A_{21}$ & $-A_{32}$&0\\
$A_{33}$ & 0& $A_{32}$ & $A_{31}$ \\
$A_{12}$ &
$[A_{11}-A_{22}]$ & 0 & $-q^{-A_{11}+A_{22}} A_{32}$ \\
$A_{23}$ &
 0 & $[A_{22}-A_{33}]$ & $A_{21} q^{A_{22}-A_{33}}$ \\
$A_{13}$ &
$-A_{23} q^{A_{11}-A_{22}}$ & $q^{-A_{22}+A_{33}} A_{12}$ &
$[A_{11} -A_{33}]$ \\
$A_{21}$ &
 0 & $-q A_{31} (q)$ & 0 ($q^{-1}$) \\
$A_{32}$ &
 $A_{31} (q^{-1})$ & 0 & 0 ($q$) \\
$A_{31}$ &
 0 ($q$) &
0 ($q^{-1}$) & 0\\
\hline
\end{tabular}
\end{center}
\end{table}

In order to obtain a realization of u$_q$(3) in terms of
the $q$-bosons described in sec. 10,
one starts with  \cite{Kul195}
$$A_{12}= a_1^\dagger a_2, \quad A_{21}= a_2^\dagger a_1,\quad
A_{23}= a_2^\dagger a_3 , \quad A_{32}= a_3^\dagger a_2 .\eqno(27.6)$$
One can easily verify that the u$_q$(3) commutation relations
involving these generators are satisfied. For example, one
has
$$[A_{12}, A_{21}]=  [N_1-N_2], \qquad
  [A_{23}, A_{32}]= [N_2 -N_3], \eqno(27.7) $$
using the identity of eq. (2.6) and the identifications
$$N_1 = A_{11}, \quad N_2= A_{22}, \quad N_3= A_{33}.\eqno(27.8) $$
One can now determine the boson realizations of $A_{13}$
and $A_{31}$ from other commutation relations, as follows
$$ A_{13}= [A_{12}, A_{23}]_q = a_1^\dagger a_3 q^{-N_2},\eqno(27.9)$$
$$ A_{31}= [A_{32}, A_{21}]_{q^{-1}}= a_3^\dagger a_1 q^{N_2}.\eqno(27.10) $$
Using eq. (2.6) once more one can verify that
the relation
$$[A_{13}, A_{31}]= [N_1 -N_3] \eqno(27.11)$$
is fulfilled by the boson images of (27.9), (27.10). It is by now
a straightforward task to verify that all commutation relations
of Table 2 are fulfilled by the boson images obtained above.

Before turning to the study of the two limits of the model, we give for
completeness some additional information on su$_q$(3):

i) The irreps of su$_q$(3) have been studied in
\cite{STK437,ST375,MSK595,MSK1031,PC4017,Que5977,Cap5942,Yu399,Yu5881,YY3025}.

ii) Clebsch-Gordan coefficients for su$_q$(3) have been given
in \cite{STK437,Yu5881,SK263,SK690}.

iii) The Casimir operators of su$_q$(3) and their eigenvalues have been
given explicitly in \cite{RP2020}. Using the Elliott quantum
numbers \cite{Ell128,Ell562,Ell557}
$$\lambda =f_1-f_2, \qquad \mu=f_2,\eqno(27.12)$$
where $f_i$ represents the number of boxes in the $i$-th line of the
corresponding Young diagram, the irreps of su$_q$(3) are labelled as
$(\lambda, \mu)$. The eigenvalues of the second order Casimir operator
then read
$$ C_2= \left[{\lambda\over 3}-{\mu\over 3}\right]^2+\left[{2\lambda\over 3}
+{\mu\over 3}+1\right]^2+\left[{\lambda\over 3}+{2\mu\over 3}+1\right]^2-2,
\eqno(27.13)$$
while the eigenvalues of the third order Casimir operator are
$$C_3=2\left[{\lambda\over 3}-{\mu \over 3}\right] \left[{2\lambda \over 3}
+{\mu\over 3}+1\right] \left[ {\lambda\over 3}+{2\mu\over 3}+1\right].
\eqno(27.14)$$
In the limit $q\to 1$ these reproduce the ordinary results for su(3):
$$C_2={2\over 3} (\lambda^2+\mu^2+\lambda \mu + 3\lambda+3\mu),\eqno(27.15)$$
$$C_3={2\over 27} (\lambda-\mu) (2\lambda+\mu+3)(\lambda+2\mu+3).\eqno(27.16)$$

iv) A different deformation of sl(3) has been studied in \cite{BDF74}.

\subsection{The su$_q$(2) limit}

We shall study the su$_q$(2) limit of the model first, since it is
technically less demanding.

So far we have managed to write a boson realization of u$_q$(3)
in terms of 3 $q$-bosons, namely $a_1$, $a_2$, $a_3$. Omitting
the generators involving one of the bosons,
one is left with an su$_q$(2) subalgebra. Omitting the generators
involving $a_3$, for example, one is left with $A_{12}$,
$A_{21}$, $N_1$, $N_2$, which satisfy usual su$_q$(2)
commutation relations if the identifications
$$ J_+= A_{12}, \quad J_-=A_{21}, \quad J_0={1\over 2} (N_1-N_2)
\eqno(27.17)$$
are made. $J_0$ alone forms an so$_q$(2) subalgebra. Therefore
the relevant chain of subalgebras is
$$ {\rm su}_q(3)\supset {\rm su}_q(2) \supset {\rm so}_q(2) .\eqno(27.18)$$

The second order Casimir operator of su$_q$(2) has been given in eq. (14.9).
Substituting the above expressions for the generators one finds
$$C_2({\rm su}_q(2))= \left[{N_1+N_2\over 2}\right] \left[{N_1+N_2\over 2}
+1\right].\eqno(27.19)$$

All of the above equations go to their classical counterparts by
allowing $q\rightarrow 1$, for which $[x]\rightarrow x$, i.e.
$q$-numbers become usual numbers. In the classical case \cite{BSR719} out
of the three bosons ($a_0$, $a_+$, $a_-$) forming su(3), one chooses
to leave out $a_0$, the boson with zero angular momentum,
 in order to be left with the su(2) subalgebra
formed by $a_+$ and $a_-$, the two bosons of angular momentum two.
The choice of the su$_q$(2) subalgebra made above  is then consistent
with the following correspondence between classical bosons and
$q$-bosons
$$ a_+\rightarrow a_1, \quad a_-\rightarrow a_2, \quad
a_0\rightarrow a_3.\eqno(27.20)$$
(We have opted in using different indices for usual bosons and
$q$-bosons  in order to avoid confusion.)

In the classical case  the states of the system are characterized by
the quantum numbers characterizing the irreducible representations
(irreps) of the algebras appearing in the classical
counterpart of the chain of eq. (27.18). For su(3) the total number
of bosons $N$ is used. For su(2) and so(2) one can use the
eigenvalues of $J^2$ and $J_0$, or, equivalently, the eigenvalues
of $a^\dagger_+ a_+ + a_-^\dagger a_-$ and $L_3= 4 J_0$, for which we use the
symbols $n_d$ (the number of bosons with angular momentum 2)
and $M$. Then the basis in the classical case can be written as \cite{BSR719}
$$|N,n_d,M> = {(a_0^\dagger)^{N-n_d}\over (N-n_d)!}
{(a^\dagger_+)^{n_d/2+M/4}\over (n_d/2+M/4)!}
{(a_-^\dagger)^{n_d/2-M/4}\over (n_d/2-M/4)!} |0>.\eqno(27.21) $$
In the quantum case for each oscillator one defines the basis as
in sec. 10. Then  the full basis in the $q$-deformed case is
$$ |N, n_d, M>_q = {(a_3^\dagger)^{N-n_d} \over [N-n_d]!}
{(a_1^\dagger)^{n_d/2+M/4}\over [n_d/2+M/4]!}
{(a_2^\dagger)^{n_d/2-M/4}\over [n_d/2-M/4]!} |0>,\eqno(27.22) $$
where $N=N_1 +N_2 +N_3$ is the total number of bosons,
$n_d = N_1 + N_2$ is the number of bosons with angular momentum 2,
and $M$ is the eigenvalue of $L=4J_0$.
$n_d$ takes values from 0 up to $N$, while for a given value
of $n_d$, $M$ takes the values $\pm 2n_d$,
$\pm 2(n_d-2)$, \dots, $\pm 2$ or 0, depending on whether $n_d$
is odd or even.
In this basis the eigenvalues of the second order Casimir
operator of su$_q$(2) are then
$$ C_2({\rm su}_q(2)) |N, n_d, M>_q = \left[{n_d\over 2} \right]
\left[{n_d\over 2}+1\right]
|N, n_d, M>_q .\eqno (27.23)$$
In the case of $N=5$ one can easily see that the spectrum will
be composed by the ground state band, consisting of states
with $M=0$, 2, 4, 6, 8, 10 and $n_d=M/2$, the first excited
band with states characterized by $M=0$, 2, 4, 6 and
$n_d=M/2+2$, and the second excited band, containing states
with $M=0$, 2 and $n_d=M/2+4$.

In the case that the Hamiltonian has the su$_q$(2)
dynamical symmetry, it can be written in terms of the Casimir
operators of the chain (27.18). Then one has
$$ H = E_0 + A C_2({\rm su}_q(2)) + B C_2({\rm so}_q(2)),\eqno(27.24)$$
where $E_0$, $A$, $B$ are constants.
Its eigenvalues are
$$ E= E_0 + A \left[{n_d \over 2}\right] \left[{n_d\over 2}+1\right] +
B M^2.\eqno(27.25)$$

Realistic nuclear spectra are characterized by strong
electric qua\-dru\-po\-le transitions among the levels of the
same band, as well as by interband transitions. In the
framework of the present toy model one can define, by analogy
to the classical case \cite{BSR719},
 quadrupole transition operators
$$Q_+= a_1^\dagger a_3 + a_3^\dagger a_2, \qquad
  Q_-= a_2^\dagger a_3 + a_3^\dagger a_1.\eqno(27.26)$$
In order to calculate  transition matrix elements of these
operators one only needs eqs. (10.7), (10.8), i.e. the action of the
$q$-boson operators on the $q$-deformed basis. The selection
rules, as in the classical case, are $\Delta M=\pm 2$,
$\Delta n_d=\pm 1$, while the corresponding matrix elements
are
$$_q<N, n_d+1, M\pm 2| Q_{\pm}| N,n_d,M>_q =
\sqrt{[N-n_d] [{n_d\over 2}\pm{M\over 4} +1]}, \eqno(27.27)$$
$$_q<N, n_d-1, M\pm 2| Q_{\pm} | N, n_d,M>_q =
\sqrt{[N-n_d+1][{n_d\over 2}\mp {M\over 4}]}.\eqno(27.28)$$
{}From these equations it is clear that both intraband and interband
transitions are possible.

In order to get a feeling of the qualitative changes in the
spectrum and the transition matrix elements resulting from the
$q$-deformation of the model, a simple calculation for
a system of 20 bosons ($N=20$) has been performed in \cite{BDFRR267}.
Two cases are distinguished:
i) $q$ real  ($q=e^{\tau}$, with $\tau$ real),
ii) $q$ a phase factor ($q=e^{i\tau}$, with $\tau$ real). The main conclusions
are:

i) When $q$ is real the spectrum is
increasing more rapidly than in the classical case, while when
$q$ is a phase the spectrum increases more slowly than in the
classical case, in agreement with the findings of the
$q$-rotator model.

ii) The transition matrix elements in the case that
$q$ is real increase more rapidly than in the classical case,
while they increase less rapidly than in the classical case when
$q$ is a phase.

iii) Transition matrix elements
are much more sensitive to $q$-de\-fo\-rma\-tion than energy spectra.
This is an interesting feature, showing that $q$-deformed
algebraic models can be much more flexible in the description of
transition probabilities than their classical counterparts.

\subsection { The so$_q$(3) limit}

The classical su(3) toy model has, in addition to the above mentioned su(2)
chain of subalgebras, an so(3) chain. However, the problem of constructing
 the su$_q$(3)$\supset$so$_q$(3) decomposition is a very di\-f\-fi\-cult one.
Since this decomposition is needed in constructing the $q$-deformed versions
of several collective models, including the Elliott model
\cite{Ell128,Ell562,Ell557,Har67}
and the su(3) limit of the IBM,
we report here the state of the art in this problem:

i) As far as the completely symmetric irreps of su$_q$(3) are concerned,
the problem has been solved by Van der Jeugt
\cite{VdJ213,VdJ131,VdJ1799,VdJ947,VdJ948}. This suffices for our
needs in the framework of the toy IBM, since only completely symmetric
su$_q$(3) irreps occur in it.

ii) Sciarrino \cite{Sci92} started from so$_q$(3) and obtained a deformed gl(3)
containing so$_q$(3) as a subalgebra. However, it was not clear how to
impose the Hopf structure on this larger algebra. Trying the other way
around, he found that by starting from a gl$_q$(3) algebra, which already
possesses the Hopf structure, one loses the Hopf structure of the
principal 3-dim subalgebra, which should have been so$_q$(3).

iii) Pan \cite{Pan257} and Del Sol Mesa {\it et al} \cite{Del1147} attacked
the problem through the use of $q$-deforming functionals of secs 10, 14.

iv) Quesne \cite{Que81} started with $q$-bosonic operators transforming as
vectors under so$_q$(3) and constructed a $q$-deformed u(3) by
tensor coupling.

v) $q$-deformed subalgebras of several $q$-deformed algebras have recently
been studied by Sciarrino \cite{Sci94}.

vi) A simplified version of the so$_q$(3) subalgebra of u$_q$(3) has been
constructed in \cite{BRRT95}. Furthermore, explicit
expressions for the irreducible vector and quadrupole tensor operators under
so$_q$(3) have been given and the matrix elements of the latter have been
calculated.

In what follows, it suffices to use the solution given in \cite{VdJ213},
since in the model under study only completely symmetric irreps
of u$_q$(3) enter. Using the notation
$$ a_+\rightarrow a_1, \quad a_-\rightarrow a_2, \quad
a_0 \rightarrow a_3, \eqno(27.29)$$
the  basis states are of the form
$$ | n_+ n_0 n_- > =
{ (a_+^\dagger)^{n_+} (a_0^\dagger)^{n_0} (a_-^\dagger)^{n_-} \over
\sqrt{[n_+]! [n_0]! [n_-]!}} |0>, \eqno(27.30)$$
with $a_i |0>=0$ and
$N_i|n_+ n_0 n_-> = n_i |n_+n_0n_->$, where $i=+,0,-$.
The principal subalgebra so$_q$(3) is  then generated by
$$L_0= N_+-N_-, \eqno(27.31)$$
$$L_+= q^{N_--{1\over 2} N_0}  \sqrt{q^{N_+} + q^{-N_+} } \quad
a_+^\dagger a_0 \quad+\quad a_0^\dagger a_- \quad q^{N_+-{1\over 2}N_0}
\sqrt{q^{N_-}+q^{-N_-}}, \eqno(27.32)$$
$$L_-= a_0^\dagger a_+ q^{N_--{1\over 2} N_0}\sqrt{q^{N_+}+q^{-N_+}}
+ q^{N_+-{1\over 2} N_0} \sqrt{q^{N_-}+q^{-N_-}} a_-^\dagger a_0,
\eqno(27.33)$$
satisfying the commutation relations
$$ [L_0, L_{\pm}] =\pm L_{\pm}, \quad  [L_+, L_-]= [2L_0].
\eqno(27.34)$$
$L_0$ alone generates then the so$_q$(2) subalgebra.
Therefore the relevant chain of subalgebras is
$$ {\rm su}_q(3)\supset {\rm so}_q(3) \supset {\rm so}_q(2) .\eqno(27.35)$$
The so$_q$(3) basis vectors can be written in terms of the vectors
of eq. (27.30) as
$$ |v(N,L,M)>= q^{-[(L+M)(L+M-1)]/4}
 \sqrt{{[N+L]!![2L+1] [L+M]! [L-M]!\over [N-L]!! [N+L+1]!}} $$
$$\sum_x q^{(2L-1)x/2} \quad s^{(N-L)/2} \quad
{|x, L+M-2x, x-M>\over
\sqrt{[2x]!![L+M-2x]![2x-2M]!!}}, \eqno(27.36)$$
where
$$s= (a_0^\dagger)^2 q^{N_++N_-+1} -\sqrt{{[2N_+][2N_-]\over [N_+]
[N_-]}} a_+^\dagger a_-^\dagger q^{-N_0-{1\over 2}}, \eqno(27.37)$$
$x$ takes values from max(0,$M$) to $[(L+M)/2]$ in steps of 1,
$L = N,$ $N-2,$ $\ldots,$ 1 or 0, $M=-L,$ $-L+1,$ $\ldots,$ $ +L$, and
$[2x]!!=[2x] [2x-2] \ldots [2]$. The action of the generators of
so$_q$(3) on these states  is given by
$$L_0 |v(N,L,M)> = M \quad |v(N,L,M)>, \eqno(27.38)$$
$$L_{\pm}\quad |v(N,L,M)>=
\sqrt{[L\mp M]  [L\pm M+1]} \quad |v(N,L,M\pm 1)>.\eqno(27.39)$$
The second order Casimir operator of so$_q$(3)
has the form
$$C_2({\rm so}_q(3))=  L^2 = L_- L_+ + [L_0] [L_0+1].\eqno(27.40)$$
Its eigenvalues in the above basis are given by
$$C_2({\rm so}_q(3))\quad |v(N,L,M)> = [L] [L+1]\quad |v(N,L,M)>.\eqno(27.41)$$

All of the above equations go to their classical counterparts by
allowing $q\rightarrow 1$, for which $[x]\rightarrow x$, i.e.
$q$-numbers become usual numbers. In the classical case \cite{BSR719}
 the states of the system are characterized by
the quantum numbers characterizing the irreducible representations
(irreps) of the algebras appearing in the classical
counterpart of the chain of eq. (27.35). For su(3) the total number
of bosons $N$ is used. For so(3) and so(2) one can use $L$ and
$M$, respectively. In \cite{BSR719}, however, the eigenvalue of
 $L_0' =2 L_0$ is used, which is $M' =2M$.

Since the rules for the decomposition of the totally symmetric
u$_q$(3) irreps into so$_q$(3) irreps are the same as in the classical
case, it is easy to verify that for a system with
 $N=6$  the spectrum will
be composed by the ground state band, consisting of states
with $M'=0$, 2, 4, 6, 8, 10, 12 and $L=N$, the first excited
band with states characterized by $M'=0$, 2, 4, 6, 8 and
$L=N-2$, the second excited band, containing states
with $M=0$, 2, 4 and $L=N-4$, and the third  excited band,
containing a state with $M'=0$ and $L=N-6$.

In the case that the Hamiltonian has the so$_q$(3)
dynamical symmetry, it can be written in terms of the Casimir
operators of the chain (27.35). Then one has
$$ H = E_0 + A C_2({\rm so}_q(3)) + B C_2({\rm so}_q(2)),\eqno(27.42)$$
where $E_0$, $A$, $B$ are constants.
Its eigenvalues are
$$ E= E_0 + A [L] [L+1] + B M^2.\eqno(27.43)$$
It is then clear that in this simple model the internal structure
of the rotational bands is not influenced by $q$-deformation.
What is changed is the position of the bandheads.

We turn now to the study of electromagnetic transitions. In the
present limit one can define, by analogy
to the classical case,  quadrupole transition operators $Q_{\pm}$
proportional to the so$_q$(3) generators $L_{\pm}$
$$Q_{\pm} = L_{\pm}.\eqno(27.44)$$
In order to calculate  transition matrix elements of these
operators one only needs eq. (27.39).
 The selection
rules, as in the classical case, are $\Delta M'=\pm 2$,
$\Delta L =0$, i.e. only intraband transitions are allowed.
The relevant matrix elements are
$$ <v(N,L,M'+2)| Q_+ | v(N,L,M')> =
\sqrt{\left[L-{M'\over 2}\right] \left[L+{M'\over 2}+1\right]}, \eqno(27.45)$$
$$ <v(N,L,M'-2)| Q_- | v(N,L, M')> =
\sqrt{\left[L+{M'\over 2}\right] \left[L-{M'\over 2}+1\right]}. \eqno(27.46)$$

In order to get a feeling of the qualitative changes in the
spectrum and the transition matrix elements resulting from the
$q$-deformation of this limit of the model,  a simple calculation
has been done in \cite{BonNii}. Again the cases of $q$ being real
or $q$ being a phase factor have been considered. The main conclusions are:

i) $q$-deformation influences only the
position of bandheads, while it leaves the internal structure of
the bands intact.

ii)  When $q$ is real the bandheads are
increasing more rapidly than in the classical case, while when
$q$ is a phase the bandheads increase more slowly than in the
classical case. This result is in qualitative agreement with the
findings of the su$_q$(2) model.

iii) Transition matrix elements in the case that
$q$ is real have values higher than in the classical case,
while they have values lower than in the classical case when
$q$ is a phase.

\section{ The question of complete breaking of symmetries and some
applications}

In the cases examined so far the $q$-deformed symmetries considered were close
to their classical counterparts, to which they reduce for $\tau=0$ ($q=1$),
since the values of $\tau$ were relatively small. One can then argue that
results similar to the ones provided by the quantum symmetries can also
be obtained from the usual Lie symmetries through the addition of suitable
perturbations. What can be  very useful is to start with one limiting
symmetry and, through large deformations, reach another limiting symmetry.
We shall refer to this as the complete breaking of the symmetry.
In this way one could hope to ``bridge'' different Lie symmetries through
the use of $q$-deformations, providing in addition new symmetries in the
regions intermediate between the existing Lie ones.

The question of complete breaking of the symmetry in the framework of the toy
IBM under study has been studied by Cseh \cite{Cseh1225}, Gupta
\cite{Gup1067}, and Del Sol Mesa {\it et al.} \cite{Del1147}.

Cseh \cite{Cseh1225} started with the su$_q$(2) (vibrational) limit (in a
form different from  the one used above) and tried to
reach the so$_q$(3) (rotational) limit. He noted that for $q$ being a
phase factor this is not possible, while for real $q$ some rotational features
are obtained, but without all of the requirements for rotational behaviour
being satisfied simultaneously.

Gupta \cite{Gup1067} started with the su$_q$(2) limit, in the form given
above. He
noted that for $q$ being a phase factor a recovery of the su(3) symmetry occurs
(see also the next paragraph),
while for real $q$, and even better
for $q$ complex ($q=e^s$ with $s=a+ib$), the so$_q$(3) limit is indeed
reached.

Del Sol Mesa {\it et al.} \cite{Del1147} considered the o$_q$(3) limit
of the model, since it corresponds to the symmetry of a $q$-deformed version
of the spherical Nilsson Hamiltonian with spin-orbit coupling term. They
found that for $q$ being a phase factor ($q=e^{i\tau}$) and for $\tau$
obtaining values in the region $0.5\leq \tau\leq 2$ the u(3) symmetry,
which is broken in the initial model because of the presence of the spin-orbit
term, is recovered. This offers a way of recovering the u(3) symmetry
alternative to the one developed for the spherical Nilsson model
\cite{BCM2841,CMQ238} and the deformed Nilsson model \cite{MDel146,CVHH303}
through the use of appropriate unitary operators.

Complex deformations have also been used in \cite{GCLG73} in
the framework of a deformed u(2) model, possessing the u(2)$\supset$u(1)
and u(2)$\supset$o(2) chains. Again it has been found that complex
deformations can bridge the two limiting symmetries.

Possible complete breaking of the symmetry has also been studied in the
framework of the $q$-deformed version of the full Interacting Boson Model
(see sec. 29).

A problem associated with complex deformations as the ones considered
above is that the energy eigenvalues become complex as well. A way to
avoid this problem has been introduced recently by Jannussis and
collaborators \cite{FJMS94,ABF605}.

Finally, the o$_q$(3) limit of the model has been used for describing the
$^{16}$O + $\alpha$ cluster states in $^{20}$Ne \cite{Cseh63}. It turns out
that an
improved description of the energy spectrum and the $\alpha$-particle
spectroscopic factors occurs for $q=e^{0.124}$.

\section{ $q$-deformation of the Interacting Boson Model (IBM)}

The Interacting Boson Model \cite{AI201,AI253,AI468}
(see \cite{IA1987,Bon88} for recent overviews) is the
most popular algebraic model of nuclear structure. It describes the
collective properties of medium-mass and heavy nuclei away from closed
shells in terms of bosons, which correspond to correlated valence fermion
pairs. In its simplest form, called IBM-1, only  $s$ ($J=0$) and $d$ ($J=2$)
bosons are used. The overall symmetry of the model is u(6), possessing
three limiting symmetries: the u(5) (vibrational) limit, corresponding
to the chain of subalgebras
$$ {\rm u}(6) \supset {\rm u}(5) \supset {\rm o}(5) \supset {\rm o}(3),
\eqno(29.1)$$
the su(3) (rotational) limit, characterized by the chain
$$ {\rm u}(6) \supset {\rm su}(3) \supset {\rm o}(3),\eqno(29.2)$$
and the o(6) ($\gamma$-unstable) limit, for which the relevant chain is
$$ {\rm u}(6) \supset {\rm o}(6) \supset {\rm o}(5) \supset {\rm o}(3).
\eqno(29.3)$$

If one of these dynamical symmetries is present, the Hamiltonian can be
written in terms of the Casimir operators of the algebras appearing in
the relevant chain. Thus the Hamiltonian can be analytically diagonalized
in the corresponding basis. This is a great advantage of IBM and its
numerous generalizations: they provide us with a large number of exactly
soluble models, the predictions of which can be directly compared to
experiment, without any need for lengthy numerical calculations.

{}From what we have already seen in sec. 27, it is worth examining if
a $q$-deformed version of the IBM has any advantages in comparison to the
standard version. In order to accomplish this, one has to construct
the $q$-analogues of the three chains mentioned above. The difficulties
associated with the su(3) chain have already been discussed in subsec. 27.3.
In what follows  we are going to focus attention on the o(6) chain, for
which the relevant construction has been carried out \cite{WY492}.
The technique used is based on the notion of complementary subalgebras,
which is explained in detail in \cite{ABS1088}, while here only final results
will be reported. We only mention here that the notion of complementary
subalgebras was introduced by Moshinsky and Quesne
\cite{MQ1631,CDQ779,Que2675}. Two subalgebras
A$_1$ and A$_2$ of a larger algebra A are complementary within a definite
irrep of A, if there is an one-to-one correspondence between all the irreps
of A$_1$ and A$_2$ contained in this irrep of A.

In the o(6) limit of IBM the Hamiltonian is
$$ H= E_0 + \beta C_2({\rm o}(5))+\gamma C_2({\rm o}(3)) + \eta C_2({\rm o}
(6)).\eqno(29.4)$$
The eigenvalues of the energy in the relevant basis have already been given
in subsec. 26.1.

Using the notion of complementarity it turns out that, instead of using
the o(6) chain mentioned above, it suffices to study the chain
$$ {\rm su}^{sd}(1,1) \otimes {\rm so}(6) \supset {\rm su}^d(1,1)
\otimes {\rm so}(5) \supset {\rm so}(3), \eqno(29.5)$$
where su$^{sd}$(1,1) is the algebra closed by the pair operators formed
out of the $s$ and $d$ bosons, while su$^d$(1,1) is the algebra closed
by the pair operators formed out of $d$ bosons alone.
(Details on the basis for symmetric irreps of su(1,1)$\otimes$o(5)
can be found in \cite{SG76}.)  The irreps of
su$^{sd}$(1,1) are characterized by the same quantum numbers as the
irreps of o(6) in the o(6) chain of the IBM, while the irreps of
su$^d$(1,1) are characterized by the same quantum numbers as the irreps
of o(5) in the o(6) limit of IBM.
Therefore in the Hamiltonian one can use the Casimir operators of the
su$^{sd}$(1,1), su$^d$(1,1) and su(2) subalgebras (the deformed versions
of which are well known, as seen in secs 14--16) instead of the
Casimir operators of o(6), o(5), o(3) respectively. Keeping the same
notation as in eq. (26.3) the final result reads
$$E(N,\sigma,\tau,\nu_{\Delta},J,M_J) = E_0 + \beta 8 \left[{\tau\over 2}
\right]_q \left[{\tau+3\over 2}\right]_q $$ $$+ \gamma 2 [J]_q [J+1]_q +\eta 8
 \left[{\sigma\over 2}\right]_q  \left[ {\sigma+4\over 2}\right]_q,
\eqno(29.6)$$
where the free parameters have been chosen so that the present equation
reduces to its classical counterpart for $q\rightarrow 1$.
For the ground state band then the analog of eq. (26.4) is
$$ E(J) = E_0+ \beta' 8 [J]_{q^{1/4}} [J+6]_{q^{1/4}} + \gamma 2 [J]_q [J+1]_q
+\eta' 8 [N]_{q^{1/2}} [N+4]_{q^{1/2}},\eqno(29.7)$$
where the identities
$$ \left[ {x \over 2} \right]_q = [ x]_{q^{1/2}} (q^{1/2}+q^{-1/2})^{-1},
\eqno(29.8)$$
$$   \left[ {x \over 4} \right]_q = [ x]_{q^{1/4}} (q^{1/2}+q^{-1/2})^{-1}
                        (q^{1/4}+q^{-1/4})^{-1}, \eqno(29.9)$$
have been used and $\beta'$, $\eta'$ are related to $\beta$, $\eta$ and $q$
in an obvious way.
We remark that the Casimir operator of su$_q^d$(1,1),
which is complementary to o(5) in the undeformed case, leads to
a term of the form $[J]_{q'} [J+6]_{q'}$ with $q'=q^{1/4}$.

In refs \cite{GL593,Gup117}
 the question has been studied if large values of the
deformation parameter can lead us from the o(6) limit to the su(3)
(rotational) or u(5) (vibrational) limits, so that complete breaking
of the symmetry, in the sense of sec. 28, could be obtained. It turns out
that for $q$ real the spectrum of the ground state band goes towards the
rotational limit, while $q$ being a phase factor leads towards the vibrational
limit. Many more detailed studies, of both spectra and electromagnetic
transition probabilities, are required before such a claim can be made.

A different method for constructing the $q$-deformed versions of the u(5)
and o(6) limits of the IBM has been used by Pan \cite{Pan1876}.
The method is based on the use of $q$-deforming functionals (see secs 10, 14).
The same method has been used in \cite{Pan257} for studying the $q$-deformed
version of the su(3)$\supset$so(3) decomposition. The final  result
for the energy eigenvalues in the o(6) case is similar to the one reported in
eq. (29.6), the main difference being that different deformation parameters
are allowed in each of the three deformed terms in the rhs. Some comparisons
of the model predictions for spectra and B(E2) values to the experimental data
have been performed \cite{Pan1876}.

It is clear that several deformed versions of the IBM can be constructed,
providing us with a large number of exactly soluble models.
In order to demonstrate their usefulness, one has to show that by deforming
the model one gets some advantages over the classical (non-deformed)
version. One way to achieve this is the use of parameter-independent
tests based on systematics of the data, like the ones used in
\cite{Bon2001,BSR63,BSR865} for the usual IBM. It is also
desirable for the deformation parameter to be associated with some
physical quantity, as in the case of the su$_q$(2) model. Much work is
still required in these directions. Some mathematical results which
can be useful in these efforts are reported below:

i) Casimir operators for su$_q$(n) have been given in \cite{Bin1133},
while the quadratic Casimir of so$_q$(5) can be found in \cite{ZGB937}.

ii) Raising and lowering operators for u$_q$(n) have been given by
Quesne \cite{Que357}.

iii) Irreps of u$_q$(m+n) in the u$_q$(m)$\oplus$u$_q$(n) basis have been
constructed in \cite{Pan1992}, while generalized $q$-bosonic operators acting
in a tensor product of $m$ Fock spaces have been constructed as double
irreducible tensors with respect to u$_q$(m)$\oplus$u$_q$(n) in
\cite{Que298,Que322}.

\section{Deformed versions of other collective models}

The Moszkowski model \cite{Moszk} is a schematic two-level model which
provides a description of the phase transition from
the vibrational regime to the rotational one. A $q$-deformed version
of the model has been developed \cite{MAP6317}
and the RPA modes in it have been discussed \cite{BBMPP895}.
Furthermore, the $q$-deformed Moszkowski model with cranking has been
studied in the mean field approximation and the relation between
$q$-deformation and temperature has been discussed \cite{PBPBM1209}.
It should be noticed
here that quantum algebraic
techniques have also been found useful in describing thermal effects in
the framework of the $q$-deformed Thouless model for superconductivity
\cite{BAMP192}.

The Lipkin--Meshkov--Glick (LMG) model \cite{LMG188}
is an exactly soluble schematic shell
model. $q$-deformed versions of the 2-level LMG model (in terms of an
su$_q$(2) algebra) \cite{AB174,LS50,AMMC831,ACMM701,AEGPL4915}
and of the 3-level LMG model (in terms of an su$_q$(3)
algebra) \cite{Brito92} have been developed.

\section{Fermion pairs as deformed bosons: approximate mapping}

We have seen so far that several quantum algebraic phenomenological models
have been proposed for the description of nuclear collective properties.
These models make use of $q$-deformed bosons, which satisfy commutation
relations differing from the standard boson commutation relations, to
which they reduce in the limit $q\rightarrow 1$.

On the other hand, it is known that vibrational nuclear spectra, which are
described in the simplest way by a pairing Hamiltonian, show anharmonicities
(see also sec. 26), described, for example, by the {\sl Anharmonic Vibrator
 Model (AVM)} \cite{DDK632}
$$ E(J)=aJ+bJ(J-2).\eqno(31.1)$$
In the framework of the single-j shell model
\cite{DK236,VD272,VKD2188,Klein39,BKL521},
which can be extended to several non-degenerate j-shells
\cite{BK87,MBK381,KM441} these anharmonicities are related to the fact
that correlated fermion pairs satisfy commutation relations
which resemble boson commutation relations but in addition include corrections
due to the presence of the Pauli principle. This fact has been
the cause for the development of boson mapping techniques
(see the recent reviews by Klein and Marshalek \cite{KM375} and Hecht
\cite{Hecht87}  and references therein), by
which the description of systems of fermions in terms of bosons
is achieved. In recent years boson mappings have attracted
additional attention in nuclear physics as a necessary tool
in providing a theoretical justification for the success of
the phenomenological Interacting  Boson Model \cite{IA1987}
 and its various extentions, in
which low lying collective states of medium and heavy mass
nuclei are described in terms of bosons.

{}From the above observations it is clear that both q-bosons
and correlated fermion pairs satisfy commutation relations
which resemble the standard boson commutation relations but
they deviate from them, due to the $q$-deformation in the former
case and to the Pauli principle in the latter.
A question is thus created: Are $q$-bosons
suitable for the approximate description of correlated fermion
pairs? In particular, is it possible to construct a boson mapping
in which correlated fermion pairs are mapped onto $q$-bosons, in a way
that the $q$-boson operators approximately satisfy the same
commutation relations as the correlated fermion pair operators?
In this section we show for the simple case of su(2) that such
a mapping is indeed possible.

\subsection{The single-j shell model}

Let us consider the single-j shell model
\cite{DK236,VD272,VKD2188,Klein39,BKL521}.
One can define fermion pair and multipole operators as
$$A_{JM}^\dagger = {1\over \sqrt{2}} \sum_{m m'} ( j m j m' | J M)
a^\dagger_{jm} a^\dagger_{jm'}, \eqno(31.2)$$
$$B_{JM}= {1\over \sqrt{2J+1}} \sum _{m m'} (j m j -m'| J M)
(-1)^{j-m'}  a^\dagger_{jm} a_{jm'}, \eqno(31.3)$$
with the following definitions
$$ A_{JM}= [ A^\dagger_{JM}]^\dagger , \quad B^\dagger_{JM}=
[B_{JM} ]^\dagger.\eqno(31.4)$$
In the above $a_{jm}^\dagger$ ($a_{jm}$) are fermion creation (annihilation)
operators and $(jm jm'|JM)$ are the usual Clebsch--Gordan
coefficients.

The pair and multipole operators given above satisfy the following
commutation relations:
$$[ A^\dagger_{JM}, A^\dagger_{J'M'} ] =0, \eqno(31.5)$$
$$[A_{JM}, A^\dagger_{J'M'}]=\delta_{JJ'} \delta_{M M'}
-2\sum_{J''} (-1)^{2j+M} \sqrt{(2J+1)(2J'+1)(2J''+1)} $$
$$(J -M J' M' | J'' M'-M) \left\{\matrix{J & J' & J''\cr
j & j & j \cr}\right\} B_{J'', M'-M}, \eqno(31.6)$$
$$[ B_{JM}^\dagger, A^\dagger_{J'M'}]= \sum_{J''} 2 \sqrt{2J'+1} (-1)^{2j-M}
(J -M J' M' | J'' M'-M)                             $$
$$\left\{\matrix{J & J' & J''\cr j & j & j\cr}\right\}
A^\dagger_{J'', M'-M} {1+(-1)^{J''} \over 2}, \eqno(31.7)$$
$$[B_{JM}, B_{J'M'}] = \sum_{J''} (-1)^{2j-J''}
[1-(-1)^{J+J'+J''}] \sqrt{2J''+1}$$ $$ (J M J' M' | J'' M+M')
\left\{\matrix{J & J' & J''\cr j & j & j \cr}\right\}
B_{J'', M+M'}, \eqno(31.8)$$
in which the curly brackets are the usual 6-j symbols.
These are the commutation relations of the so(2(2j+1)) algebra.

\subsection{Fermion pairs of zero angular momentum}

In the present subsection we will restrict ourselves to fermion pairs
coupled to angular momentum zero. The relevant commutation relations take
the form
$$[ A_0, A_0^\dagger] = 1-{N_F\over\Omega}, \qquad
  [{N_F\over 2}, A^\dagger_0]= A^\dagger_0, \qquad
  [{N_F\over 2}, A_0]= -A_0, \eqno(31.9)$$
where $N_F$ is the number of fermions, $2\Omega=2j+1$ is the
size of the shell, and $$B_0=N_F/\sqrt{2\Omega}. \eqno(31.10)$$
 With the identifications
$$J_+= \sqrt{\Omega} A^\dagger_0, \quad J_-=\sqrt{\Omega} A_0, \quad
J_0= {N_F-\Omega\over 2}, \eqno(31.11)$$
eqs (31.9) take the form of the usual su(2) commutation relations
$$[J_+, J_-]= 2J_0, \quad [J_0, J_+]=J_+, \quad [J_0, J_-]=-J_-.
\eqno(31.12)$$
An exact boson mapping of the su(2) algebra is given in \cite{BKL521,KM375}
$$A_0^\dagger= a_0^\dagger \sqrt{1-{n_0\over \Omega}}, \quad
  A_0=\sqrt{1-{n_0\over \Omega}} a_0, \quad
N_F=2 n_0, \eqno(31.13)$$
where $a_0^\dagger$ ($a_0$) are boson creation (annihilation) operators
carrying angular momentum zero and $n_0$ is the number of these
bosons.

The simplest pairing Hamiltonian one can consider has the form
$$H=-G\Omega A^\dagger_0 A_0. \eqno(31.14)$$
The Casimir operator of su(2) can be written as
$$\{A^\dagger_0, A_0\} + {\Omega\over 2} \left(1-{N_F\over \Omega}\right)^2 =
{\Omega\over 2}+1, \eqno(31.15)$$
while the pairing energy takes the form
$${E\over (-G\Omega)} = {N_F\over 2} -{N_F^2\over 4\Omega} +
{N_F\over 2\Omega}.\eqno(31.16) $$

Our aim is to check if there is a boson mapping
for the operators $A^\dagger_0$, $A_0$ and $N_F$
in terms of $q$-deformed bosons, having the following
properties:

i) The mapping is simpler than the one of eq. (31.13), i.e. to
each fermion pair operator $A^\dagger_0$, $A_0$ corresponds a bare
$q$-boson operator and not a boson operator accompanied by
a square root (the Pauli reduction factor).

ii) The commutation relations (31.9) are satisfied up to a
certain order.

ii) The pairing energies of eq. (31.16) are reproduced up to the
same order.

\subsection{The $q$-deformed oscillator: A story of failure}

In the case of the $q$-deformed harmonic oscillator (sec. 10),
the commutation relation
$$ [a, a^\dagger] = [N+1] -[N] \eqno(31.17)$$
for $q$ being a phase can be written as
$$[a, a^\dagger] = {\cos{(2N+1)\tau\over 2}\over\cos{\tau\over 2}}.
\eqno(31.18)$$

In physical situations $\tau$ is expected to be small (i.e.
of the order of 0.01).
Therefore in eq. (31.18) one can take Taylor expansions
of the functions appearing there and thus find an expansion of the
form
$$[a, a^\dagger] = 1 -{\tau^2\over 2} (N^2+N) +{\tau^4\over 24}
(N^4+2 N^3-N)-\ldots .\eqno(31.19)$$
We remark that the first order corrections contain not only a term
proportional to $N$, but in addition a term proportional to $N^2$,
which is larger than $N$. Thus one cannot make the simple mapping
$$A_0\rightarrow a,\quad A^\dagger_0\rightarrow a^\dagger,\quad N_F\rightarrow
2N, \eqno(31.20)$$
because then one cannot get the first of the commutation
relations (31.9) correctly up to
the first order of the corrections. The same problem appears
in the case that $q$ is real as well. In addition, by making the
simple mapping of eq. (31.20) the pairing Hamiltonian can be written
as
$${H\over -G \Omega} = a^\dagger a = [N].\eqno(31.21)$$
In the case of small $\tau$, one can take Taylor expansions
of the functions appearing in the definition of the $q$-numbers
(eq. (2.2) or (2.3)) and thus obtain the following expansion
$$ [N]= N \pm {\tau^2\over 6} (N-N^3)+{\tau^4\over 360}
(7N-10N^3+3 N^5) $$ $$\pm {\tau^6\over 15120} (31N-49N^3+21N^5-3N^7)+
\ldots, \eqno(31.22)$$
where the upper (lower) sign corresponds to $q$ being a phase factor (real).
We remark that while the first order corrections in eq. (31.16)
are proportional to $N_F^2$ and $N_F$, here the first order
corrections are proportional to $N$ and $N^3$. Thus neither the
pairing energies can be reproduced correctly by this mapping.

\subsection{The $Q$-deformed oscillator: A story of success}

In the case of the $Q$-oscillator of sec. 11, however,
the commutation relation among the bosons is
$$[b, b^\dagger] = Q^N.\eqno(31.23)$$
Defining $Q=e^T$ this can be written as
$$[b, b^\dagger]= 1 + TN + {T^2  N^2\over 2} +{T^3 N^3\over 6} +\ldots.
\eqno(31.24)$$
We remark that the first order correction is proportional to $N$.
Thus, by making the boson mapping
$$A_0^\dagger\rightarrow b^\dagger, \quad
A_0\rightarrow b, \quad N_F\rightarrow 2N ,\eqno(31.25)$$
one can satisfy the first commutation relation of eq. (31.9) up to the first
order of the corrections by determining $T=-2/\Omega$.

We should now check if the pairing energies (eq. (31.16)) can be found
correctly up to the same order of approximation when this
mapping is employed. The pairing Hamiltonian in this case takes
the form
$${ H\over -G\Omega}= b^\dagger b = [N]_Q.\eqno(31.26)$$

Defining $Q=e^T$ it is instructive to construct the expansion of
the $Q$-number of eq. (6.1) in powers of $T$. Assuming that $T$ is
small and taking Taylor expansions in eq. (6.1) one finally has
$$[N]_Q= N+{T\over 2} (N^2 -N) +{T^2\over 12} (2N^3-3N^2+1) +
{T^3\over 24} (N^4-2N^3+N^2) +\ldots\eqno(31.27)$$
 Using the
 value of the deformation
parameter $T=-2/\Omega$, determined above from the requirement
that the commutation relations are satisfied up to first order
corrections, the pairing energies become
$${E\over -G \Omega}= N- {N^2-N\over \Omega} +
{2N^3-3N^2+1\over 3\Omega^2} - {N^4-2N^3+N^2\over 3\Omega^3} +
\ldots. \eqno(31.28)$$
The first two terms in the rhs of eq. (31.28), which correspond to the
leading term plus the first order corrections, are exactly equal to
the pairing energies of eq. (31.16), since $N_F\rightarrow 2N$.
 We therefore conclude that
through the boson mapping of eq. (31.25) one can both satisfy the
fermion pair commutation relations of eq. (31.9) and reproduce the
pairing energies of eq. (31.16) up to the first order corrections.

The following comments are also in place:

i) By studying the spectra of the two versions of the $q$-deformed
harmonic oscillator, given in eqs.  (10.10) and (11.10), one can easily
draw the following conclusions: when compared to the usual oscillator
spectrum, which is equidistant, the spectrum of the $q$-oscillator
is getting shrunk for $q$ being a phase, while the spectrum of the
$Q$-oscillator gets shrunk when $T<0$. In a similar way, the spectrum
of the $q$-oscillator gets expanded for $q$ real, while the spectrum of
the $Q$-oscillator gets expanded for $T>0$. In physical situations
(secs 19--23)  it has been found that the physically interesting results
are gotten with $q$ being a phase. Thus in the case of the
$Q$-oscillator it is the case $T<0$ the one which corresponds to the
physically
interesting case. As we have already seen, it is exactly for
$T=-2/\Omega<0$ that the present mapping gives the fermion pair
results.

ii) It should be recalled that the pairing model under discussion
is studied under the assumptions  that the degeneracy of
the shell is large ($\Omega >>1$), that the number of particles
is large ($N>>1$), and that one stays away from the center of the
shell ($\Omega -N=O(N)$). The accuracy of the
present mapping in reproducing the pairing energies has been checked
in \cite{Bon101}, where results for $\Omega=11$
(the size of the nuclear fp major shell), $\Omega=16$
(the size of the nuclear sdg major shell) and $\Omega=22$ (the
size of the nuclear pfh major shell) are reported, along with
results for the case $\Omega=50$ (as an
example of a large shell). In all cases good agreement
between the classical pairing model results and the
$Q$-Hamiltonian of eq. (31.26) is obtained up to the point
at which about 1/4 of the shell is filled. The deviations
observed near the middle of the shell are expected, since there the
expansion used  breaks down.

 We have thus shown that an approximate  mapping of the fermion
pairs coupled to angular momentum zero in a single-j shell
onto suitably defined $q$-bosons (the $Q$-bosons) is possible. The su(2)
commutation relations are satisfied up to the first order corrections, while
at the same time the eigenvalues of a simple pairing Hamiltonian are correctly
reproduced up to the same order. The small parameter of the
expansion, which is $T$ (where $Q=e^T$), turns out to be negative
and inversely proportional to the size of the shell.

The present results are an indication that suitably defined $q$-bosons
could be used for approximately describing systems of correlated fer\-mi\-ons
under certain conditions in a simplified way.
The construction of $q$-bosons
which would {\sl exactly} satisfy the fermion pair su(2) commutation
relations  will be undertaken in the following section.

\section { Fermion pairs as deformed bosons: exact mapping}

{}From the contents of the previous section,
the following question is created: Is it possible to construct
a generalized deformed oscillator (as in sec. 12) using deformed bosons in such
a way that the spectrum of the oscillator will exactly correspond to the
pairing energy in the single-j shell model, while the
commutation relations of the deformed bosons will exactly correspond to
the commutation relations of the correlated fermion pairs in the
single-j shell under discussion? In this section  we show that such an
oscillator can indeed be constructed \cite{BD2781}
by using the method of sec. 12.

\subsection{An appropriate generalized deformed oscillator}

We apply the procedure of sec. 12 in the case of the pairing in a
single-j shell mentioned before. The boson number is half
the fermion number, i.e. $N=N_F/2$. Then eq. (31.16) can be
written as
$$ {E\over -G \Omega} = N -{N^2\over \Omega} +{N\over
\Omega}.\eqno(32.1)$$
One can then use a generalized deformed  oscillator with structure function
$$ F(N) = a^\dagger a =N-{N^2\over \Omega}+{N\over \Omega}.\eqno(32.2)$$
In addition one has
$$ F(N+1) = a a^\dagger  = N+1 -{(N+1)^2\over \Omega}+{N+1\over\Omega}.
\eqno(32.3)$$
What we have constructed is a boson mapping for the operators
$A_0$, $A_0^\dagger$, $N_F$:
$$A_0\rightarrow a, \quad A_0^\dagger\rightarrow a^\dagger, \quad
N_F\rightarrow 2N. \eqno(32.4)$$
{}From eq. (32.2) it is clear that this mapping gives the
correct pairing energy. In addition one has
$$[a, a^\dagger]= F(N+1)-F(N)=1-{2N\over \Omega}, \eqno(32.5)$$
in agreement to the first commutation relation of eq. (31.9). Thus the
correct commutation relations
are also obeyed. (The last two commutation relations of eq. (31.9)
are satisfied because of (12.1).)

As we have already seen in the previous section,
an exact hermitian boson mapping for the su(2) algebra is known
to have the form of eq. (31.13). In this mapping the Pauli principle effects
are carried by the square roots accompanying the ordinary boson
operators, while in the mapping of eq. (32.4) the Pauli
principle effects are ``built in'' the deformed bosons.

The generalized oscillator obtained here has energy spectrum
$$E_N = {1\over 2} (F(N)+F(N+1))=N+{1\over 2}-{N^2\over \Omega},
\eqno(32.6)$$
which is the spectrum of an anharmonic oscillator.

\vfill\eject
\subsection{Related potentials}

The classical potential giving the same spectrum, up to first
order perturbation theory, can be easily determined (see also
subsecs 13.1, 13.2). The potential
$$V(x)=\kappa x^2 +\lambda x^4 , \eqno(32.7)$$
is known to give in first order perturbation theory the
spectrum
$$E_n = \kappa (2n+1) +\lambda (6n^2 +6 n +3)
= (2\kappa+6\lambda) (n+{1\over 2}) + 6\lambda n^2. \eqno(32.8)$$
Comparing eqs. (32.6) and (32.8) one finds
$$\kappa = {1\over 2} (1+{1\over \Omega}), \quad \lambda=
-{1\over 6\Omega}.\eqno(32.9)$$
Then the classical potential giving the same spectrum, up to
first order perturbation theory, as the generalized oscillator
determined here, is
$$V(x)={1\over 2} (1+{1\over \Omega}) x^2 -{1\over 6\Omega}
x^4. \eqno(32.10)$$
It is therefore demonstrated that the Pauli principle effects
in a single-j shell with pairing interaction are equivalent
to an $x^4$ anharmonicity.

The generalization of the results obtained in this section  for the
pairing hamiltonian to any anharmonic oscillator is
straightforward. For example, the potential
$$V(x)=\kappa x^2 + \lambda x^4 + \mu x^6 +\xi x^8, \eqno(32.11)$$
is known to give up to first order perturbation theory the
spectrum
$$E_n =\kappa (2n+1)+\lambda (6n^2 +6n+3)$$
$$+\mu
(20 n^3+30 n^2 + 40 n+15) +\xi (70 n^4+140 n^3+350 n^2 + 280 n
+ 105), \eqno(32.12)$$
which can be rewritten in the form
$$E_n =(n+(n+1)) (\kappa +5\mu) + (n^2+(n+1)^2) (3\lambda+70\xi)$$
$$ + (n^3+(n+1)^3) (10\mu) + (n^4+(n+1)^4) (35 \xi).
\eqno(32.13)$$
Taking into account eq. (12.11),  from eq. (32.13) one gets
$${F(N)\over 2} = (\kappa +5\mu) n + (3\lambda+70 \xi) n^2
+(10\mu) n^3 + (35\xi) n^4.\eqno(32.14)$$
For $\mu=\xi=0$ and $\kappa$, $\lambda$ given from eq. (32.9),
the results for the pairing problem are regained.

It is worth mentioning at this point that the energy spectrum
of the generalized oscillator corresponding to the pairing
correlations (eq. (32.6))  can be rewritten as
$$E_N= {2\over \Omega} \left(-{1\over 8} +{\Omega+1\over 2}
\left(N+{1\over 2}\right)
-{1\over 2} \left(N+{1\over 2}\right)^2\right). \eqno(32.15)$$
On the other hand, it is known that for the modified P\"oschl--Teller
potential (see also subsec. 13.2)
$$ V(x)= D \tanh^2(x/R), \eqno(32.16)$$
the energy spectrum is given by \cite{Haar75}
$$ E_N = {\hbar^2\over m R^2} \left(-{1\over 8} +
{1\over 2}\sqrt{8mDR^2/\hbar^2
+1}\quad \left(N+{1\over 2}\right) -{1\over 2} \left(N+{1\over 2}\right)^2
\right).\eqno(32.17)$$
It is thus clear that the energy spectrum of the generalized oscillator
studied here
can be obtained from the modified P\"oschl--Teller potential for
special values of the potential depth $D$.

It is also worth remarking that the ``structure function'' $F(N)$
of the generalized oscillator obtained here (eq. (32.2)) can be
written as
$$F(N)= {N\over \Omega} (\Omega+1-N), \eqno(32.18)$$
which is similar to the one of the para-fermionic oscillator
of Ohnuki and Kamefuchi \cite{OK82} (see also secs 12, 18).

In summary, we have constructed a generalized deformed oscillator
 which satisfies the same commutation relations as fermion
pair and multipole operators of zero angular momentum in a
single-j shell, and, in addition, reproduces the pairing
energy of this shell exactly. We have thus demonstrated that
an exact hermitian boson mapping of a system of angular-momentum-zero
fermion pairs in terms of bare deformed bosons can be constructed,
while in the usual case the ordinary bosons  are accompanied
by square roots due to the Pauli principle effects.
The oscillator corresponding to the pairing problem has a spectrum
which can be reproduced up to first order perturbation theory by a
harmonic oscillator with an $x^4$ anharmonicity. The construction
of a generalized deformed oscillator corresponding to any
anharmonic oscillator has also been achieved.

The results obtained in this section indicate that deformed bosons
might be a convenient tool for describing systems of fermion
pairs under certain conditions. The generalisation
of the results obtained here to fermion pairs of nonzero
angular momentum, which will allow for a fuller treatment of
the single-j shell in terms of deformed bosons, is a very interesting problem.

\vfill\eject
\section{The seniority scheme}

In the previous two sections we have seen how correlated fermion pairs of zero
angular momentum can be mapped onto deformed bosons. It is however known that
pairs of non-zero angular momentum play an important role in the formation of
nuclear properties.
In the present section a first step in the direction of describing the
$J\neq 0$ pairs in terms of deformed bosons is taken.

\subsection{Uncovering a dynamical symmetry}

In the usual formulation of the theory of pairing in a single-j
shell \cite{Heyde90}, fermion pairs of angular momentum $J=0$ are created
by the pair creation operators
$$ S^\dagger = {1\over \sqrt{\Omega}} \sum_{m>0} (-1)^{j+m}  a^\dagger_{jm}
a^\dagger_{j-m}, \eqno(33.1)$$
where $a^\dagger_{jm}$ are fermion creation operators and $2\Omega=2j+1$
is the degeneracy of the shell. In addition, pairs of nonzero
angular momentum are created by the $\Omega-1$ operators
$$ B^\dagger_J=\sum_{m>0} (-1)^{j+m} (j m j -m | J 0) a^\dagger_{jm}
a^\dagger_{j-m}, \eqno(33.2)$$
where $( j m j -m | J 0)$ are the usual Clebsch Gordan coefficients.
The fermion number operator is defined as
$$ N_F = \sum_m a^\dagger_{jm} a_{jm} = \sum_{m>0} (a^\dagger_{jm} a_{jm} +
a^\dagger_{j-m} a_{j-m} ). \eqno(33.3)$$
As we have already seen, the $J=0$ pair creation and annihilation operators
satisfy the  commutation relation
$$ [S, S^\dagger] = 1-{N_F\over \Omega}, \eqno(33.4)$$
while the pairing Hamiltonian is
$$ H = -G \Omega S^\dagger S.\eqno(33.5)$$
The seniority $V_F$ is defined as the number of fermions not
coupled  to $J=0$. If only pairs of $J=0$ are present (i.e. $V_F=0$),
the eigenvalues of the Hamiltonian are (as already seen in eq. (31.16))
$$ E(N_F,V_F=0) = -G \Omega \left( {N_F\over 2} + {N_F\over 2\Omega}
-{N_F^2\over 4\Omega}\right). \eqno(33.6)$$
For non-zero seniority the eigenvalues of the Hamiltonian are
$$ E(N_F,V_F)= -{G\over 4} (N_F-V_F) (2\Omega-N_F-V_F+2).\eqno(33.7)$$
We denote the operators $N_F$, $V_F$ and their eigenvalues by the
same symbol for simplicity.

In subsec. 31.4 it has been proved that the behaviour of the $J=0$ pairs
can be described, up to first order corrections, in terms of
$Q$-bosons. In particular, making the mapping
$$ S^\dagger \rightarrow b^\dagger, \quad\quad S\rightarrow b, \quad\quad
N_F\rightarrow 2 N, \eqno(33.8)$$
the relevant pairing Hamiltonian of eq. (33.5)  becomes
$$ H(N,V=0) = -G \Omega b^\dagger b = -G\Omega [N]_Q,\eqno(33.9)$$
which coincides with eq. (33.6) up to
first order corrections in the small parameter, which is identified
as $T=-2/\Omega$. Furthermore, the $Q$-bosons satisfy the commutation
relation of eq. (31.24),
which coincides with eq. (33.4) up to first order corrections in the
small parameter, which is, consistently with the above finding,
identified as $T=-2/\Omega$. Therefore the fermion pairs of $J=0$
can be approximately described as $Q$-bosons, which correctly reproduce
both the pairing energies and the commutation relations up to
first order corrections in the small parameter.

For the case of nonzero seniority, one observes that eq. (33.7)
can be written as
$$ E(N_F,V_F)= G\Omega \left({V_F\over 2} + {V_F\over 2\Omega} -
{V_F^2\over 4\Omega} \right) -G \Omega \left({N_F\over 2} +
 {N_F\over 2\Omega} -
{N_F^2\over 4\Omega} \right), \eqno(33.10)$$
i.e. it can be separated into two parts, formally identical to
each other. Since the second part (which corresponds to the $J=0$
pairs) can be adequately described by the $Q$-bosons $b$, $b^\dagger$,
and their number operator $N$,
 as we have already seen, it is reasonable to  assume that the first
 part can also be described in terms of some $Q$-bosons $d$,
$d^\dagger$, and their number operator $V$ (with $V_F\rightarrow 2V$),
satisfying commutation relations similar to eqs (11.2) and (11.3):
$$ [V,d^\dagger]= d^\dagger, \quad\quad [V,d]=-d,\quad\quad d d^\dagger
-Q d^\dagger d = 1. \eqno(33.11)$$
 From the
physical point of view this description means that a set of $Q$-bosons
is used for the $J=0$ pairs and another set for the $J\neq 0$ pairs.
The latter is reasonable, since in the context of this theory
the angular momentum value of the $J\neq 0$ pairs is not used explicitly.
The $J\neq 0$ pairs are just counted separately from the $J=0$
pairs. A Hamiltonian giving the same spectrum as in eq. (33.10),
up to first order corrections in the small parameter,
 can then be written as
$$ H(N,V) =G \Omega ([V]_Q -[N]_Q). \eqno(33.12)$$
Using eq. (31.27) it is easy to see that this expression agrees
to eq. (33.10) up to first order corrections in the small parameter
$T=-2/\Omega$.

Two comments concerning eq. (33.12) are in place:

i) In the classical theory states of maximum seniority (i.e.
states with $N=V$) have zero energy. This is also holding for the
Hamiltonian of eq. (33.12) to all orders in the deformation
parameter.

ii) A landmark of the classical theory is that $E(N,V)-E(N,V=0)$
is independent of $N$. This also holds for eq. (33.12) to all orders
in the deformation parameter.

Knowing the Schwinger realization of the su$_q$(2) algebra in terms
of $q$-bosons (sec. 15), one may wonder if the
operators used here close an algebra. It is easy to see that
the operators $b^+d$, $d^+ b$ and $N-V$ do not close an algebra.
Considering, however, the operators (see \cite{CJ711} with $p=1$)
$$ J_+ = b^\dagger Q^{-V/2} d, \quad\quad J_-= d^\dagger Q^{-V/2} b,
\quad\quad
J_0= {1\over 2} (N-V), \eqno(33.13)$$
one can easily see that they satisfy the commutation relations
\cite{CJ711,Jan91}
$$ [J_0, J_{\pm}] = \pm J_{\pm} , \quad\quad J_+ J_- - Q^{-1} J_- J_+
= [2 J_0] _Q .\eqno(33.14)$$
Using the transformation
$$ J_0 = \tilde J_0, \quad\quad J_+ = Q^{(1/2)(J_0-1/2)}
\tilde J_+, \quad\quad J_-= \tilde J_- Q^{(1/2)(J_0-1/2)}, \eqno(33.15)$$
one goes to the usual su$_q$(2) commutation relations
$$ [\tilde J_0, \tilde J_{\pm}] = \pm \tilde J_{\pm}, \quad\quad
[\tilde J_+, \tilde J_-] = [2 \tilde J_0], \eqno(33.16)$$
where $q^2=Q$. One can thus consider eq. (33.14) as a rewriting of the
algebra su$_q$(2), suitable for boson realization in terms of $Q$-bosons.

It is clear that $N+V$ is the first order Casimir operator of
the  u$_Q$(2) algebra formed above (since it commutes with all the
generators given in eq. (33.13)),
 while $N-V$ is the first order
Casimir operator of its u$_Q$(1) subalgebra, which is generated
by $J_0$ alone. Therefore the Hamiltonian of eq. (33.12) can be
expressed in terms of the Casimir operators of the algebras
appearing in the chain u$_Q$(2) $\supset$ u$_Q$(1) as
$$ E(N,V) = G\Omega ( \left[{C_1({\rm u}_Q(2))-C_1({\rm u}_Q(1))\over 2}
\right]_Q  -  \left[{C_1({\rm u}_Q(2))+C_1({\rm u}_Q(1))\over 2}\right]_Q),
\eqno(33.17)$$
  i.e. the Hamiltonian has a u$_Q$(2) $\supset$ u$_Q$(1) dynamical symmetry.

\subsection{Comparison to experiment}

In the construction given above we have shown that $Q$-bosons can be
used for the approximate description of correlated fermion pairs in a
single-j shell.  The results obtained in the $Q$-formalism agree
to  the classical (non-deformed) results up to first order corrections
in the small parameter. However, the $Q$-formalism contains in
addition  higher order terms. The question is then born if these
additional terms are useful or not. For answering this question,
the simplest comparison with experimental data which can be made
concerns the classic example of the neutron pair separation energies
of the Sn isotopes, used by Talmi \cite{Talmi1,Talmi85}.

In Talmi's formulation  of the pairing theory, the energy of the
states with zero seniority is given by
$$ E(N)_{cl} = N V_0 + {N(N-1)\over 2} \Delta, \eqno(33.18)$$
where $N$ is the number of fermion pairs and $V_0$, $\Delta$
are constants. We  remark that this expression is the same
as the one in eq. (33.6), with the identifications
$$\Delta/(2 V_0) =-1/\Omega, \quad\quad \Delta=2G,
 \quad\quad N_F=2N. \eqno(33.19)$$
 The neutron pair separation energies are given by
$$ \Delta E(N+1)_{cl} =E(N+1)_{cl}-E(N)_{cl}=
V_0 \left( 1+ {\Delta\over V_0} N\right).\eqno(33.20)$$
Thus the neutron pair separation energies are expected to decrease
linearly with increasing $N$. (Notice from eq. (33.19) that $\Delta/V_0<0$,
since $\Omega>0$.) A similar  linear decrease is predicted also by the
Interacting Boson Model \cite{IA1987}.

In our formalism the neutron pair separation energies are given by
$$ \Delta E(N+1)_Q= -G \Omega ([N+1]_Q -[N]_Q) = -G \Omega Q^N =
-G \Omega e^{T N}.\eqno(33.21)$$
Since, as we have seen, $T$ is expected to be $-2/\Omega$, i.e.
negative and small, the neutron pair separation energies are
expected to fall exponentially with increasing $N$, but the small
value of $T$ can bring this exponential fall very close to a linear
one.

The neutron pair separation energies of the even Sn isotopes
from $^{104}$Sn to $^{130}$Sn (i.e. across the whole sdg neutron
shell) have been fitted in \cite{BDF1299}  using both theories.
 Furthermore, in \cite{BDF1299} a fit
 of the logarithms of the energies has been performed, since eq. (33.21)
predicts a linear decrease of the logarithm of the energies with
increasing $N$.
Both fits give almost identical results. Eq. (33.21) (in which the
free parameters are $G\Omega$ and $T$),
gives a better result than eq. (33.20) (in which the free
parameters are $V_0$ and $\Delta/V_0$)
for every single isotope,
without introducing any additional parameter, indicating that
the higher order terms can be useful.

One should, however, remark that $^{116}$Sn lies in the middle
of the sdg neutron shell. Fitting the isotopes in the lower half
of the shell ($^{104}$Sn to $^{116}$Sn) and the isotopes in the
upper half of the shell ($^{118}$Sn to $^{130}$Sn) separately,
one finds that both theories give indistinguishably good results
in both regions. Therefore $Q$-deformation can be understood as
expressing higher order correlations which manifest themselves
in the form of particle-hole asymmetry.  It is also known that a strong
subshell closure exists at N=64 (which corresponds to $^{114}$Sn).
The presence of this subshell closure can also affect the neutron pair
separation energies in a way similar to the one shown by the data.

In \cite{BDF1299} a fit of the neutron pair
separation energies of the Pb isotopes from $^{186}$Pb to $^{202}$Pb
has also been attempted. In
this case both theories give indistinguishably good fits. This
result is in agreement with the Sn  findings,
since all of these Pb
isotopes lie in the upper half of the pfh neutron shell. Unfortunately,
no neutron pair separation energy data exist for Pb isotopes in
the lower part of the pfh neutron shell.

Concerning the values of $T$ obtained in the case of the Sn isotopes
($T=-0.0454$, $=-0.0447$),
one observes that they are slightly
smaller than the value ($T=-0.0488$) which would have been obtained
by considering the neutrons up to the end of the sdg shell as lying
in a single-j shell. This is, of course, a very gross approximation
which should not be taken too seriously, since it ignores the fact that most
properties of nuclei can be well accounted for by the valence nucleons alone,
without being affected by the closed core. In the case of the Pb isotopes
mentioned above, however, the best fit was obtained with
$T=-0.0276$, which is again slightly smaller than the value of
$T=-0.0317$ which corresponds to considering all the neutrons up to
the end of the pfh shell as lying in a single-j shell.

In summary, we have shown that pairing in a single-j shell
can be described, up to first order corrections, by two $Q$-oscillators,
one describing the $J=0$ pairs and the other corresponding
to the $J\neq 0$ pairs, the deformation parameter $T=\ln Q$ being
related to the inverse of the size of the shell.
 These two oscillators can be used for
forming an su$_Q$(2) algebra. A Hamiltonian giving the correct
pairing energies up to first order corrections in the small
parameter can be
written in terms of the Casimir operators of the algebras appearing
in the u$_Q$(2) $\supset$ u$_Q$(1) chain, thus exhibiting a
quantum algebraic dynamical symmetry. The additional terms introduced
by the $Q$-oscillators serve in improving the description of the
neutron pair separation energies of the Sn isotopes, with no extra
parameter introduced.

In the previous section a generalized deformed oscillator describing the
correlated fermion pairs of $J=0$ {\sl exactly} has been
introduced. This generalized deformed oscillator is the same as
the one giving the same spectrum as the Morse potential (sec. 35), up
to a shift in the energy spectrum. The use of two generalized deformed
oscillators for the description of $J=0$ pairs and $J\neq 0$ pairs
in a way similar to the one of the present section is a straightforward task,
while the construction out of them
of a closed algebra analogous to the su$_Q$(2)
obtained here is an open problem. The extension of the ideas presented
here to the case of the BCS theory is an interesting open problem.

\subsection{Other approaches}

A $q$-deformed version of the pairing theory was {\sl assumed} by
Petrova \cite{Petr92} and Shelly Sharma \cite{Sha904},
with satisfactory results when compared to experimental
data. The present construction offers some justification for this
assumption, since in both cases the basic ingredient is the modification
of eq. (33.4).  It should be noticed, however, that the deformed version of
eq. (33.4)  considered in \cite{Petr92,Sha904} is different from
the one obtained  here (eq. (31.24)). A basic difference is that the deformed
theory of \cite{Petr92,Sha904}  reduces to the classical
theory for $q\rightarrow 1$, so that $q$-deformation is introduced in order
to describe additional correlations, while in the present formalism
the $Q$-oscillators involved for $Q\rightarrow 1$ reduce to usual
harmonic oscillators, so that $Q$-deformation is introduced in order
to attach to the oscillators the anharmonicity needed by the energy
expression (eq. (33.6)).

Continueing along the same line Shelly Sharma and Sharma \cite{SSS2323}
derived Random Phase Approximation (RPA) equations for the pairing
vibrations of nuclei differing by two nucleons in comparison to the initial
one and applied their method to the study of the $0^+$ states of the
Pb isotopes, which offer a good example of pairing vibrations in
nonsuperconducting nuclei. Furthermore, using deformed quasi-particle pairs
coupled to zero angular momentum they developed a deformed version of the
quasi-boson approximation for $0^+$ states in superconducting nuclei and
tested it against a schematic two-level shell model \cite{SSS2323}. Another
deformed two-level shell model has been developed by Avancini and
Menezes \cite{AM6261}.

\section{Anisotropic quantum harmonic oscillators with rational ratios of
frequencies}

3-dim anisotropic harmonic oscillators with rational ratios of frequencies
(RHOs) are of
current interest because of their relevance as possible underlying symmetries
of superdeformed and hyperdeformed nuclei \cite{Mot522,Rae1343}.
In particular, it is thought \cite{NT533,JK321}
that superdeformed nuclei correspond to a ratio of frequencies of 2:1, while
hyperdeformed nuclei correspond to a 3:1 ratio. In addition they have been
recently connected \cite{RZ599,ZRM61}
to the underlying geometrical structure in the Bloch--Brink $\alpha$-cluster
model \cite{BB247},
and possibly to the interpretation of the observed shell structure
in atomic clusters \cite{Martin25},
especially after the realization that large deformations
can occur in such systems \cite{BL4130}.
The 2-dim RHO is also of interest, since its
single particle level spectrum characterizes the underlying symmetry of
``pancake'' nuclei \cite{Rae1343}.

RHOs are examples of maximally superintegrable systems \cite{Hiet87,Eva5666}
in N dimensions. {\sl Superintegrable systems} in N dimensions have more than
N independent integrals (constants of motion). {\sl Maximally superintegrable
systems} in N dimensions have 2N$-1$ independent integrals.

The two-dim \cite{JH641,Dem1349,Con273,Con1297,Cis870,CLP6685,GKM925}
 and three-dim
\cite{DZ1203,Mai1004,Ven190,MV57,RD1323,BCD1401,Naz154} anisotropic harmonic
oscillators have been the subject of several investigations, both at the
classical and the quantum mechanical level.
The special cases with frequency
ratios 1:2 \cite{Holt1037,BW2215} and 1:3 \cite{FL325} have also been
considered. While
at the classical level it is clear that the su(N) or sp(2N,R) algebras can
be used for the description of the N-dimensional anisotropic oscillator, the
situation at the quantum level, even in the two-dimensional case, is not as
simple.

In this section we are going to prove that a generalized deformed u(2)
algebra is the symmetry algebra of the two-dimensional anisotropic quantum
harmonic oscillator, which is the oscillator describing the single-particle
level spectrum of ``pancake'' nuclei, i.e. of triaxially deformed nuclei
with $\omega_x >> \omega_y$, $\omega_z$ \cite{Rae1343}. The method can be
extended to the 3-dim RHO in a rather straightforward way.

\vfill\eject
\subsection{ A deformed u(2) algebra}

Let us consider the system described by the Hamiltonian:
$$H=\frac{1}{2}\left( {p_x}^2 + {p_y}^2 + \frac{x^2}{m^2}
+ \frac{y^2}{n^2} \right), \eqno(34.1)$$
where $m$ and $n$ are two natural numbers mutually prime ones, i.e. their
great common divisor is $\gcd (m,n)=1$.

We define the creation and annihilation operators \cite{JH641}
$$ a^\dagger=\frac{x/m - i p_x}{\sqrt{2}}, \qquad a =
\frac{x/m + i p_x}{\sqrt{2}}, \eqno(34.2)$$
$$ b^\dagger=\frac{y/n - i p_y}{\sqrt{2}}, \qquad b=\frac{y/n + i p_y}
{\sqrt{2}}.\eqno(34.3)$$
These operators satisfy the commutation relations:
$$ \left[ a,a^\dagger \right] = \frac{1}{m}, \quad \left[
b,b^\dagger \right] = \frac{1}{n}, \quad \mbox{other commutators}=0.
\eqno(34.4)$$

Further defining
$$ U=\frac{1}{2} \left\{ a, a^\dagger \right\}, \qquad W=\frac{1}{2} \left\{ b,
b^\dagger \right\},\eqno(34.5)$$
one can consider the enveloping algebra generated by
the operators:
$$ S_+= \left(a^\dagger\right)^m \left(b\right)^n,\qquad S_-= \left(a\right)^m
\left(b^\dagger\right)^n, \eqno(34.6)$$
$$ S_0= \frac{1}{2}\left( U - W \right), \qquad H=U+W. \eqno(34.7)$$
These genarators satisfy the following relations:
$$\left[ S_0,S_\pm \right]=\pm S_\pm, \quad \left[H,S_i\right]=0,
\quad \mbox{for}\quad i=0,\pm,\eqno(34.8) $$
and
$$ S_+S_- = \prod\limits_{k=1}^{m}\left( U - \frac{2k-1}{2m} \right)
\prod\limits_{\ell=1}^{n}\left( W + \frac{2\ell-1}{2n} \right),\eqno(34.9) $$
$$ S_-S_+ = \prod\limits_{k=1}^{m}\left( U + \frac{2k-1}{2m} \right)
\prod\limits_{\ell=1}^{n}\left( W - \frac{2\ell-1}{2n} \right).\eqno(34.10) $$
The fact that the operators $S_i$, $i=0, \pm$ are integrals of motion has
been already realized in \cite{JH641}.

The above relations mean that the harmonic oscillator of eq. (34.1)
is described by the enveloping algebra of the
generalization of the u(2) algebra formed by the generators $S_0$, $S_+$, $%
S_-$ and $H$, satisfying the commutation relations of eq. (34.8) and
$$\left[S_-,S_+\right] = F_{m,n} (H,S_0+1) - F_{m,n} (H,S_0), \eqno(34.11)$$
where
$$\quad F_{m,n}(H,S_0)= \prod\limits_{k=1}^{m}\left(
H/2+S_0 - \frac{2k-1}{2m} \right) \prod\limits_{\ell=1}^{n}\left( H/2-S_0 +
\frac{2\ell-1}{2n} \right). \eqno(34.12)$$
In the case of $m=1$, $n=1$ this algebra is the usual u(2) algebra, and the
operators $S_0,S_\pm$ satisfy the commutation relations of the ordinary u(2)
algebra, since in this case one easily finds that
$$ [S_-, S_+]=-2 S_0.\eqno(34.13)$$
In the rest of the cases, the algebra is a deformed version of u(2), in
which the commutator $[S_-,S_+]$ is a polynomial of $S_0$ of order $m+n-1$.
In the case with $m=1$, $n=2$ one has
$$[S_-,S_+]= 3 S_0^2 - H S_0 -{\frac{H^2}{4}} +{\frac{3}{16}},\eqno(34.14)$$
i.e. a polynomial quadratic in $S_0$ occurs, while in the case of $m=1$, $n=3
$ one finds
$$[S_-, S_+]= -4 S_0^3 + 3 H S_0^2 -{\frac{7}{9}} S_0 -{\frac{H^3}{4}} + {%
\frac{H}{4}},\eqno(34.15)$$
i.e. a polynomial cubic in $S_0$ is obtained.

\subsection{ The representations}

The finite dimensional representation modules  of this algebra can be found
using the concept of the generalized deformed oscillator (sec. 12), in a
method similar to the one used in \cite{BDK3407,BDK3700} for the study of
quantum
superintegrable systems. The operators:
$$ {\cal A}^\dagger= S_+, \quad {\cal A}= S_-, \quad {\cal N}%
= S_0-u, \quad u=\mbox{ constant}, \eqno(34.16)$$
where $u$ is a constant to be determined, are the generators of a deformed
oscillator algebra:
$$ \left[ {\cal N} , {\cal A}^\dagger \right] = {\cal A}^\dagger, \quad \left[
{\cal N} , {\cal A} \right] = -{\cal A}, \quad {\cal A}^\dagger{\cal A}
=\Phi( H, {\cal N} ), \quad {\cal A}{\cal A}^\dagger =\Phi( H, {\cal N}+1 ).
\eqno(34.17)$$
The structure function $\Phi$ of this algebra is determined by the function $%
F_{m,n}$ in eq. (34.12):
$$ \Phi( H,
{\cal N} )= F_{m,n} (H,{\cal N} +u )= $$
$$ = \prod\limits_{k=1}^{m}\left( H/2+%
{\cal N} +u - \frac{2k-1}{2m} \right) \prod\limits_{\ell=1}^{n}\left( H/2-%
{\cal N} - u + \frac{2\ell-1}{2n} \right).\eqno(34.18$$
The deformed oscillator corresponding to the structure function of eq.
(34.18) has an energy dependent Fock space of dimension $N+1$ if
$$\Phi(E,0)=0, \quad \Phi(E, N+1)=0, \quad \Phi(E,k)>0,
\quad \mbox{for} \quad k=1,2,\ldots,N. \eqno(34.19)$$
The Fock space is defined by:
$$H\vert E, k > =E \vert E, k >, \quad {\cal N} \vert E, k >= k \vert E, k
>,\quad a\vert E, 0 >=0, \eqno(34.20)$$
$${\cal A}^\dagger \vert E, k> = \sqrt{\Phi(E,k+1)} \vert E, k+1>,
\quad {\cal %
A} \vert E, k> = \sqrt{\Phi(E,k)} \vert E, k-1>. \eqno(34.21)$$
The basis of the Fock space is given by:
$$\vert E, k >= \frac{1}{\sqrt{[k]!}} \left({\cal A}^\dagger\right)^k\vert E,
0 >, \quad k=0,1,\ldots N,\eqno(34.22) $$
where the ``factorial'' $[k]!$ is defined by the recurrence relation:
$$[0]!=1, \quad [k]!=\Phi(E,k)[k-1]! \quad . \eqno(34.23)$$
Using the Fock basis we can find the matrix representation of the deformed
oscillator and then the matrix representation of the algebra of eqs (34.8),
(34.12).
The solution of eqs (34.19) implies
the following pairs of permitted values for the energy eigenvalue $E$ and
the constant $u$:
$$E=N+\frac{2p-1}{2m}+\frac{2q-1}{2n} , \eqno(34.24)$$
where $p=1,2,\ldots,m$, $\qquad q=1,2,\ldots,n$, and
$$u=\frac{1}{2}\left( \frac{2p-1}{2m}-\frac{2q-1}{2n} -N \right),
\eqno(34.25)$$
the corresponding structure function being given by:
$$ \Phi(E,x)=\Phi^{N}_{(p,q)}(x)=
\prod\limits_{k=1}^{m}\left( x +
\displaystyle \frac{2p-1}{2m}- \frac{2k-1}{2m} \right)
\prod\limits_{\ell=1}^{n}\left( N-x+ \displaystyle \frac{2q-1}{2n} + \frac{%
2\ell-1}{2n}\right) $$
$$ =\displaystyle\frac{1}{m^m n^n} \displaystyle\frac{
\Gamma\left(mx+p\right) }{\Gamma\left(mx+p-m\right)} \displaystyle \frac{%
\Gamma\left( (N-x)n + q + n \right)} {\Gamma\left( (N-x)n + q \right)},
\eqno(34.26)$$
where $\Gamma(x)$ denotes the usual Gamma-function.
In all these equations one has $N=0,1,2,\ldots$, while the dimensionality of
the representation is given by $N+1$. Eq. (34.24)  means that there
are $m\cdot n$ energy eigenvalues corresponding to each $N$ value, each
eigenvalue having degeneracy $N+1$. (Later we shall see that the degenerate
states corresponding to the same eigenvalue can be labelled by an ``angular
momentum''.)

It is useful to show at this point that a few special cases are in agreement
with results already existing in the literature.

i) In the case $m=1$, $n=1$ eq. (34.26) gives
$$ \Phi(E,x)= x(N+1-x),\eqno(34.27)$$
while eq. (34.24) gives
$$ E=N+1,\eqno(34.28)$$
in agreement with Sec. IV.A of \cite{BDK3700}.

ii) In the case $m=1$, $n=2$ one obtains for $q=2$
$$\Phi(E,x)= x(N+1-x)\left(N+{3\over 2}-x\right), \qquad E=N+{5\over 4},
\eqno(34.29)$$
while for $q=1$ one has
$$\Phi(E,x)= x(N+1-x)\left(N+{1\over 2}-x\right), \qquad E=N+{3\over 4}.
\eqno(34.30)$$
These are in agreement with the results obtained in Sec. IV.F of \cite{BDK3700}
for the Holt potential (for $\delta =0$).

iii) In the case $m=1$, $n=3$ one has for $q=1$
$$\Phi(E,x)=x(N+1-x)\left(N+{1\over 3}-x\right) \left(N+{2\over 3}-x\right),
\qquad E=N+{2\over 3},\eqno(34.31)$$
while for $q=2$ one obtains
$$\Phi(E,x)=x(N+1-x)\left(N+{2\over 3}-x\right) \left(N+{4\over 3}-x\right),
\qquad E=N+1,\eqno(34.32)$$
and for $q=3$ one gets
$$\Phi(E,x)=x(N+1-x)\left(N+{4\over 3}-x\right) \left(N+{5\over 3}-x\right),
\qquad E=N+{4\over 3}.\eqno(34.33)$$
These are in agreement with the results obtained in Sec. IV.D of \cite{BDK3700}
for the Fokas--Lagerstrom potential.

In all of the above cases we remark that the structure function has
 forms  corresponding to various versions of the  generalized deformed
parafermionic algebra of eq. (18.1), the relevant conditions of eq. (18.2)
being satisfied in all cases.
It is easy to see that the obtained algebra corresponds to this of the
generalized parafermionic oscillator in all cases with frequency
ratios $1:n$.

The energy formula can be corroborated by using the
corresponding Schr\"{o}dinger equation. For the Hamiltonian of eq. (34.1)
the eigenvalues of the Schr\"{o}dinger equation are given
by:
$$E=\frac{1}{m}\left(n_x+\frac{1}{2}\right)+ \frac{1}{n}\left(n_y+%
\frac{1}{2}\right),\eqno(34.34) $$
where $n_x=0,1,\ldots$ and $n_y=0,1,\ldots$. Comparing eqs (34.24) and
(34.34) one concludes that:
$$N= \left[n_x/m\right]+\left[n_y/n\right],\eqno(34.35)$$
where $[x]$ is the integer part of the number $x$, and
$$p=\mbox{mod}(n_x,m)+1, \quad q=\mbox{mod}(n_y,n)+1. \eqno(34.36)$$

The eigenvectors of the Hamiltonian can be parametrized by the
dimensionality of the representation $N$, the numbers $p,q$, and the number $%
k=0,1,\ldots,N$. $k$ can be identified as $[n_x/m]$. One then has:
$$H\left\vert
\begin{array}{c}
N \\
(p,q)
\end{array}
, k \right>= \left(N+\displaystyle
\frac{2p-1}{2m}+\frac{2q-1}{2n} \right)\left\vert
\begin{array}{c}
N \\
(p,q)
\end{array}
, k \right>,\eqno(34.37) $$
$$S_0 \left\vert
\begin{array}{c}
N \\
(p,q)
\end{array}
, k \right>= \left( k+ \displaystyle
\frac{1}{2} \left( \frac{2p-1}{2m}- \frac{2q-1}{2n} -N \right) \right)
\left\vert
\begin{array}{c}
N \\
(p,q)
\end{array}
, k \right>, \eqno(34.38)$$
$$S_+\left\vert
\begin{array}{c}
N \\
(p,q)
\end{array}
, k \right> = \sqrt{ \Phi^N_{(p,q)}(k+1)} \left\vert
\begin{array}{c}
N \\
(p,q)
\end{array}
, k +1\right>,\eqno(34.39)$$
$$S_-\left\vert
\begin{array}{c}
N \\
(p,q)
\end{array}
, k \right> = \sqrt{ \Phi^N_{(p,q)}(k)} \left\vert
\begin{array}{c}
N \\
(p,q)
\end{array}
, k -1\right>. \eqno(34.40)$$

\subsection{ The ``angular momentum'' quantum number}

It is worth noticing that the operators $S_0,S_\pm$ do not correspond to a
generalization of the angular momentum, $S_0$ being the operator
corresponding to the Fradkin operator $S_{xx}-S_{yy}$ \cite{Hig309,Lee489}.
The corresponding ``angular momentum'' is defined by:
$$L_0=-i\left(S_+-S_-\right).\eqno(34.41) $$
The ``angular momentum'' operator commutes with the Hamiltonian:
$$\left[ H,L_0 \right]=0. \eqno(34.42)$$
Let $\vert \ell> $ be the eigenvector of the operator $L_0$ corresponding to
the eigenvalue $\ell$. The general form of this eigenvector can be given by:
$$\vert \ell > = \sum\limits_{k=0}^N \frac{i^k c_k}{\sqrt{[k]!}} \left\vert
\begin{array}{c}
N \\
(p,q)
\end{array}
, k \right>. \eqno(34.43)$$

In order to find the eigenvalues of $L_0$ and the coefficients $c_k$ we use
the Lanczos algorithm \cite{Lan255}, as formulated in \cite{Flo331}. From
eqs (34.39) and (34.40) we find
$$L_0|\ell >=\ell |\ell >=\ell \sum\limits_{k=0}^N\frac{i^kc_k}{\sqrt{[k]!}}%
\left|
\begin{array}{c}
N \\
(p,q)
\end{array}
,k\right\rangle = $$
$$=\frac 1i\sum\limits_{k=0}^{N-1}\frac{i^kc_k\sqrt{\Phi _{(p,q)}^N(k+1)}}{%
\sqrt{[k]!}}\left|
\begin{array}{c}
N \\
(p,q)
\end{array}
,k+1\right\rangle -\frac 1i\sum\limits_{k=1}^N\frac{i^kc_k\sqrt{\Phi
_{(p,q)}^N(k)}}{\sqrt{[k]!}}\left|
\begin{array}{c}
N \\
(p,q)
\end{array}
,k-1\right\rangle
\eqno(34.44)$$
{}From this equation we find that:
$$c_k=(-1)^k 2^{-k/2}H_k(\ell /\sqrt{2})/{\cal N},
\quad {\cal N}^2= \sum\limits_{n=0}^N 2^{-n}H_n^2(\ell /\sqrt{2})/n!
\eqno(34.45)$$
where the function $H_k(x)$ is a generalization of the ``Hermite''
polynomials (see also \cite{BDEF150,KZ121}), satisfying the recurrence
relations:
$$ H_{-1}(x)=0,\quad H_0(x)=1, \eqno(34.46)$$
$$ H_{k+1}(x)=2xH_k(x)-2\Phi _{(p,q)}^N(k)H_{k-1}(x), \eqno(34.47)$$
and the ``angular momentum'' eigenvalues $\ell $ are the roots of the
polynomial equation:
$$ H_{N+1}(\ell /\sqrt{2})=0. \eqno(34.48)$$
Therefore for a given value of $N$ there are $N+1$ ``angular momentum''
eigenvalues $\ell $, symmetric around zero (i.e. if $\ell $ is an ``angular
momentum'' eigenvalue, then $-\ell $ is also an ``angular momentum''
eigenvalue). In the case of the symmetric harmonic oscillator ($m/n=1/1$)
these eigenvalues are uniformly distributed and differ by 2. In the general
case the ``angular momentum'' eigenvalues are non-uniformly distributed. For
small values of $N$ analytical formulae for the ``angular momentum''
eigenvalues can be found \cite{BDEF150}. Remember that to each value of $N$
correspond $m\cdot n$ energy levels, each with degeneracy $N+1$.

In order to have a formalism corresponding to the one of the isotropic
oscillator, let us introduce  for every $N$ and
$(m,n,p,q)$ an ordering of the ``angular momentum'' eigenvalues
$$ \ell_\mu ^{L,m,n,p,q}, \quad \mbox{where} \quad L=N
\quad \mbox{and} \quad \mu=-L,-L+2,\ldots,L-2,L,\eqno(34.49)$$
by assuming that:
$$ \ell_\mu ^{L,m,n,p,q} \le \ell_\nu^{L,m,n,p,q} \quad \mbox{if} \quad
\mu < \nu, \eqno(34.50)$$
the corresponding eigenstates being given by:
$$\AVector{L}{\mu}{m,n,p,q}=
\sum\limits_{k=0}^L
\frac{(-i)^k H_k(\ell_\mu^{L,m,n,p,q)} /\sqrt{2}) }
{{\cal N} \sqrt{2^{k/2}[k]! }}
\CVector{N}{(p,q)}{k}
=
\sum\limits_{k=0}^L d_{k+1} \CVector{N}{(p,q)}{k}
\eqno(34.51)$$
The above vector elements constitute  the analogue corresponding
to the  basis of ``sphe\-rical harmonic'' functions of the usual oscillator.
The calculation of the ``angular momentum'' eigenvalues
of eq. (34.49)
and the coefficients $d_1,d_2,\ldots,d_{L+1}$
 in the expansion of eq. (34.51) is a
quite difficult task. The existence of  general analytic expressions for
these quantities is not obvious. The first few ``angular momentum'' eigenvalues
are given by:
$$\ell^{1,m,n,p,q}_{\pm 1}=
\pm
\sqrt{
\frac{1}{m^m n^n} \frac{ \Gamma (m+p) }{\Gamma (p) }
\frac{ \Gamma(n+q) }{\Gamma (q) }  }, \eqno(34.52)$$
and
$$\ell^{2,m,n,p,q}_{0}=0,\eqno(34.53)$$
$$ \ell^{2,m,n,p,q}_{\pm 2}=
\pm
\sqrt{
\frac{1}{m^m n^n}
\left( \frac{ \Gamma (m+p) }{\Gamma (p) }
\frac{ \Gamma(2n+q) }{\Gamma (n+q) }
+\frac{ \Gamma (2m+p) }{\Gamma (m+p) }
\frac{ \Gamma(n+q) }{\Gamma (q) } \right)
  }
\eqno(34.54)$$
For $L>2$ the analytic expressions of the angular momentum eigenvalues and
the coefficients $d_k$ are longer, but their calculation
is a straightforward task. Numerical results for these quantities in the cases
of frequency ratios 1:2 and 1:3 are given in \cite{BDKLu2}.

After working out a few examples (see \cite{BDKLu2} for details) one finds out
the following points:

i) In the basis described by eqs. (34.16)-(34.19) it is a trivial matter to
distinguish the states belonging to the same irrep for any $m:n$ ratio,
while in the Cartesian basis this is true only in the 1:1 case.

ii) In the 1:2 case the irreps have degeneracies 1, 1, 2, 2, 3,
3, 4, 4, \dots, i.e. ``two copies'' of the u(2) degeneracies 1, 2, 3, 4,
\dots are obtained.

iii) In the 1:3 cases the degeneracies are 1, 1, 1, 2, 2, 2, 3, 3, 3, \dots,
i.e. ``three copies'' of the u(2) degeneracies are obtained.

iv) It can be easily seen that the 1:n case corresponds to ``n copies'' of
the u(2) degeneracies.

v) Cases with both $m$, $n$ different from unity show more complicated
degeneracy patterns, also correctly reproduced by the above formalism. In
the 2:3 case, for example, the degeneracy pattern is 1, 1, 1, 1, 1, 2, 1, 2,
2, 2, 2, 3, 2, 3, 3, \dots.

vi) The only requirement for each energy eigenvalue to correspond to one
irrep of the algebra is that $m$ and $n$ have to be mutually prime numbers.
If $m$ and $n$ possess a common divisor other than 1, then some energy
eigenvalues will correspond to sums of irreps, i.e. to reducible
representations.

vii) The difference between the formalism used in
\cite{DZ1203,Ven190,MV57,Naz154}
and the one used here is that in the former case for given $m$ and $n$
appropriate
operators have to be introduced separately for each set of $(p,q)$ values,
while in the present case only one set of operators is introduced.

\subsection{Multisections of the isotropic oscillator}

In \cite{RSP1950} the concept of bisection of an isotropic harmonic oscillator
has been introduced. One can easily see that multisections (trisections,
tetrasections, \dots) can be introduced in a similar way. The degeneracies
of the various anisotropic oscillators can then be obtained from these
of the isotropic oscillator by using appropriate multisections.

Using the Cartesian notation $(n_x, n_y)$ for the states of the isotropic
harmonic oscillator we have the following list:

N=0: (00)

N=1: (10) (01)

N=2: (20) (02) (11)

N=3: (30) (03) (21) (12)

N=4: (40) (04) (31) (13) (22)

N=5: (50) (05) (41) (14) (32) (23),
\hfill\break
where $N=n_x+n_y$.  The corresponding degeneracies are 1, 2, 3, 4, 5, 6, \dots,
i.e. these of u(2).

A bisection can be made by choosing only the states with $n_y$=even. Then the
following list is obtained:

N=0: (00)

N=1: (10)

N=2: (20) (02)

N=3: (30) (12)

N=4: (40) (04) (22)

N=5: (50) (14) (32).

The degeneracies are 1, 1, 2, 2, 3, 3, \dots, i.e. these of the anisotropic
oscillator with ratio of frequencies 1:2. The same degeneracies are obtained
by choosing the states with $n_y$=odd. Therefore a {\bf bisection} of the
isotropic oscillator, distinguishing states with mod$(n_y,2)=0$ and states
with mod$(n_y,2)=1$, results in two interleaving sets of levels of the
1:2 oscillator.

By analogy, a {\bf trisection} can be made by distinguishing states with
mod$(n_y,3)=0$, or mod$(n_y,3)=1$, or mod$(n_y,3)=2$. One can easily see
that in this case three interleaving sets of states of the 1:3 oscillator,
having degeneracies 1, 1, 1, 2, 2, 2, 3, 3, 3, \dots, occur.

Similarly a {\bf tetrasection} can be made by distinguishing states with
mod$(n_y,4)=0$, or mod$(n_y,4)=1$, or mod$(n_y,4)=2$, or mod$(n_y,4)=3$.
The result is four interleaving sets of states of the 1:4 oscillator,
having degeneracies 1, 1, 1, 1, 2, 2, 2, 2, 3, 3, 3, 3, \dots.

By bisecting $n_x$ and trisecting $n_y$ one is left  with six interleaving
sets of states with degeneracies 1, 1, 1, 1, 1, 2, 1, 2, 2, 2, 2, 3, 2, 3,
3, \dots, i.e. degeneracies of the 2:3 oscillator.

By bisecting (or trisecting, tetresecting, etc) both $n_x$ and $n_y$ one is
obtaining the original u(2) degeneracies of the isotropic oscillator.

It is therefore clear that the degeneracies of all $m:n$ oscillators
can be obtained from these of the isotropic oscillator by appropriate
multisections. In particular:

i) The degeneracies of the $1:n$ oscillator can be obtained from these
of the 1:1 (isotropic) oscillator by $n$-secting $n_y$ or $n_x$.

ii) The degeneracies of the $m:n$ oscillator can be obtained from these
of the 1:1 oscillator by $m$-secting $n_x$ and $n$-secting $n_y$.

\subsection{ Connection to W$_3^{(2)}$ }

For the special case $m = 1$, $n=2$ it should be noticed that the  deformed
algebra received here coincides with the finite W algebra  W$_3^{(2)}$ \cite
{Tj1,Tj2,Tj3,Tj4}. The commutation relations of the W$_3^{(2)}$ algebra are
$$ [H_W, E_W]= 2 E_W, \qquad [H_W, F_W]= -2 F_W, \qquad [E_W,F_W]= H^2_W +
C_W, \eqno(34.55)$$
$$[C_W,E_W]=[C_W,F_W]=[C_W,H_W]=0,\eqno(34.56)$$
while in the $m=1$, $n=2$ case one has the relations
$$[{\cal N}, {\cal A}^{\dagger}]= {\cal A}^{\dagger}, \qquad [{\cal N},
{\cal A%
}]= -{\cal A}, \qquad [{\cal A},{\cal A}^{\dagger}]= 3 S_0^2 -{\frac{H^2}{4}}
- H S_0 +{\frac{3 }{16}},\eqno(34.57)$$
$$ [H, {\cal A}^{\dagger}]= [H, {\cal A}]= [H, S_0] =0,\eqno(34.58)$$
with $S_0= {\cal N}+u$ (where $u$ a constant). It is easy to see that the
two sets of commutation relations are equivalent by making the
identifications
$$F_W= \sigma {\cal A}^{\dagger}, \qquad E_W= \rho {\cal A}, \qquad H_W= -2
S_0 + k H, \qquad C_W= f(H), \eqno(34.59)$$
with
$$\rho \sigma = {\frac{4}{3}}, \qquad k={\frac{1}{3}}, \qquad f(H)=
-{\frac{4}{9%
}} H^2 +{\frac{1}{4}}.\eqno(34.60)$$

\subsection{Discussion}

In conclusion, the two-dimensional anisotropic quantum harmonic oscillator
with rational ratio of frequencies equal to $m/n$, is described dynamically
by a deformed version of the u(2) Lie algebra, the order of this algebra
being $m+n-1$. The representation modules of this algebra can be generated
by using the deformed oscillator algebra. The energy eigenvalues are
calculated by the requirement of the existence of finite dimensional
representation modules. An ``angular momentum'' operator useful for
labelling degenerate states has also been constructed.  The algebras
obtained in the special cases with $1:n$ ratios are shown to correspond to
generalized parafermionic oscillators.
In the special case
of $m:n=1:2$ the resulting algebra has been identified as the finite W
algebra W$_3^{(2)}$. Finally, it is demonstrated how the degeneracies
of the various $m:n$ oscillators can be obtained from these of the
isotropic oscillator by appropriate multisections.

The extension of the present method to the three-dimensional anisotropic
quantum harmonic oscillator is already receiving attention, since it is of
clear interest in the study of the symmetries underlying the structure of
superdeformed and hyperdeformed nuclei \cite{Mot522,Rae1343}.

\section{The use of quantum algebras in molecular structure}

The techniques developed in this article can be applied in very similar
ways in describing properties of diatomic and polytomic molecules. A brief
list will be given here.

1) Rotational spectra of diatomic molecules have been described in terms of
the su$_q$(2) model \cite{Iwao363,BRRS300,ZS17,CY254,Est5614}.
 As in the case of nuclei, $q$ is a phase factor
($q=e^{i\tau}$). In molecules $\tau$ is of the order of 0.01.
The use of the su$_q$(2) symmetry leads to a partial summation of the Dunham
expansion describing the rotational--vibrational spectra of diatomic
molecules \cite{BRRS300}. Molecular backbending (bandcrossing) has also been
described in this framework \cite{MRM1115}. Rotational spectra of symmetric
top molecules have also been considered \cite{Chang1400,KN221} in the
framework of the su$_q$(2) symmetry.

2) Vibrational spectra of diatomic molecules have been described in terms of
$q$-deformed anharmonic oscillators having the
su$_q$(1,1) \cite{BAR403} or the u$_q$(2) $\supset$ o$_q$(2) \cite{BRF221}
symmetry, as well as in terms of generalized deformed oscillators
similar to the ones used in sec. 26 \cite{BD75,CGY192,CGY183}.
These results, combined with 1), lead
to the full summation of the Dunham expansion \cite{BAR403,BRF221}.
A two-parameter deformed anharmonic oscillator with u$_{qp}$(2) $\supset$
o$_{qp}$(2) symmetry has also been considered \cite{ZZH1053}.

3) The physical content of the anharmonic oscillators mentioned in 2)
has been clarified by constructing WKB equivalent potentials (WKB-EPs)
\cite{BDKJMP,BDK191} and
classical equivalent potentials \cite{BDK6153}, similar to the
ones of sec. 13. The results have been corroborated by the study of the
relation between su$_q$(1,1) and the anharmonic oscillator with  $x^4$
anharminicities \cite{NQ1699}.
The WKB-EP corresponding to the su$_q$(1,1) anharmonic
oscillator has been connected to a class of Quasi-Exactly Soluble Potentials
(QESPs) \cite{BDM199}.

4) Generalized deformed oscillators
giving the same spectrum as the Morse potential \cite{BD150} and the modified
P\"oschl--Teller potential \cite{Das2261},  as well as a deformed oscillator
containing them as special cases \cite{Jan233,Jan180} have also been
constructed.
In addition,  $q$-deformed versions of the Morse potential have been given,
either by using the so$_q$(2,1) symmetry \cite{CG941} or by solving a
$q$-deformed Schr\"odinger equation for the usual Morse potential
\cite{DD015}.

5) A $q$-deformed version of the vibron model for diatomic molecules has been
constructed \cite{ABS1088,CL317,Pan47,GC3123}, in a way similar to that
described in sec. 29.

6) For vibrational spectra of polyatomic molecules a model of $n$ coupled
generalized deformed oscillators has been built \cite{BD3611}, containg the
approach of Iachello and Oss \cite{IO2976,IO500} as a special case.

7) Quasi-molecular resonances in the systems $^{12}$C+$^{12}$C and
$^{12}$C+$^{16}$O have been described in terms of a $q$-deformed oscillator
plus a rigid rotator \cite{CY325}.

A review of several of the above topics, accompanied by a detailed and
self-contained introduction to quantum algebras, has been given by Raychev
\cite{RayAQC}.

\section{Outlook}

Nobody likes binding himself by statements concerning the future. However,
we attempt to give here a partial list of open problems, roughly following
the order of the material in this review:

1) The list of physical systems which can be classified under a generalized
deformed su(2) symmetry (sec. 17) or under a generalized deformed
parafermionic oscillator scheme (sec. 18) can be enlarged. Self-similar
potentials and isospectral oscillator Hamiltonian systems could probably
be related to these symmetries.

2) The description of B(E2) values in terms of the su$_q$(2) model
attempted so far (sec. 21) takes into account only the kinematical deformation
effects. In order to take into account dynamical deformation effects, one
has to build a larger algebra, of which the quadrupole operators will be
members. (This will then probably be a deformed version of an su(3)
algebra.) These quadrupole operators should behave as irreducible tensors
of rank 2 under su$_q$(2).

3) The su$_q$(2) prediction about B(E2) values increasing with increasing
angular momentum $J$, supported by the predictions of other models as well
(see sec. 21), requires further testing against detailed experimental data.

4) The construction of Clebsch-Gordan coefficients for the subclass of
generalized deformed su(2) algebras for which this could be possible
(sees secs 17, 25) is an open problem.

5) The su$_q$(3) $\supset$ so$_q$(3) decomposition for su$_q$(3) irreps
other than the completely symmetric ones (see subsec. 27.3 for the current
state of the art) remains an open problem, the
solution of which is necessary for developing a deformed version of the
su(3) limit of the Interacting Boson Model.

6) Realizations of multi-level shells models in terms of deformed bosons
(see secs 31--33 for some one- and two-level cases) should be further
pursued.

7) The symmetry algebras of the various 3-dim anisotropic harmonic oscillators
with rational
ratios of frequencies should be worked out, since they are of interest in
relation to superdeformed and hyperdeformed nuclei, and possibly to
deformed atomic clusters (see sec. 34 for references). A deformed u(3)
algebra should occur in this case, which could serve as the basis for
building a deformed analog of the Elliott model
\cite{Ell128,Ell562,Ell557,Har67}
suitable for superdeformed nuclei.

8) In molecular physics (sec. 35) the study of vibrations of highly symmetric
polyatomic molecules (including fullerenes) by these techniques is of
interest.

\bigskip
{\bf Acknowledgements}

\medskip
The authors are grateful to Constanca Provid\^encia for a careful reading of
the manuscript.
One of the authors (DB) has been supported by CEC under contract
ERBCHBGCT930467.

\end{document}